%% file: ANA-HMBS-2024-45-PAPER.tex
\pdfoutput=1
\documentclass[PAPER, atlasdraft=false, cernpreprint, texlive=2023, UKenglish, texmf, orcidlogo]{atlasdoc}
\usepackage{atlaspackage}
\usepackage{atlasbiblatex}

\usepackage{atlasphysics}
\graphicspath{{logos/}{figures/}}

\usepackage{ANA-HMBS-2024-45-PAPER-defs}
\usepackage{multirow}

\AtlasTitle{Search for a heavy charged Higgs boson decaying into a $W$ boson and a Higgs boson in final
states with leptons and $b$-jets in $\sqrt{s} = 13\, \TeV$ $pp$ collisions with the ATLAS detector}

\AtlasAbstract{
This article presents a search for a heavy charged Higgs boson produced in association with a top quark and a
bottom quark, and decaying into a $W$ boson and a $125\,\GeV$ Higgs boson $h$. The search is performed in final
states with one charged lepton, missing transverse momentum, and jets using proton--proton collision data at
$\sqrt{s} = 13\,\TeV$ recorded with the ATLAS detector during Run 2 of the LHC at CERN.
This data set corresponds to a total integrated luminosity of 140\,fb$^{-1}$. The search is
conducted by examining the reconstructed invariant mass distribution of the $Wh$ candidates for evidence of a
localised excess in the charged Higgs boson mass range from $250\,\GeV$ to $3\,\TeV$. No significant excess of data
over the expected background is observed and $95\%$ confidence-level upper limits between $2.8\,$pb and $1.2\,$fb
are placed on the production cross-section times branching ratio for charged Higgs bosons decaying into $Wh$.
}

\PreprintIdNumber{CERN-EP-2024-290}

\AtlasJournalRef{JHEP 02 (2025) 143}
\AtlasDOI{10.1007/JHEP02(2025)143}

\hypersetup{pdftitle={ATLAS document},pdfauthor={The ATLAS Collaboration}}

\begin{document}

\maketitle

\tableofcontents

\section{Introduction}\label{sec:intro}

The observation of the Higgs boson at the Large Hadron Collider (LHC) was a great success of the ATLAS
and CMS collaborations~\cite{HIGG-2012-27,CMS-HIG-12-028}. Following its discovery, numerous
studies have been performed to establish whether it is a Standard Model (SM) particle or
rather the first observed physical state of an extended scalar sector.

Searches for an extended scalar sector are crucial as numerous models of new physics beyond the SM
require additional scalar states. For example, two Higgs doublets~\cite{Branco:2011iw,Gunion:2002zf} are required in the minimal supersymmetric extension
of the SM, while Higgs triplets~\cite{PhysRevD.22.2860,PhysRevD.22.2227,LAZARIDES1981287,CHANOWITZ1985105,PhysRevD.42.1673} are required in
models with a type-II seesaw mechanism. In addition,
an extended scalar sector can modify the electroweak phase transition and facilitate
baryogenesis~\cite{Cohen:1993nk, Turok:1990zg}, enhance vacuum stability, provide a dark matter
candidate~\cite{Abdallah:2015ter} or provide a solution to the strong CP problem (i.e. predict axions)~\cite{Kim:2008hd}.
In short, extending the scalar sector provides solutions to some of the open questions in the SM.

Various theories predicting an extended scalar sector postulate also the existence of at least one set of charged
Higgs bosons in addition to the observed neutral one, such as models that add a second doublet or one or more triplets
to the scalar sector. The main production and decay modes of these new particles are strongly model dependent. For example, in the
alignment limit of the two-Higgs-doublet model (2HDM)~\cite{Bernon:2015qea}, the dominant production mode, for charged Higgs boson
masses larger than the sum of the top and the bottom quark masses, is expected to be in association with a top quark and a bottom quark
($tbH^\pm$),\footnote{The notation $tbH^\pm$ is used to represent the $\bar{t} bH^{+}$ and $t \bar{b} H^{-}$ processes. In general, the difference between particles and antiparticles is to be understood from the context.}
while the dominant charged Higgs boson decay modes are via $H^{\pm}\rightarrow tb$ or $H^{\pm}\rightarrow \tau^\pm\nu$.
However, there are also several models such as the next-to-minimal two-Higgs-doublet model (N2HDM)~\cite{PhysRevD.89.075009,Muhlleitner:2016mzt},
the three-Higgs-doublet Model (3HDM)~\cite{Keus:2013hya} or the Georgi-Machacek model~\cite{GEORGI1985463} in which other
decay and production modes become important. The studies presented in this article search for charged Higgs bosons decaying via
$H^\pm\rightarrow W^\pm h$, where $h$ is a Higgs boson with mass $m_{h}=125\,\GeV$. This decay mode is predicted to have significant branching
ratios by various extended scalar sector models~\cite{Biekotter:2019drs,Boto:2021qgu,Ghosh:2019qie,Keus:2015hva}.

The ATLAS and CMS collaborations searched for charged Higgs bosons in proton--proton ($pp$) collisions at $\sqrt{s} = 7, 8$ and $13\, \TeV$
with data samples corresponding to integrated luminosities ranging from $2.9$ up to $140$ fb$^{-1}$, probing the mass range below the
top-quark mass in the $\tau^\pm\nu$~\cite{HIGG-2012-09,HIGG-2012-22,HIGG-2013-30,CMS-HIG-11-019,CMS-HIG-14-023,HIGG-2016-11},
$cs$~\cite{HIGG-2012-10,CMS-HIG-13-035}, and $cb$~\cite{CMS-HIG-16-030,ATLAS:2023bzb}
decay modes, as well as above the top-quark mass in the $\tau^\pm\nu$~\cite{HIGG-2015-11} and $tb$~\cite{HDBS-2018-51,CMS-HIG-18-004,CMS-HIG-18-015} decay modes.
Searches for $H^\pm \rightarrow W^{\pm} Z$ decays have been performed in the vector-boson-fusion (VBF) production
mode~\cite{HIGG-2014-13,CMS-HIG-16-027,CMS-HIG-20-017}. Searches for doubly-charged Higgs bosons have also been performed~\cite{CMS-HIG-12-005,EXOT-2018-34,HDBS-2019-06,CMS-HIG-20-017,STDM-2018-32}.
Charged Higgs boson decays via $H^{\pm}\rightarrow W^{\pm}h$ have been so far not yet searched for by either the ATLAS or CMS collaborations.

\begin{figure}[t]
\centering
\subfloat[]{\includegraphics[width=0.425\textwidth]{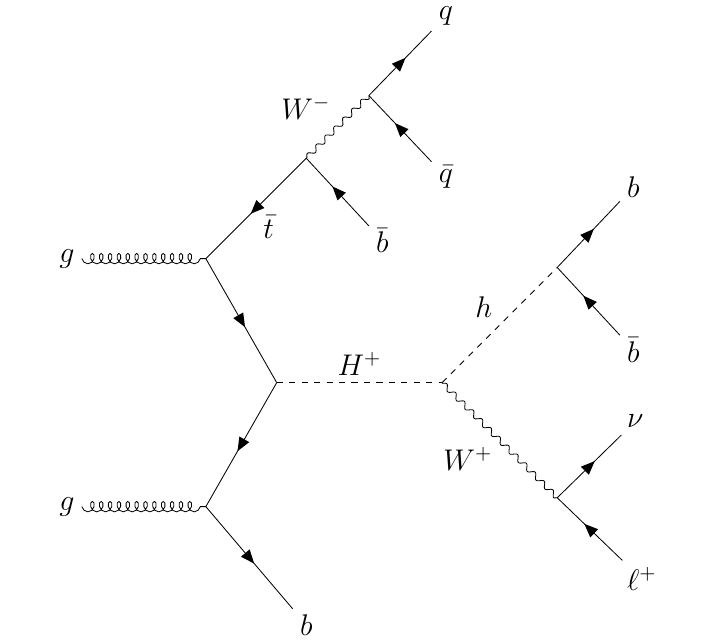}}
\subfloat[]{\includegraphics[width=0.425\textwidth]{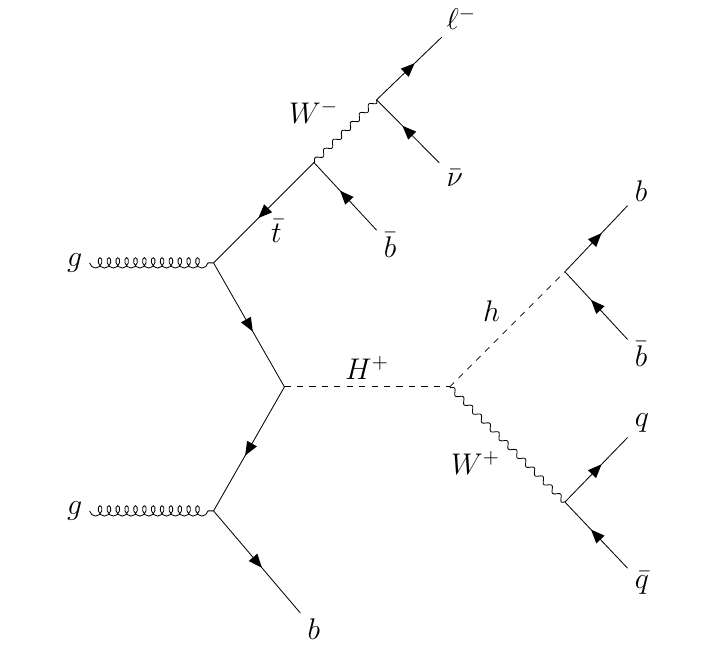}}
\caption{Representative lowest-order Feynman diagrams of $pp\rightarrow \bar{t} bH^{+}$ production and subsequent decays via
(a) $H^{+}\rightarrow W^{+}h\rightarrow \ell^{+}\nu b\bar{b}$ and (b) $H^{+}\rightarrow W^{+}h\rightarrow q\bar{q}b\bar{b}$.}
\label{fig:FeynmanDiagrams_Hplus_Wh}
\end{figure}

This article describes a first search for a charged Higgs boson produced in association with a top quark
and a bottom quark with subsequent decays of the charged Higgs boson via $H^{\pm}\rightarrow W^{\pm}h\rightarrow \ell^\pm\nu b\bar{b}$ or
$H^{\pm}\rightarrow W^{\pm}h\rightarrow q\bar{q}b\bar{b}$. The search is performed in events that
are consistent with the final state $\ell^\pm\nu b\bar{b}b\bar{b}q\bar{q}$ (with $\ell = e, \mu$),
where the charged lepton can originate either from the decay chain of the charged Higgs boson or of the associated top quark.
Representative lowest-order Feynman diagrams of these processes are shown in Figure~\ref{fig:FeynmanDiagrams_Hplus_Wh}.

To ensure high sensitivity to both low- and high-mass resonances, two different analysis
techniques are used. At low charged Higgs boson masses, when the final state particles have a relatively low
Lorentz-boost, the decay products of the neutral Higgs boson and hadronically decaying $W$ boson are
reconstructed via individual small-radius jets (such decays are referred to as `resolved'). At high
charged Higgs boson masses, when the final state particles have a relatively large Lorentz-boost, the neutral Higgs boson
and the hadronically decaying $W$ boson are reconstructed as single large-radius jets (such decays are
referred to as `merged').

The search for charged Higgs bosons is performed by probing for a localised excess of events in the invariant
mass distribution of the reconstructed $\ell^\pm\nu b\bar{b}$ and $q\bar{q}b\bar{b}$ systems.
This is achieved through a simultaneous profile likelihood fit on the invariant mass distribution
obtained in selected signal and control regions. The signal and control regions are defined based on requirements on kinematic
properties of the final-state particles and event-level quantities. Multivariate analysis techniques
are used to improve the background rejection and to reconstruct the decays of the charged Higgs boson
candidates. The major backgrounds are modelled using simulation, while their normalisations are determined by a profile-likelihood fit to data.
The results are presented as upper limits at $95\%$ confidence level (CL) on the production cross-section times branching ratio
$\sigma\left(pp\rightarrow tb H^\pm\right)\times \mathcal{B} \left( H^\pm \rightarrow W^\pm h \right) \times \mathcal{B}\left(h\rightarrow b\bar{b}\right)$
of the signal process. In this article, charged Higgs boson mass hypotheses are probed in a range from $250\,\GeV$ to $3\,\TeV$.

%
\section{ATLAS detector}\label{sec:detector}

%
%
%

%
\newcommand{\AtlasCoordFootnote}{%
ATLAS uses a right-handed coordinate system with its origin at the nominal interaction point (IP)
in the centre of the detector and the \(z\)-axis along the beam pipe.
The \(x\)-axis points from the IP to the centre of the LHC ring,
and the \(y\)-axis points upwards. Polar coordinates \((r,\phi)\) are used in the transverse plane,
\(\phi\) being the azimuthal angle around the \(z\)-axis. The pseudorapidity is defined in terms of
the polar angle \(\theta\) as \(\eta = -\ln \tan(\theta/2)\) and is equal to the rapidity
$ y = \frac{1}{2} \ln \left( \frac{E + p_z}{E - p_z} \right) $ in the relativistic limit.
Angular distance is measured in units of \(\Delta R \equiv \sqrt{(\Delta y)^{2} + (\Delta\phi)^{2}}\).}

The ATLAS detector~\cite{PERF-2007-01} at the LHC covers nearly the entire solid angle around the collision point.\footnote{\AtlasCoordFootnote}
It consists of an inner tracking detector surrounded by a thin superconducting solenoid, electromagnetic and hadronic calorimeters,
and a muon spectrometer incorporating three large superconducting air-core toroidal magnets.

The inner-detector system (ID) is immersed in a \qty{2}{\tesla} axial magnetic field
and provides charged-particle tracking in the range \(|\eta| < 2.5\).
The high-granularity silicon pixel detector typically provides four measurements per track, the first hit generally being in the
insertable B-layer (IBL) installed before Run~2~\cite{ATLAS-TDR-19,PIX-2018-001}. It is followed by the SemiConductor Tracker (SCT),
which usually provides eight measurements per track. These silicon detectors are complemented by the transition radiation tracker (TRT),
which enables radially extended track reconstruction up to \(|\eta| = 2.0\). The TRT also provides electron identification information
based on the fraction of hits (typically 30 in total) above a higher energy-deposit threshold corresponding to transition radiation.

The calorimeter system covers the pseudorapidity range \(|\eta| < 4.9\).
Within the region \(|\eta|< 3.2\), electromagnetic calorimetry is provided by barrel and
endcap high-granularity lead/liquid-argon (LAr) calorimeters,
with an additional thin LAr presampler covering \(|\eta| < 1.8\)
to correct for energy loss in material upstream of the calorimeters.
Hadronic calorimetry is provided by the steel/scintillator-tile calorimeter,
segmented into three barrel structures within \(|\eta| < 1.7\), and two copper/LAr hadronic endcap calorimeters.
The solid angle coverage is completed with forward copper/LAr and tungsten/LAr calorimeter modules
optimised for electromagnetic and hadronic energy measurements respectively.

The muon spectrometer (MS) comprises separate trigger and high-precision tracking chambers measuring the deflection of muons
in a magnetic field generated by the superconducting air-core toroidal magnets. The field integral of the toroids ranges
between \num{2.0} and \qty{6.0}{\tesla\metre} across most of the detector. Three layers of precision chambers, each consisting
of layers of monitored drift tubes, cover the region \(|\eta| < 2.7\), complemented by cathode-strip chambers in the forward region,
where the background is highest. The muon trigger system covers the range \(|\eta| < 2.4\) with resistive-plate chambers in the barrel,
and thin-gap chambers in the endcap regions.

The luminosity is measured mainly by the LUCID--2~\cite{LUCID2} detector that records Cherenkov light produced in the quartz windows of
photomultipliers located close to the beampipe.

Events are selected by the first-level trigger system implemented in custom hardware,
followed by selections made by algorithms implemented in software in the high-level trigger~\cite{TRIG-2016-01}.
The first-level trigger accepts events from the \qty{40}{\MHz} bunch crossings at a rate below \qty{100}{\kHz},
which the high-level trigger further reduces in order to record complete events to disk at about \qty{1}{\kHz}.

A software suite~\cite{SOFT-2022-02} is used in data simulation, in the reconstruction
and analysis of real and simulated data, in detector operations, and in the trigger and data acquisition
systems of the experiment.


%
\section{Data and simulated event samples}\label{sec:Samples}

The $pp$ collision data at $\sqrt{s} = 13\,\TeV$ used in the analysis were recorded with the ATLAS detector between 2015 and 2018, and correspond
to a total integrated luminosity of $140.0\pm1.2$\,fb$^{-1}$~\cite{DAPR-2021-01}.
The data are required to satisfy criteria that ensure that the detector was in good operating condition~\cite{DAPR-2018-01}.
Monte Carlo (MC) simulation samples were used to model the background and signal processes, as well as to derive
modelling uncertainties. The MC simulation samples were processed using either the \GEANT-based simulation of the
ATLAS detector geometry and response~\cite{SOFT-2010-01,GEANT4}
or fast simulation~\cite{ATL-PHYS-PUB-2010-013}, where the \GEANT simulation of the calorimeter response
is replaced by a detailed parameterisation of shower shapes. The simulated events were reconstructed using the same
algorithms as were used for the data events.

The signal process, i.e. the associated production of a charged Higgs boson, a bottom quark and a top quark,
was simulated using the matrix element (ME) generator \MGNLO[2.7.3]~\cite{MG5} in the 4-flavour scheme (4FS) at
next-to-leading-order (NLO) accuracy in QCD and the \NNPDF[3.0nlo]~\cite{NNPDF} set of parton distribution
functions (PDFs). \Pythia[8.244]~\cite{Sjostrand:2014zea} with the A14 set~\cite{ATL-PHYS-PUB-2014-021} of tuned parameters is used to decay the charged Higgs boson and to
model the parton shower (PS), hadronisation, and underlying event. The renormalisation and factorisation scales $\mu_R$ and $\mu_F$ were set
to $\frac{1}{3} \sum_{i} \sqrt{m_{i}^2 + p_\mathrm{T, i}^2}$, where $i$ runs over all final state particles used in the matrix
element calculation. The signal process is simulated using the FeynRules~\cite{Alloul:2013bka} model \texttt{2HDMtypeII}~\cite{Degrande:2014vpa}
using a narrow-width approximation. The choice of model is expected to have only small impact on the results of this search as long as
the narrow-width approximation is valid. While specific models may alter the cross-section times branching ratio values, they typically
do not affect the event kinematics of the $tbH^\pm$ process.
Seventeen signal samples were generated covering a mass range between $250~\GeV$ and $3~\TeV$.\footnote{In the mass range between $250\,\GeV$
and $400\,\GeV$, the signal samples are produced in $50\,\GeV$ steps, while in the ranges from $400\,\GeV$ to $1000\,\GeV$ and $1000\,\GeV$
to $2000\,\GeV$ step sizes of $100\,\GeV$ and $200\,\GeV$ were chosen. In addition, signal masses of $2500\,\GeV$ and $3000\,\GeV$ are
considered.} Fast detector simulation was employed for mass points below $500\,\GeV$ and the \GEANT-based simulation of the ATLAS detector
was used otherwise.\footnote{The decision to employ \GEANT-based simulation for masses above $500\,\GeV$ was motivated by the
onset of the merged analysis at this mass point and the limitations of the fast detector simulation to accurately describe the properties of
large-radius jets, which are essential to the merged analysis.} In the simulation of the signal processes, only the $H^{+}$ decay into $W^{+}h$ and
Higgs boson decaying to pairs of $b$-quarks were considered assuming a Higgs boson mass of $m_{h} = 125\,\GeV$. Other decay modes of the
$125\,\GeV$ Higgs boson were neglected, as their contributions to the signal and control regions (cf. Section~\ref{sec:AnalysisStrategy}) were an
order of magnitude lower than those for the $h\rightarrow b\bar{b}$ decay.

The production of top-quark pair (\ttbar) events was modelled using the \POWHEGBOX[v2]~\cite{Frixione:2007nw,Nason:2004rx,Frixione:2007vw,Alioli:2010xd}
generator in the five-flavour scheme (5FS) to calculate the ME at NLO accuracy in QCD, and the \NNPDF[3.0nlo] PDF set.
The \hdamp\ parameter\footnote{The \hdamp\ parameter is a resummation damping factor and one of the
parameters that control the matching of \POWHEG MEs to the parton shower, effectively regulating the high-\pt\ radiation against
which the \ttbar\ system recoils.} was set to $1.5\,m_{t}$~\cite{ATL-PHYS-PUB-2016-020}, where $m_{t}$ is the top-quark mass.
The top-quark decays are modelled using \MADSPIN~\cite{Frixione:2007zp,Artoisenet:2012st}.
The PS, hadronisation, and underlying event were modelled with the \Pythia[8.230] generator using the A14 set of
tuned parameters and the \NNPDF[2.3lo]~\cite{Ball:2012cx} PDF set. The top-quark pair events are normalised to the state-of-the-art cross-section
prediction calculated with \TOPpp[2.0]~\cite{Beneke:2011mq,Cacciari:2011hy,Baernreuther:2012ws,Czakon:2012zr,Czakon:2012pz,Czakon:2013goa,Czakon:2011xx}
at next-to-next-to-leading order (NNLO) in QCD, including the resummation of next-to-next-to-leading logarithmic (NNLL) soft-gluon terms.

The $t\bar{t}h$ sample was generated at NLO accuracy in QCD using the \POWHEGBOX[v2] generator in the 5FS, and the \NNPDF[3.0nlo] PDF set.
The \hdamp\ parameter was set to $\frac{3}{4}\cdot \left( 2m_{t} + m_{h} \right) = 352.5\,\GeV$ and the events were showered with \Pythia[8.230],
which used the A14 set of tuned parameters and the \NNPDF[2.3lo] PDF set. The $t\bar{t}h$ production cross-section is calculated at NLO
accuracy in both QCD and electroweak (EW) using \MGNLO, as reported in Ref.~\cite{deFlorian:2016spz}.
The production of $t\bar{t}V$ (with $V = W^\pm$ or $Z$) events was modelled using the \MGNLO[2.3.3] generator, which provides MEs at
NLO in QCD with the \NNPDF[3.0nlo] PDF set. The events were interfaced to \Pythia[8.210] using the A14 set of tuned parameters and the
\NNPDF[2.3lo] PDF set.

The associated production of a top quark and $W$ boson ($Wt$), and the $s$- and $t$-channel single-top-quark production were
modelled with the \POWHEGBOX[v2] generator at NLO in QCD using the 5FS and the \NNPDF[3.0nlo] set of PDFs. The diagram-removal (DR)
scheme~\cite{Frixione:2008yi} was used to remove interference and overlap with the production of top-quark pairs.
The events were interfaced to \Pythia[8.230], which used the A14 set of tuned parameters and the \NNPDF[2.3lo] set of PDFs.

Rare processes including top-quarks, such as $tZq$, $tWZ$, $thjb$, $tWh$, and $t\bar{t}t\bar{t}$, were also simulated
and accounted for, even though their contribution to any analysis region is lower than $1\%$ of the total background yields.
The \MGNLO generator and the \NNPDF PDF set were used to calculate the MEs for these processes. The events were interfaced to
\Pythia[8.2] using the A14 set of tuned parameters and the \NNPDF[2.3lo] set of PDFs. The MEs of the $tZq$ process was calculated
at leading-order (LO) accuracy in QCD, while the MEs of the other four processes were calculated at NLO accuracy in QCD.

A sample of $V$+jets events was simulated using \Sherpa[2.2.11]~\cite{Bothmann:2019yzt} with the \NNPDF[3.0nnlo]~\cite{NNPDF}
PDF set. The ME was calculated based on the \textsc{Comix}~\cite{Gleisberg:2008fv} and \OPENLOOPS~\cite{Buccioni:2019sur,Cascioli:2011va,Denner:2016kdg}
libraries at NLO accuracy in QCD for diagrams with up to two additional parton emissions, and
LO accuracy in QCD for diagrams with three, four or five additional parton emissions.
The \MEPSatNLO prescription~\cite{Hoeche:2011fd,Hoeche:2012yf,Catani:2001cc,Hoeche:2009rj} was used to merge the
ME and the \Sherpa PS~\cite{Schumann:2007mg}, which is based on a set of tuned parameters
developed by the \Sherpa authors. The $V$+jets event sample was normalised to match cross-section
predictions at NNLO accuracy in QCD calculated with \textsc{FEWZ}~\cite{Anastasiou:2003ds}.

Diboson ($VV$) events with decays into semileptonic final states were simulated using
\Sherpa[2.2.1], while events with decays into fully leptonic final states were simulated using \Sherpa[2.2.2].
Both samples include off-shell effects and Higgs boson contributions where appropriate. Diagrams with up
to one additional emission were calculated at NLO accuracy in QCD, while diagrams with two or three parton emissions
were described at LO accuracy. The ME calculations were matched and merged with the \Sherpa PS using the
\MEPSatNLO prescription. Virtual QCD corrections for the ME at NLO accuracy were provided by the \OPENLOOPS library.
Loop-induced diboson processes initiated via the $gg$ production mode were simulated at LO in QCD for
diagrams with up to one additional parton emission in the ME using \OPENLOOPS in \Sherpa[2.2.2].
For electroweak $VVjj$ production, the calculation of the ME was performed in the $G_{\mu}$-scheme~\cite{Denner:2003iy}
to describe the pure electroweak interactions at the electroweak scale. All diboson events were generated using
the \NNPDF[3.0nnlo] PDF set, along with the \Sherpa PS.

Finally, the production of a SM Higgs boson in association with a vector boson ($Vh$) was simulated using
\POWHEGBOX[v2], interfaced with \Pythia[8.212] for PS and non-perturbative effects. The \POWHEG prediction
is accurate to NLO in QCD for the production of $Vh$ plus one jet. The loop-induced $gg \rightarrow Zh$
process was generated separately at LO. The \PDFforLHC[15] PDF set~\cite{Butterworth:2015oua} and the AZNLO set of tuned
parameters~\cite{STDM-2012-23} of \Pythia[8.212] were used. The $gg \rightarrow Zh$ production cross-section was calculated at
NLO accuracy in QCD, including the resummation of next-to-leading logarithmic (NLL) soft-gluon terms~\cite{Harlander:2014wda}.
For the generation of $Vh$ events, the Higgs boson mass was set to $125\,\GeV$.\par

All simulated event samples include the effect of multiple $pp$ interactions in the same and neighbouring bunch
crossings (pile-up) by overlaying simulated minimum-bias events on each generated signal and background event.~\footnote{An average of 34 interactions per bunch
crossing were observed during Run 2 data taking.} The
minimum-bias events were simulated with the single-, double- and non-diffractive $pp$ processes of \Pythia[8.186]
using the A3 set of tuned parameters~\cite{ATL-PHYS-PUB-2016-017} and the \NNPDF[2.3lo] PDF set. \GEANT-based
simulations of the ATLAS detector were used for the production of the background samples (unless otherwise stated).
For all samples produced with \MADGRAPH  and \POWHEGBOX, the \EVTGEN[1.6.0] programme~\cite{Lange:2001uf} was used to
model the decays of bottom and charm hadrons. Simulated events were corrected to compensate for differences between data
and simulations regarding the energy (or momentum) scale and resolution of leptons and jets, the efficiencies for the
reconstruction, identification, isolation and triggering of leptons, and the tagging efficiency for heavy-flavour jets.

A summary of MC generators and programs used to model the signal and background processes is provided in Table~\ref{tab:mcsamples}.

\begin{table}[h!]
\caption{Overview of the simulation tools used to generate signal and background processes, and to model the
underlying event and parton shower (UEPS). The PDF sets are also summarised. The perturbative accuracy (in QCD and if
relevant in EW corrections) of the total cross-section is stated for each process. Alternative event generators
and configurations used to estimate systematic uncertainties are discussed in Section~\ref{sec:Systematics}.}
\centering
\small
\renewcommand{\arraystretch}{1.15}
\resizebox{1.0\textwidth}{!}{
\begin{tabular}{c c c c c}
\hline
\hline
Process            & Matrix element     & UEPS                & PDF set  &  Perturbative accuracy  \\
&                    &                     &          &  of total cross-section \\
\hline
$tbH^\pm(\rightarrow W^\pm h\rightarrow \ell^\pm\nu b\bar{b}, q\bar{q}b\bar{b})$ & \MGNLO[2.7.3]  & \Pythia[8.244] & \NNPDF[3.0nlo]  & NLO (QCD) \\
\hline
$\ttbar$ + jets                                       & \POWHEGBOX[v2] & \Pythia[8.230] & \NNPDF[3.0nlo]  & NNLO+NNLL (QCD)\\
\hline
$t\bar{t}W^\pm$                                     & \MGNLO[2.3.3]  & \Pythia[8.210] & \NNPDF[3.0nlo]  & NNLO (QCD) and NLO (EW) \\
$t\bar{t}Z$                                         & \MGNLO[2.3.3]  & \Pythia[8.210] & \NNPDF[3.0nlo]  & NLO+NNLL (QCD) \\
$t\bar{t}h$                                         & \POWHEGBOX[v2] & \Pythia[8.230] & \NNPDF[3.0nlo]  & NLO (QCD) and NLO (EW) \\
\hline
single top quark ($s$- and $t$-channels)             & \POWHEGBOX[v2] & \Pythia[8.230] & \NNPDF[3.0nlo]  & NLO (QCD)  \\
single top quark ($Wt$-channel)                      & \POWHEGBOX[v2] & \Pythia[8.230] & \NNPDF[3.0nlo]  & approx.~NNLO (QCD) \\
$tZq$                                                 & \MGNLO[2.3.3]  & \Pythia[8.210] & \NNPDF[3.0lo]   & NLO  (QCD)   \\
$tWZ$                                                 & \MGNLO[2.3.3]  & \Pythia[8.212] & \NNPDF[3.0nlo]  & NLO  (QCD)   \\
\hline
$t\bar{t}t\bar{t}$                                    & \MGNLO[2.6.2]  & \Pythia[8.230] & \NNPDF[3.1nlo]  & NLO (QCD)    \\
$thjb$                                                & \MGNLO[2.6.2]  & \Pythia[8.230] & \NNPDF[3.0nlo]  & NLO (QCD)    \\
$tWh$                                                 & \MGNLO[2.6.2]  & \Pythia[8.235] & \NNPDF[3.0nlo]  & NLO (QCD)    \\
$q\bar{q} \to Wh$                                     & \POWHEGBOX[v2] & \Pythia[8.212] & \PDFforLHC[15]  & NNLO (QCD) and NLO (EW) \\
$q\bar{q} \to Zh$                                     & \POWHEGBOX[v2] & \Pythia[8.212] & \PDFforLHC[15]  & NNLO (QCD) and NLO (EW) \\
$gg \to Zh$                                           & \POWHEGBOX[v2] & \Pythia[8.212] & \PDFforLHC[15]  & NLO + NLL (QCD) \\
\hline
$W^\pm \to \ell^\pm\nu$, $Z\to \ell^\pm\ell^\mp$                      & \multicolumn{2}{c}{\Sherpa[2.2.11]} & \NNPDF[3.0nnlo] & NNLO (QCD)     \\
$qg/q\bar{q} \to VV \to \ell^\pm\ell^\mp/\ell^\pm\nu/\nu\nu + q\bar{q}$   & \multicolumn{2}{c}{\Sherpa[2.2.1]}  & \NNPDF[3.0nnlo] & NLO (QCD)   \\
$qg/q\bar{q} \to VV \to \ell^\pm\ell^\mp\ell^\pm\ell^\mp/\ell^\pm\nu\ell^\pm\ell^\mp/\ell^\pm\ell^\mp\nu\nu/\ell^\pm\nu\nu\nu$   & \multicolumn{2}{c}{\Sherpa[2.2.2]}  &  \NNPDF[3.0nnlo] & NLO (QCD)  \\
$gg \to VV$                                           & \multicolumn{2}{c}{\Sherpa[2.2.2]}  & \NNPDF[3.0nnlo] & NLO (QCD)  \\
$VVjj$                                                & \multicolumn{2}{c}{\Sherpa[2.2.2]}  & \NNPDF[3.0nnlo] & LO (QCD)   \\
\hline
\hline
\end{tabular}}
\label{tab:mcsamples}
\end{table}


%
\section{Event reconstruction}\label{sec:Reco}


Charged-particle tracks are reconstructed in the ID. They are required to have a transverse momentum (\pT) larger than
$\SI{500}{\MeV}$, $|\eta| < 2.5$, and at least seven hits in the pixel and SCT detectors.
A maximum of one (two) of the expected hits may be missing from the pixel (SCT) detector sensors,
and no more than one hit may be shared with other tracks~\cite{PERF-2015-08}. Collision vertices are reconstructed from at least two ID
tracks~\cite{ATL-PHYS-PUB-2015-026}. Among all vertices, the one with the highest $\pt^2$ sum of associated tracks is chosen to be
the primary vertex (PV) of the event. The properties of ID tracks are calculated relative to the PV.

Electrons are reconstructed from ID tracks originating from the PV that are matched to clusters of energy deposits in the
electromagnetic calorimeter~\cite{EGAM-2018-01}. The compatibility of the track and the PV is satisfied by a requirement on the
transverse impact parameter significance $|d_0|/\sigma_{d_0}$ < 5, and on the longitudinal impact parameter
$|z_0 \sin\theta|<0.5\,\unit{mm}$. Electron candidates must satisfy requirements on the electromagnetic shower shapes, track
quality, and track-cluster matching, using a likelihood-based approach~\cite{EGAM-2018-01}, where the \textit{Tight} operating point
is used for this study. Electrons are also required to have a \pt larger than $27~\GeV$
and $|\eta| < 2.47$, with the transition region between the barrel and endcap
electromagnetic calorimeters, $1.37 < |\eta| < 1.52$, being excluded. Finally, to reduce contributions from
hadrons mimicking electron signatures or non-prompt electrons from heavy-flavour decays or photon conversions,
a multivariate classifier is used. This classifier considers the energy deposits and charged-particle
tracks in a cone around the electron direction and information from secondary vertices~\cite{EGAM-2019-01}.

Muon reconstruction~\cite{MUON-2018-03} is based on matching MS tracks to ID tracks.  A combined fit is then performed incorporating the
information from the ID, MS and the energy deposits in the calorimeter system. Similar to electrons, muon candidates have to satisfy selection
requirements on the impact parameters: $|d_0|/\sigma_{d_0} < 3$ and $|z_0| < 0.5\,\unit{mm}$. Muon candidates are required to
have a minimum $p_\mathrm{T}$ of $27\,\GeV$ and lie within $|\eta| < 2.5$. Furthermore, they are required to satisfy
the \textit{Medium} identification operating point. However, candidates with $p_\mathrm{T} > 300\,\GeV$ must satisfy
tighter identification requirements in the MS to improve the $p_\mathrm{T}$ resolution~\cite{MUON-2018-03}.
To reduce contributions from non-prompt muons from heavy-flavour decays, muon candidates are required to be isolated
in the ID system using the \textit{TightTrackOnly} operating point~\cite{MUON-2018-03}. A muon is considered to be isolated
if the $\pt$ sum within a cone around the combined track is smaller than 0.06 times the muon’s transverse momentum, $p^{\mu}_\mathrm{T}$.
The size of the isolation cone is $\Delta R=\min(0.3, 10\,\unit{GeV}/p^{\mu}_\mathrm{T})$ for $p^{\mu}_\mathrm{T} < 50\,\GeV$,
and remains constant at $\Delta R = 0.2$ for $p^{\mu}_\mathrm{T} > 50\,\GeV$.

Three jet types are reconstructed, using the anti-$k_{t}$~\cite{AntiKTAlgorithm} algorithm as implemented in the \textsc{FastJet}
package~\cite{FastjetEntry}: small-radius (denoted small-$R$) jets, large-radius (denoted large-$R$) jets, and variable-radius jets.
The small-$R$ jets are built using a radius parameter of $R=0.4$ and particle-flow objects as
input~\cite{PERF-2015-09}. They are required to have $\pt>25\,\GeV$
and $|\eta|<2.5$. To reduce the contamination from jets originating from pile-up interactions, a selection requirement
on a multivariate classifier is applied to the selected jets. This classifier is based on calorimeter and tracking information
and is applied to jets with $\pT < 60\,\GeV$ and $|\eta| < 2.4$~\cite{PERF-2014-03}.
Large-$R$ jets are used to reconstruct high-momentum Higgs or $W$-boson candidates, for which the hadronic
decay products are emitted with small angular separation. These jets are built using a radius parameter of $R = 1.0$ and
topological calorimeter clusters with noise suppression as input~\cite{PERF-2014-07}. The clusters are locally calibrated~\cite{PERF-2011-03}
before being combined into jets. Trimming~\cite{Krohn:2009th} is used to minimise contributions from initial-state
radiation, pile-up interactions or the underlying event.  This is done by reclustering the constituents of the initial jet, using
the $k_{t}$ algorithm~\cite{Catani:1993hr,Ellis:1993tq}, into subjets with a radius parameter of $R^\mathrm{sub}$ = 0.2 and then removing any subjet
with a $p_\mathrm{T}$ less than $5\%$ of the $p_\mathrm{T}$ of the parent jet~\cite{ATL-PHYS-PUB-2015-033}. The trimmed
large-$R$ jets are required to have $p_\mathrm{T} > 250\,\GeV$ and $|\eta| < 2.0$.
The momenta of both the small-$R$ and large-$R$ jets are corrected for energy losses in passive material and
for the non-compensating response of the calorimeter. Small-$R$ jets are also corrected for the average
additional energy due to pile-up interactions~\cite{JETM-2018-05,JETM-2018-02}.
A third type of jets is clustered from ID tracks using a variable radius (VR) parameter that shrinks with increasing \pT of the studied
proto-jet~\cite{ATL-PHYS-PUB-2017-010}. VR track-jets are used in this analysis to identify decays of boosted Higgs bosons
into a pair of bottom quarks. The VR track-jets must contain at least two ID tracks compatible with the PV and must have $\pT > 7~\GeV$ as
well as $|\eta| < 2.5$.

Small-$R$ jets containing $b$-hadrons are identified ($b$-tagged) using the DL1r $b$-tagging algorithm~\cite{FTAG-2019-07}
based on a deep neural network that combines information from displaced tracks and reconstructed
secondary and tertiary vertices inside jets. A jet is $b$-tagged if the response value of the DL1r algorithm
exceeds a predefined threshold. Four operating points are defined with efficiencies of $60\%$, $70\%$, $77\%$,
and $85\%$ for $b$-jets, as measured in simulated $t\bar{t}$ events. These operating points divide the DL1r response
score distribution into five intervals. The lower edge of the lowest interval corresponds to a $b$-tagging
efficiency of $100\%$, and the upper edge of the highest interval corresponds to an efficiency of $0\%$. These
intervals are referred to as pseudo-continuous operating points. The number of $b$-tagged jets per event is evaluated
at a fixed $b$-tagging efficiency of $77\%$. Applying the $b$-tagging algorithm at this operating point reduces the
number of light-flavour and gluon jets, and jets containing $c$-hadrons, by a factor of 192 and 5.6,
respectively~\cite{FTAG-2019-07}. The pseudo-continuous operating points are used as input to the machine learning
algorithms that are designed to reconstruct the charged Higgs boson's decay chain. For this purpose, a
score, $w_\mathrm{DL1r}$, is defined for each jet as the number of pseudo-continuous operating points the jet satisfies,
where zero corresponds to failing and four to satisfying all operating points.

Boosted $h\rightarrow b\bar{b}$ decays are identified exploiting the kinematics of the large-$R$ jet as well as
the flavour-tagging information of up to three VR track-jets that are spatially matched via ghost-association~\cite{PERF-2017-04}
to the reconstructed large-$R$ jet~\cite{ATL-PHYS-PUB-2020-019}. The identification algorithm is based on a feed-forward
neural network which is trained using the probabilities of the $b$-, $c$- and light-flavour hypotheses of the three
leading VR track-jets\footnote{If there are fewer than three associated track-jets with $p_\mathrm{T} > 7\,\GeV$,
the inputs corresponding to any missing subjets are replaced with the mean input values.} and the $p_\mathrm{T}$ and $\eta$ of the
large-$R$ jet. The neural network is trained to separate boosted $h\rightarrow b\bar{b}$ jets from boosted top-quark jets and jets
arising from multijet processes. The network maps the input vector to a three-dimensional output layer. The three output nodes
quantify the probabilities for a large-$R$ jet to correspond to the signal class, and to either of the two background classes.
The three output nodes of the neural network are combined into a single discriminant:

\begin{equation*}
D_{\mathrm{Xbb}} = \ln\frac{p_{\mathrm{Higgs}}}{f_{\mathrm{top}}\cdot p_{\mathrm{top}} + (1-f_{\mathrm{top}})\cdot p_{\mathrm{multijet}}},
\end{equation*}

where $f_\mathrm{top}$ determines the fraction of top-quark jets, which is set to $f_\mathrm{top} = 0.25$~\cite{ATL-PHYS-PUB-2020-019}. Furthermore,
$p_{\mathrm{Higgs}}$, $p_{\mathrm{top}}$, and $p_{\mathrm{multijet}}$ are the probabilities for the Higgs boson jet, top-quark jet, and
multijet hypotheses. An operating point that corresponds to a selection efficiency of $60\%$ for large-$R$ jets containing
$h\rightarrow b\bar{b}$ decays is chosen for this analysis. The $h\rightarrow b\bar{b}$ tagging algorithm reduces contributions
from multijets and boosted top-quark jets by a factor of 92 and 31, respectively~\cite{ATL-PHYS-PUB-2020-019}, as
measured in simulated $t\bar{t}$ events.

Electrons, muons, and jets are reconstructed and identified independently. This can lead to ambiguous identifications when these objects
are spatially close to each other. Therefore, an overlap removal procedure is applied to uniquely identify these objects.
First, the closest small-$R$ jet within a cone of size $\Delta R = 0.2$ around an electron is removed. Furthermore, a small-$R$ jet
with fewer than three associated tracks is removed if the jet is within a cone of $\Delta R = 0.2$ around a selected muon.
Finally, electrons and muons are discarded if they are within a cone of size $\Delta R = \min(0.4, 0.04 + 10\,\GeV/p^\ell_\mathrm{T})$
around the axis of any surviving jet, where $p^\ell_\mathrm{T}$ is the transverse momentum of the electron or muon.
The latter requirement reduces the background contribution from semileptonic decays of heavy-flavour hadrons.

The missing transverse momentum (with magnitude \met) is computed as the negative vector sum of the $p_\mathrm{T}$ of
selected electrons, muons and jets, plus a track-based soft term, i.e.\ all tracks compatible with the PV
and not associated with any lepton or jet used in the \met\ calculation~\cite{JETM-2020-03}.


%
\section{Analysis strategy and event selection}\label{sec:AnalysisStrategy}


This analysis selects events consistent with the $\ell^\pm\nu b\bar{b}b\bar{b}q\bar{q}$ final state. The $pp$ collision events
are retained for further analysis using single-lepton triggers~\cite{TRIG-2018-05,TRIG-2018-01}.
The transverse momentum thresholds range from $24\,\GeV$ to $26\,\GeV$ for single-electron
triggers and from $20\,\GeV$ to $26\,\GeV$ for single-muon triggers, depending on the data-taking period.
The trigger-level lepton is required to match within $\Delta R = 0.07$ $(0.1)$ a reconstructed electron (muon)
with $\pt > 27$~\GeV. \par

Two different analysis strategies are applied to ensure high sensitivity to both low- and high-mass
resonances. One analysis strategy targets the `resolved' event topology, in which the final state
objects are well separated from each other. The second analysis strategy targets the `merged' event
topology, in particular events containing hadronic decays of strongly boosted Higgs
and $W$ bosons. These decays are reconstructed using large-$R$ jets.\par

Events are required to contain one prompt electron or muon with $\pt > 27$~\GeV and a missing transverse momentum
of $\met > 30\,\GeV$. Events with an additional lepton with $\pt > 10$~\GeV that satisfies the \textit{Medium} (\textit{Tight})
identification criteria for muons (electrons) are vetoed to reduce contributions from top-quark pair events and $Z$+jets production.
In addition, events for the resolved categories are required to contain at least
five small-$R$ jets of which at least two have to be $b$-tagged. For the merged categories,
events are required to contain at least one large-$R$ jet. Among all large-$R$ jets in the
event, exactly one has to be identified as a $h\rightarrow b\bar{b}$ candidate using
the boosted $h\rightarrow b\bar{b}$ identification technique described previously.\par

This analysis targets two charged Higgs boson decay channels: $H^{\pm}\rightarrow W^{\pm}h\rightarrow \ell^\pm\nu b\bar{b}$ (referred to
as the $\ell^\pm\nu b\bar{b}$ channel) and $H^{\pm}\rightarrow W^{\pm}h\rightarrow q\bar{q}b\bar{b}$ (referred to as the $q\bar{q}b\bar{b}$ channel).
Leptons from charged Higgs boson decays generally tend to have higher momenta and are more centrally located in the detector compared with those from
top-quark decays. This allows to exploit the differences in the event kinematics to separate the two decay channels. The $\ell^\pm\nu b\bar{b}$ and
$q\bar{q}b\bar{b}$ channels are created to be mutually exclusive using distinct selection criteria. Dedicated
analysis techniques including machine learning are employed to reconstruct the targeted charged Higgs boson decay modes. These
techniques are detailed in Sections~\ref{sec:ResolvedReco} and~\ref{sec:BoostedReco} for the resolved and merged analyses,
respectively. \par

To fully reconstruct the $H^{\pm}\rightarrow W^{\pm}h\rightarrow \ell^\pm\nu b\bar{b}$ decay chain, the four-vector of the
neutrino has to be determined. The $p_{x}$ and $p_{y}$ components of the neutrino momentum are
directly obtained by the \MEX and \MEY, while the longitudinal component of the neutrino momentum is calculated by applying
an on-shell $W$-boson mass constraint to the charged lepton plus neutrino system. This approach leads to a quadratic equation, which
provides either two, one, or zero real solutions. If it does not have an existing real solution, the missing momentum vector
$\vec{p}^\mathrm{miss}_\mathrm{T}$ is rotated until a real solution is found. If this procedure leads to ambiguities, the
rotation which provides the minimal change in the $\vec{p}^\mathrm{miss}_\mathrm{T}$ is chosen. If two real solutions are obtained,
the solution with the smallest $|p^\nu_{z}|$ is used~\cite{TOPQ-2012-18}.\par

Both the resolved and merged analyses are applied to all events, and the same events can be selected by either analysis.
Hence, the final search result is reported using the analysis that is most sensitive to the specific mass hypothesis being
tested (cf. Section~\ref{sec:result}).

\subsection{Reconstruction and classification of resolved charged Higgs boson decays}\label{sec:ResolvedReco}

\begin{figure}[t]
\centering
\includegraphics[width=0.525\textwidth]{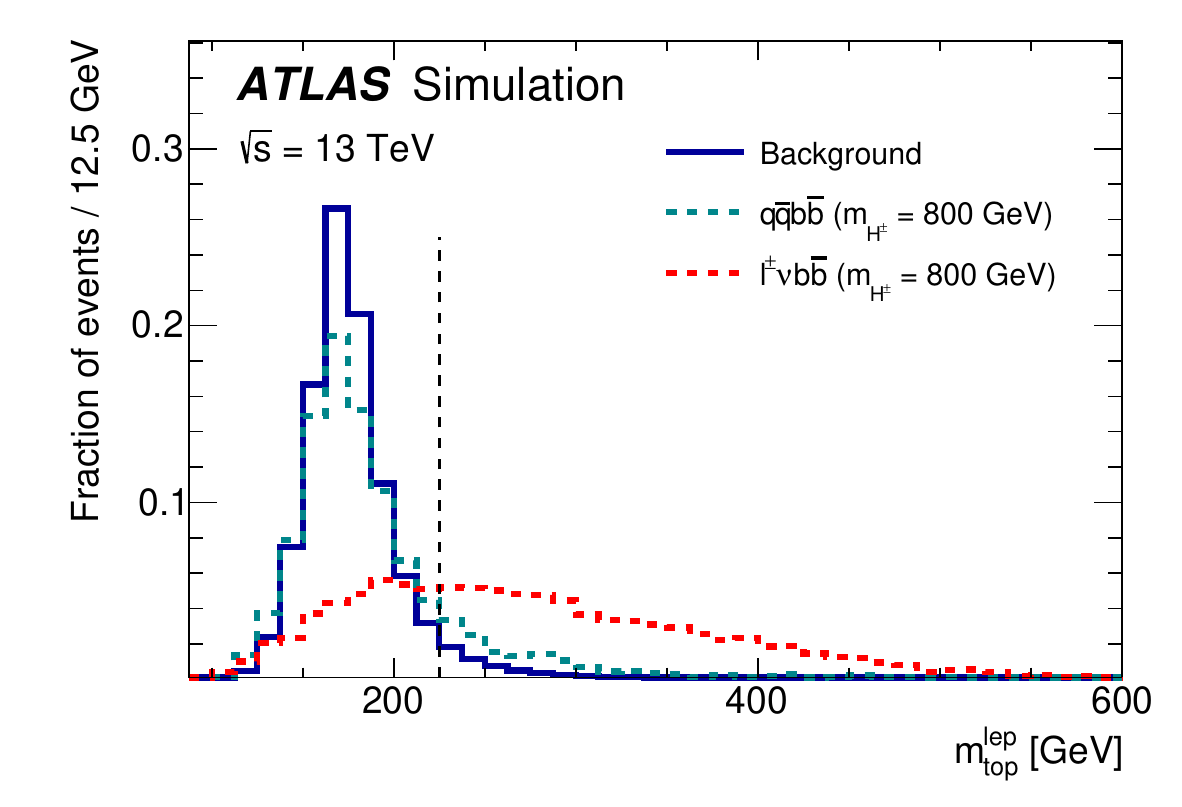}
\caption{Distributions of the $m^\mathrm{lep}_\mathrm{top}$ observable for $H^\pm\rightarrow W^\pm h\rightarrow \ell^\pm\nu b\bar{b}$
and $H^\pm\rightarrow W^\pm h\rightarrow q\bar{q}b\bar{b}$ decays as well as the sum of backgrounds after the preselection requirements
of the resolved analysis. The distributions are presented for a representative charged Higgs boson mass of $m_{H^\pm}=800\,\GeV$.
All distributions are normalised to unit area. The dashed vertical line indicates the threshold of the selection requirement on
$m^\mathrm{lep}_\mathrm{top}$ used to define the $\ell^\pm\nu b\bar{b}$ and $q\bar{q}b\bar{b}$ analysis regions.}
\label{fig:mTopLep_Distribution}
\end{figure}

For low charged Higgs boson masses, the $\ell^\pm\nu b\bar{b}$ and $q\bar{q}b\bar{b}$ decay channels
produce identical detector signatures, as indicated in Figure~\ref{fig:FeynmanDiagrams_Hplus_Wh}.
To differentiate between them, events are classified based on a requirement on the
reconstructed leptonic top-quark mass, $m^\mathrm{lep}_\mathrm{top}$. This observable is calculated from the four-vector sum
of a selected $b$-tagged jet, the charged lepton, and the neutrino candidate. Since several $b$-tagged jets are present in the event,
the selected $b$-tagged jet is chosen to minimise $\left|m_{\ell\nu j}- 172.5\,\GeV\right|$, where $m_{\ell\nu j}$ is the
invariant mass of the combined $b$-tagged jet, charged lepton, and neutrino system. While $m^\mathrm{lep}_\mathrm{top}$ is
distributed around the top-quark pole mass for events containing a leptonically decaying top quark (as in the $q\bar{q}b\bar{b}$
channel), it exhibits broader distributions at higher values for true $\ell^\pm\nu b\bar{b}$ events.
Consequently, events with $m^\mathrm{lep}_\mathrm{top} > 225\,\GeV$ are classified into the
$\ell^\pm\nu b\bar{b}$ analysis channel, while those with $m^\mathrm{lep}_\mathrm{top} \leq 225\,\GeV$ are
classified as $q\bar{q}b\bar{b}$ candidates. Distributions of the $m^\mathrm{lep}_\mathrm{top}$ observable
are presented in Figure~\ref{fig:mTopLep_Distribution} for a representative charged Higgs boson mass. \par

The accuracy of the classification requirement on $m^\mathrm{lep}_\mathrm{top}$ varies as a function of the
charged Higgs boson mass. For a mass of $250\,\GeV$, around $40\%$ ($45\%$) of the signal events are correctly
classified into the $q\bar{q}b\bar{b}$ ($\ell^\pm\nu b\bar{b}$) analysis channel. The success rates
increase with increasing charged Higgs boson mass, reaching values around $75\%$ ($90\%$) for
the $q\bar{q}b\bar{b}$ ($\ell^\pm\nu b\bar{b}$) analysis channel.

The charged Higgs boson decays are reconstructed either via a charged
lepton, a neutrino candidate, and two small-$R$ jets (for the $\ell^\pm\nu b\bar{b}$ decay mode) or via four
small-$R$ jets (for the $q\bar{q}b\bar{b}$ decay mode). Reconstructing the decay of the charged Higgs
boson is challenging due to the large number of objects produced in association with it. To address
this challenge, sets of boosted decision trees (BDTs) are used to identify the correct decay products
of the charged Higgs boson. This allows the four-momentum, and thus the invariant mass, of the heavy
scalar to be reconstructed. One set of BDTs is applied to events in the $\ell^\pm\nu b\bar{b}$ category,
and another set of BDTs is applied to events in the $q\bar{q}b\bar{b}$ category. The BDTs are trained
to distinguish between the correct pairings of the final state objects, i.e. leptons and jets, labelled as signal,
and the incorrect pairings labelled as background.\footnote{To determine whether a pairing is correct,
generator-level particles (i.e. the decay products of the Higgs and $W$ bosons) are spatially matched to the
reconstructed objects. A dijet system is considered correctly associated with the generator-level
Higgs or $W$ boson if its angular distance to the combined four-vectors of the decay products of the
Higgs or $W$ boson is smaller than 0.3. The matching between the charged lepton + neutrino system
and the corresponding generator-level particles is based on the same criteria.} All available signal samples
are used to train the BDTs. However, the BDT performance remains stable when adding or removing
individual mass points.\par

The BDTs are implemented into the analysis using the TMVA package~\cite{TMVA}. Both sets of BDTs contain a total of
400 decision trees, using the Gradient Boost algorithm with a learning rate of 0.1 and a maximum depth of five. \par

The BDTs dedicated to reconstruct the charged Higgs boson in the resolved $\ell^\pm\nu b\bar{b}$ category are trained
on seven input features built from the four-vectors of the charged lepton, the
neutrino candidate and the two jets used to construct the $W$ and Higgs boson candidates. These
features are the invariant mass of the Higgs boson candidate ($m_{j_1j_2}$), the azimuthal angular difference and
the pseudorapidity difference between the Higgs boson candidate and the $W$-boson candidate ($\Delta\Phi(\ell\nu,j_1j_2)$ and $|\Delta\eta(\ell\nu,j_1j_2)|$),
the ratio of the Higgs boson transverse momentum to the invariant mass of the reconstructed charged Higgs
boson candidate ($p_\mathrm{T}^{j_1j_2}/m_{\ell\nu j_1j_2}$), the ratio of the $W$-boson transverse momentum to the
invariant mass of the reconstructed charged Higgs boson candidate ($p_\mathrm{T}^{\ell\nu}/m_{\ell\nu j_1 j_2}$),
and the pseudo-continuous $b$-tagging intervals of the two jets used to build the Higgs boson
candidate ($w^{j_1}_{DL1r}$ and $w^{j_2}_{DL1r}$). \par

These BDTs are iteratively applied to all possible lepton, neutrino, and dijet pairings of events in
the resolved $\ell^\pm\nu b\bar{b}$ category. The pairing with the largest BDT score, $w_\mathrm{BDT}^\mathrm{max}$,
is used to construct the four-momentum vector (and hence the invariant mass) of the charged Higgs
boson candidate. Furthermore, the $w_\mathrm{BDT}^\mathrm{max}$ distribution is used to define signal and
control regions, motivated by the fact that the dominant backgrounds tend to have lower
$w_\mathrm{BDT}^\mathrm{max}$ values than the signal process.\par

The BDTs dedicated to the resolved $H^{\pm}\rightarrow W^{\pm}h\rightarrow q\bar{q} b\bar{b}$
decay are trained on ten input features built from the four-vectors of the four jets used to construct
the $W$ and Higgs boson candidates. These features are the invariant mass of the Higgs boson and $W$-boson
candidates ($m_{j_1j_2}$ and $m_{j_3j_4}$), the azimuthal angular difference and the pseudorapidity
difference between the Higgs boson candidate and the $W$-boson candidate ($\Delta\Phi(j_1j_2,j_3j_4)$
and $|\Delta\eta(j_1j_2,j_3j_4)|$), the ratio of the Higgs boson candidate's transverse momentum to the
invariant mass of the reconstructed charged Higgs boson candidate ($p_\mathrm{T}^{j_1j_2}/m_{j_1j_2j_3j_4}$),
the ratio of the $W$-boson transverse momentum to the invariant mass of the reconstructed charged Higgs
boson candidate ($p_\mathrm{T}^{j_3j_4}/m_{j_1 j_2 j_3j_4}$), and the pseudo-continuous $b$-tagging intervals of the
four jets ($w^{j_1}_{DL1r}$, $w^{j_2}_{DL1r}$, $w^{j_3}_{DL1r}$, and $w^{j_4}_{DL1r}$).\par

These BDTs are iteratively applied to all possible four-jet pairings of events in the resolved $q\bar{q} b\bar{b}$
category. The pairing with the largest $w_\mathrm{BDT}^\mathrm{max}$ value is used to construct the
four-momentum vector and the invariant mass of the charged Higgs boson candidate. Again, selection
requirements on the $w_\mathrm{BDT}^\mathrm{max}$ observable are used to define signal and control regions.\par

To protect against potential biases due to overtraining, a two-fold cross-validation is employed. Events are randomly divided into two
equal-sized subsamples, $A$ and $B$. Two independent boosted decision trees are trained on the two subsamples. The BDTs trained on sample $A$
are evaluated with sample $B$ and vice versa. Half of the data are analysed with the BDTs trained on sample $A$, and the other half with the
BDTs trained on sample $B$. Finally, the output distributions from both BDTs are merged for both simulated and collision data. This approach
results in four sets of BDTs: two for the $\ell^\pm\nu b\bar{b}$ channel and two for the $q\bar{q} b\bar{b}$ channel.\par

The sets of BDTs perform well in reconstructing the targeted final states. At a charged
Higgs boson pole mass of $500\,\GeV$, the final state is reconstructed correctly in about $55\%$
($45\%$) of the time for the $H^{\pm}\rightarrow W^{\pm}h\rightarrow \ell^\pm\nu b\bar{b}$ ($H^{\pm}\rightarrow W^{\pm}h\rightarrow q\bar{q} b\bar{b}$)
decay mode\footnote{This reconstruction efficiency is defined as the ratio of the number of correctly reconstructed
$\ell^\pm\nu b\bar{b}$ ($q\bar{q} b\bar{b}$) decays to the number of true $\ell^\pm\nu b\bar{b}$ ($q\bar{q} b\bar{b}$) events
satisfying the preselection and classification requirements.}. The success rates increase with increasing charged Higgs boson
mass values, reaching around $90\%$ for both decay modes at masses above $1\,\TeV$. The reconstruction efficiency drops to
around $12\%$ for the lowest considered Higgs boson masses, as the decay products tend to fail the kinematic selection requirements
and the charged Higgs boson decay cannot be fully reconstructed.

\begin{figure}[t]
\centering
\subfloat[]{\includegraphics[width=0.5\textwidth]{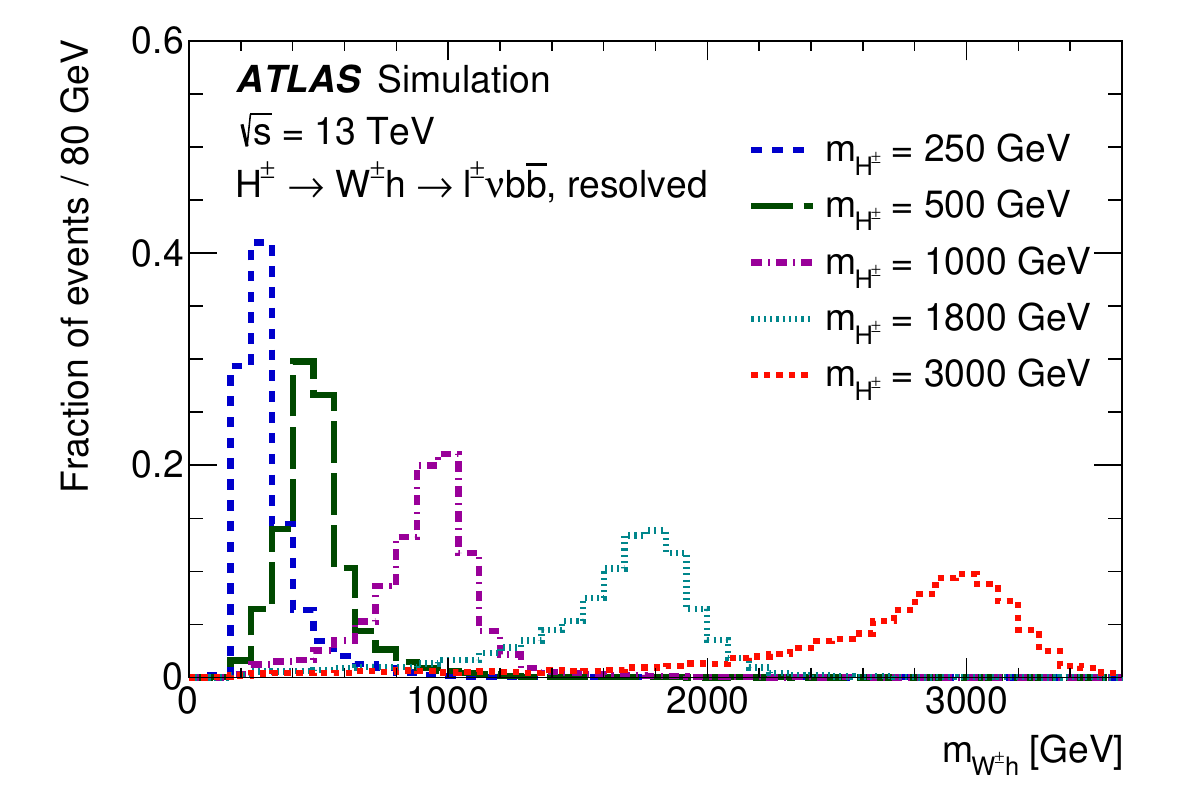}}
\subfloat[]{\includegraphics[width=0.5\textwidth]{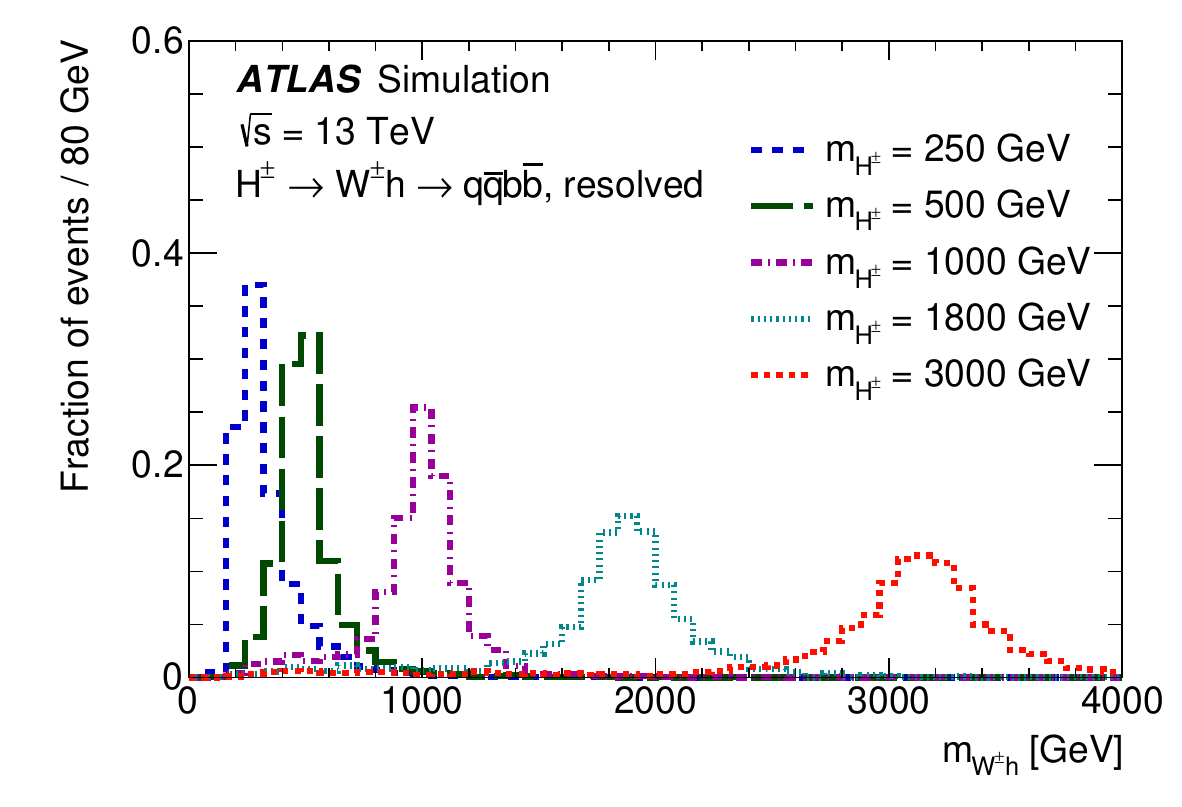}}
\caption{Distributions of the invariant mass of the (a) $H^\pm\rightarrow W^\pm h\rightarrow \ell^\pm\nu b\bar{b}$ and (b) $H^\pm\rightarrow W^\pm h\rightarrow q\bar{q}b\bar{b}$
signal hypotheses presented for a selection of different charged Higgs boson pole masses. The final states are reconstructed by
applying boosted decision trees to events satisfying the resolved preselection requirements. All distributions are normalised
to unit area.}
\label{fig:MassRecoPlots}
\end{figure}

The invariant masses of the reconstructed charged Higgs bosons are determined with a resolution\footnote{The mass resolution is
determined by fitting the convolution of a Gaussian and an exponential function (i.e. a Bukin function~\cite{BUKIN}) through the
distributions of $(m_{W^\pm h} - m_{W^\pm h}^\mathrm{truth})/m_{W^\pm h}^\mathrm{truth}$, where $m_{W^\pm h}$ and
$m_{W^\pm h}^\mathrm{truth}$ represent the reconstructed and generator-level invariant masses of the $W^\pm h$ system,
respectively. The extracted mass resolution values are given by one standard deviation of the Gaussian component.} below
$10\%$ for both decay chains and all event categories. The corresponding invariant mass distributions of the reconstructed
final states are shown in Figure~\ref{fig:MassRecoPlots}, separated by decay mode and for different signal mass hypotheses.\par

Finally, events are further categorised according to the overall number of jets ($j$), and the number of $b$-tagged jets ($b$)
in the event. In this context, four exclusive categories are defined: $5j3b$, $5j\geq4b$, $\geq6j3b$, and $\geq6j\geq4b$. \par

\subsection{Reconstruction and classification of merged charged Higgs boson decays}\label{sec:BoostedReco}

For sufficiently high charged Higgs boson masses, the ambiguities between the $\ell^\pm\nu b\bar{b}$ and $q\bar{q}b\bar{b}$
decay channels are largely reduced due to their distinct detector signatures. Thus, the events in the merged analysis are
classified based on the number and properties of selected large-$R$ jets.
The $q\bar{q}b\bar{b}$ category requires one large-$R$ jet tagged as a $h\rightarrow b\bar{b}$ candidate and another large-$R$ jet with a
mass $m_{J}$ within the $W$-boson mass window, $50\,\GeV < m_{J} < 110\,\GeV$ ($N^{W-\mathrm{tags}} = 1$). Events lacking a second large-$R$ jet with
mass around the $W$ boson pole mass ($N^{W-\mathrm{tags}} = 0$)  are classified as $\ell^\pm\nu b\bar{b}$. Consequently, the charged Higgs boson decays are
reconstructed using a charged lepton, a neutrino candidate, and one large-$R$ jet (for the $\ell^\pm\nu b\bar{b}$ channel) or two
large-$R$ jets (for the $q\bar{q}b\bar{b}$ channel). Events are further categorised based on the number of additional
$b$-tagged jets ($b$), considering only small-$R$ jets that are spatially separated from any large-$R$ jet used
in the analysis. Two exclusive categories are defined, $0b$ and $\geq1b$, based on the number of $b$-tagged small-$R$ jets.\par

A neural network (NN) algorithm is used to further distinguish between the signal and background processes.
Its architecture is sequential with three (two) fully connected dense layers of 128 nodes for the merged $\ell^\pm\nu b\bar{b}$
($q\bar{q}b\bar{b}$) category. The NNs are implemented with the \textsc{Python}-based deep learning library, \textsc{Keras}~\cite{KERAS}. The networks
are trained and optimised separately for the $\ell^\pm\nu b\bar{b}$ and $q\bar{q}b\bar{b}$ decay modes, due to differences in the event
kinematics. Both networks use the \textsc{Adam} optimiser~\cite{ADAM} to minimise a Binary Cross Entropy loss function and
seven input features. All signal samples with charged Higgs boson masses greater than or equal to $1200\,\GeV$ are used to train
the NNs. Lower mass hypotheses were excluded from the training to optimise the NN performance for the high-mass hypotheses.
Consequently, the NNs tend to assign lower scores to lower mass hypotheses due to their inherent mass dependence.\par

The NN for the merged $\ell^\pm\nu b\bar{b}$ category is trained on input features built from the leptonically decaying $W$ boson ($W^\mathrm{lep}$) obtained
from the four-vector sum of the charged lepton and the neutrino candidates and the large-$R$ jet used to construct the
boosted Higgs boson candidate.  These features are the angular separation between the charged lepton and the Higgs boson
candidate ($\Delta R\left(\ell,h\right)$), the azimuthal angular difference and the pseudorapidity difference between the Higgs
boson candidate and the reconstructed $W$ boson ($\Delta\phi\left(W^\mathrm{lep},h\right)$ and $|\Delta\eta\left(W^\mathrm{lep},h\right)|$),
the ratio of the Higgs boson transverse momentum to the invariant mass of the reconstructed charged Higgs boson candidate ($p_{\mathrm{T},h}/m_{Wh}$),
the ratio of the $W$-boson transverse momentum to the invariant mass of the reconstructed charged Higgs boson candidate ($p_{\mathrm{T},W^\mathrm{lep}}/m_{Wh}$),
the reconstructed leptonic top-quark mass ($m^\mathrm{lep}_\mathrm{top}$), which is calculated in the same way as for the resolved
analysis, and the ratio of the $W$-boson transverse momentum to the transverse momentum sum of all the decay products of the
charged Higgs boson candidate ($p_{\mathrm{T},W^\mathrm{lep}}/\left(p_{\mathrm{T},W^\mathrm{lep}} + p_{\mathrm{T},h}\right)$).\par

The NN for the merged $q\bar{q} b\bar{b}$ category is trained on input features built from the charged lepton and neutrino candidates and the two
large-$R$ jets used to reconstruct the hadronically decaying $W$-boson ($W^\mathrm{had}$) and the Higgs boson candidates. Similar to the
NN trained for the merged $\ell^\pm\nu b\bar{b}$ category, the $\Delta R\left(\ell,h\right)$, $\Delta\phi\left(W^\mathrm{had},h\right)$, $|\Delta\eta\left(W^\mathrm{had},h\right)|$, $p_{\mathrm{T},h}/m_{Wh}$, and $p_{\mathrm{T},W}/m_{Wh}$
observables are used in the training. In addition, the angular separation
between the charged lepton and the hadronically decaying $W$-boson candidate ($\Delta R\left(\ell,W^\mathrm{had}\right)$), and
the ratio of the transverse momentum of the leptonically decaying $W$-boson candidate to the transverse momentum sum of the leptonically decaying
$W$-boson candidate, the hadronically decaying $W$-boson candidate and the Higgs boson candidate ($p_{\mathrm{T},W^\mathrm{lep}}/\left(p_{\mathrm{T},W^\mathrm{lep}} + p_{\mathrm{T},W^\mathrm{had}} + p_{\mathrm{T},h}\right)$),
are used as input features. \par

The same two-folded cross-validation approach as used in the BDT training is employed in the NN training to mitigate
any bias due to overtraining. \par

\begin{figure}[t]
\centering
\subfloat[]{\includegraphics[width=0.5\textwidth]{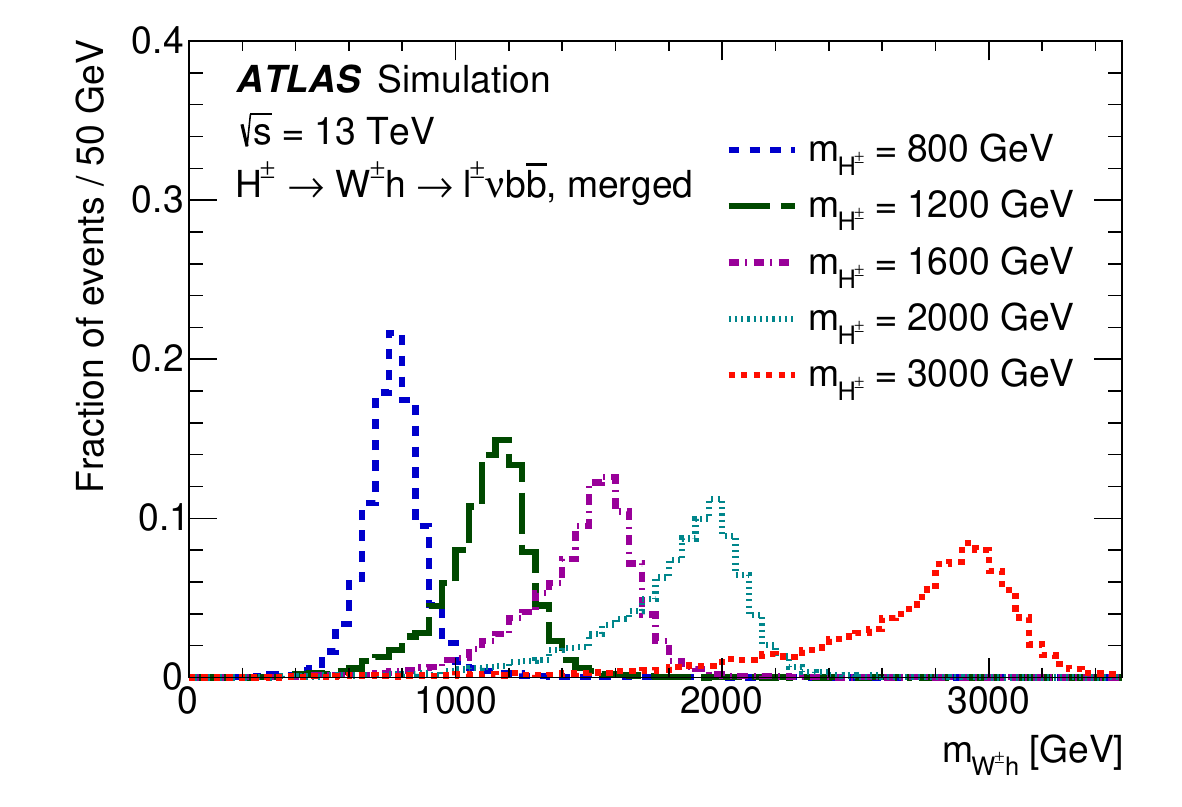}}
\subfloat[]{\includegraphics[width=0.5\textwidth]{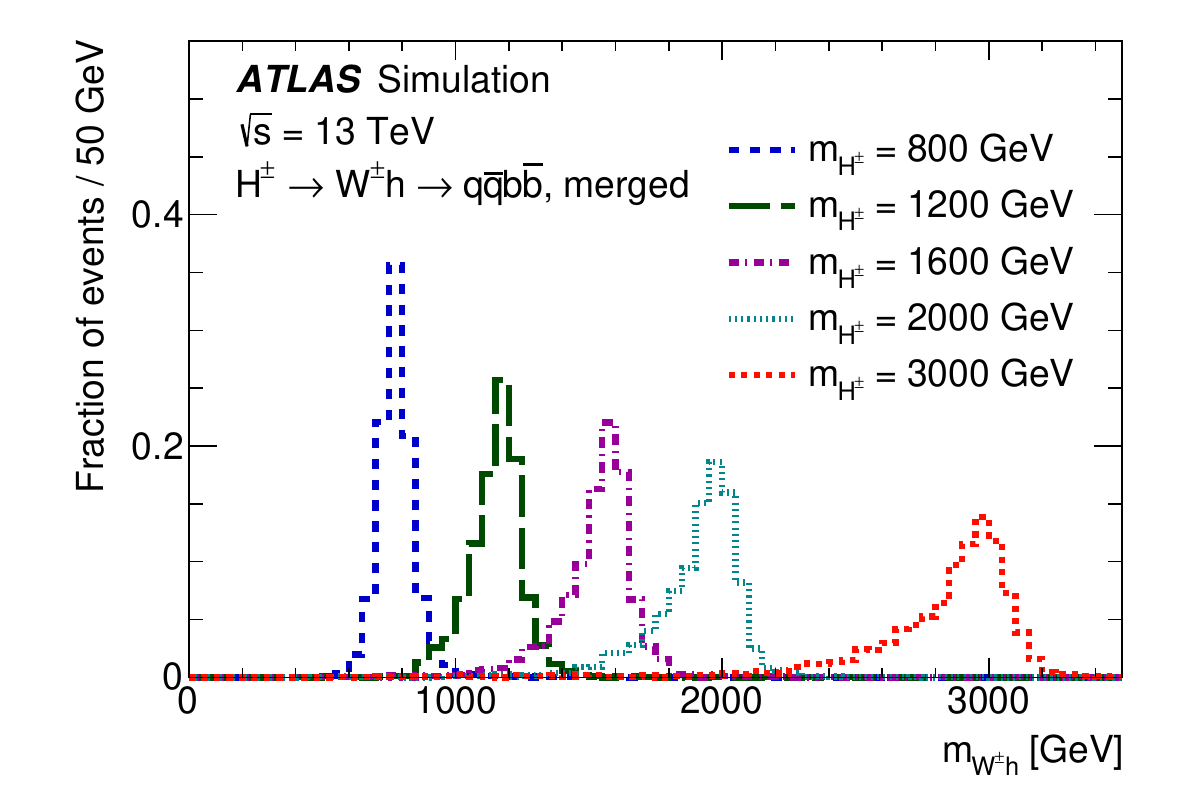}}
\caption{Distributions of the invariant mass of the (a) $H^\pm\rightarrow W^\pm h\rightarrow \ell^\pm\nu b\bar{b}$ and (b) $H^\pm\rightarrow W^\pm h\rightarrow q\bar{q}b\bar{b}$
signal hypotheses presented for a selection of different charged Higgs boson pole masses. The final states are reconstructed
using either a charged lepton, a neutrino candidate, and one large-$R$ jet (for the $\ell^\pm\nu b\bar{b}$ channel) or via two large-$R$
jets (for the $q\bar{q}b\bar{b}$ channel) for events entering the merged analysis regions. All distributions are
normalised to unit area.}
\label{fig:MassRecoPlotsMerged}
\end{figure}

The invariant mass of the charged Higgs boson candidates is reconstructed in all event categories with a resolution below $9\%$ for
the $\ell^\pm\nu b\bar{b}$ category and below $6\%$ for the $q\bar{q}b\bar{b}$ category. The corresponding charged Higgs boson invariant
mass distributions are shown in Figure~\ref{fig:MassRecoPlotsMerged}, separated by decay mode and for various signal mass hypotheses.\par

\subsection{Definition of signal and control regions}\label{sec:SRs_and_CRs}

The topological and kinematic preselection requirements of the resolved and merged event categories and the classification requirements used to
distinguish the $\ell^\pm\nu b\bar{b}$ and $q\bar{q}b\bar{b}$ decay chains are summarised in Table~\ref{tab:selKinEvt}. The signal and control regions are
finally defined by applying additional selection criteria on top of the preselection and event classification requirements.

{
\def\arraystretch{1.35}
\begin{table}[t]
\caption{
Topological and kinematic selections for each channel and category as described in the text.
Events are further classified according to the number of $b$-tagged jets in the events.
}
\label{tab:selKinEvt}
\begin{center}
\resizebox{1.0\textwidth}{!}{
\begin{tabular}{l| c| c | c | c}
\hline\hline
& \multicolumn{2}{c|}{Resolved analysis}                                                                                        & \multicolumn{2}{c}{Merged analysis}                                                         \\
\hline
Decay channel                                                                                 & $\ell^\pm\nu b\bar{b}$ &  $q\bar{q}b\bar{b}$                                                                                              & $\ell^\pm\nu b\bar{b}$                                                               &  $q\bar{q}b\bar{b}$ \\
\hline
\multirow{4}{*}{Preselection}                                                                 & \multicolumn{4}{c}{$N^\ell = 1$} \\
& \multicolumn{4}{c}{$\met > 30\,\GeV$} \\
& \multicolumn{2}{c|}{$N^{\mathrm{small-}R\,\mathrm{jets}} \geq 5$}                                                   & \multicolumn{2}{c}{$N^{\mathrm{large-}R\,\mathrm{jets}} \geq 1$} \\
& \multicolumn{2}{c|}{$N^{b\mathrm{-tags}} \geq 2$}                                                                   & \multicolumn{2}{c}{$N^{h\mathrm{-tags}} = 1$}  \\
\hline
\multirow{2}{*}{Classification requirement}                                                   & $m_\mathrm{top} > 225\,\GeV$ & $m_\mathrm{top} < 225\,\GeV$ & $N^{W-\mathrm{tags}} = 0$ & $N^{W-\mathrm{tags}} = 1$ \\
& \multicolumn{2}{c|}{$5j3b$, $5j\geq4b$, $\geq6j3b$, and $\geq6j\geq4b$}                                              & \multicolumn{2}{c}{$0b$ and $\geq1b$} \\
\hline
\hline
\end{tabular}}
\end{center}
\end{table}
}


The signal and control regions of the resolved analysis are obtained based on selection requirements on the
$w^\mathrm{max}_\mathrm{BDT}$ observable. A dedicated signal and control region is defined for each of the four
$5j3b$, $5j\geq4b$, $\geq6j3b$, and $\geq6j\geq4b$ categories. Hence, eight regions are used in the resolved
$\ell^\pm\nu b\bar{b}$ channel:
\begin{itemize}
\item \textit{Signal regions (SRs)} are defined by $w^\mathrm{max}_\mathrm{BDT} \geq 0.7$,
\item \textit{Control regions (CRs)} are defined in the range of $-0.5 \leq w^\mathrm{max}_\mathrm{BDT} < 0.5$.
\end{itemize}

Twelve regions are used in the resolved $q\bar{q} b\bar{b}$ channel:
\begin{itemize}
\item \textit{High-purity signal regions} are defined by $w^\mathrm{max}_\mathrm{BDT} \geq 0.9$,
\item \textit{Low-purity signal regions} are defined by $0.0 \leq w^\mathrm{max}_\mathrm{BDT} < 0.9$ for events with exactly three $b$-tagged jets and
by $0.6 \leq w^\mathrm{max}_\mathrm{BDT} < 0.9$ for events with at least four $b$-tagged jets,
\item  \textit{Control regions} are defined in the range of $-0.5 \leq w^\mathrm{max}_\mathrm{BDT} < 0.0$ for events with exactly three $b$-tagged jets and
by $-0.5 \leq w^\mathrm{max}_\mathrm{BDT} < 0.6$ for events with at least four $b$-tagged jets.
\end{itemize}

Events enter the signal regions of the merged analysis, if the mass of the $h\rightarrow b\bar{b}$ tagged large-$R$ jet
($m_{h}$) is in a window around the Higgs boson pole mass, $95\,\GeV \leq m_{h} < 140\,\GeV$. The signal regions are further
split based on the NN output score. A dedicated signal region is defined for each of the $0b$ and $\geq1b$
categories. Hence, six regions are used for the merged $\ell^\pm\nu b\bar{b}$ channel:
\begin{itemize}
\item \textit{High-NN-score signal regions} are defined by $w_\mathrm{NN} \geq 0.83$,
\item \textit{Medium-NN-score signal regions} are defined by $0.4\leq w_\mathrm{NN} < 0.83$,
\item \textit{Low-NN-score signal regions} are defined by $w_\mathrm{NN}$ < 0.4.
\end{itemize}

In the merged $q\bar{q}b\bar{b}$ channel four regions are used:
\begin{itemize}
\item \textit{High-NN-score signal regions} are defined by $w_\mathrm{NN} \geq 0.2$ ($w_\mathrm{NN} \geq 0.1$) for events with (without) additional $b$-tagged jets,
\item \textit{Low-NN-score signal regions} are defined by $w_\mathrm{NN} < 0.2$ ($w_\mathrm{NN} < 0.1$) for events with (without) additional $b$-tagged jets.
\end{itemize}

Furthermore, two sets of sideband control regions are defined in the merged event categories by inverting the selection
requirement on $m_{h}$. \textit{Low-mass sidebands} ($50\,\GeV \leq m_{h}< 95\,\GeV$) are defined to constrain the $W$+jets
background, while \textit{high-mass sidebands} ($140\,\GeV \leq m_{h}< 250\,\GeV$) are defined to target backgrounds
containing boosted top quarks. Again, a dedicated region is defined per $0b$ and $\geq1b$ category.
The various signal and control regions are summarised in Table~\ref{tab:AnalysisRegionsFullTable}.\par

\begin{figure}[!h]
\centering
\subfloat[]{\includegraphics[width=0.455\textwidth]{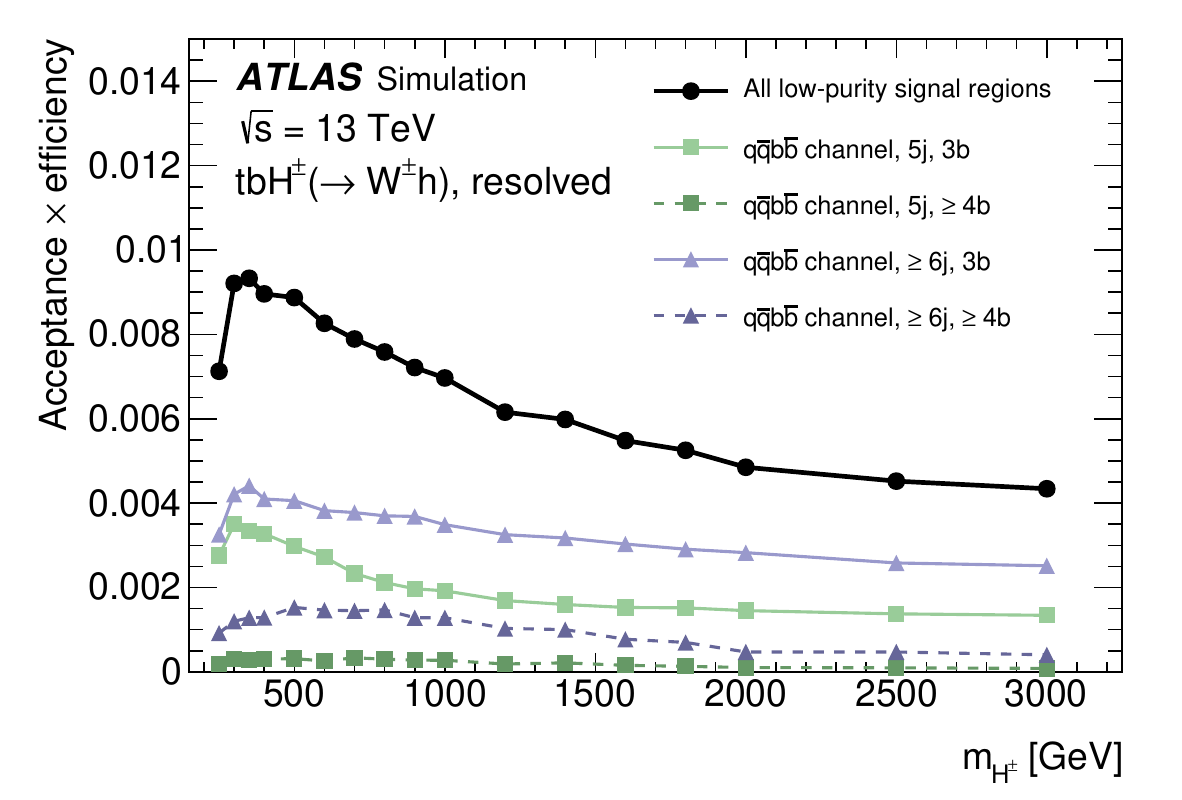}}
\subfloat[]{\includegraphics[width=0.455\textwidth]{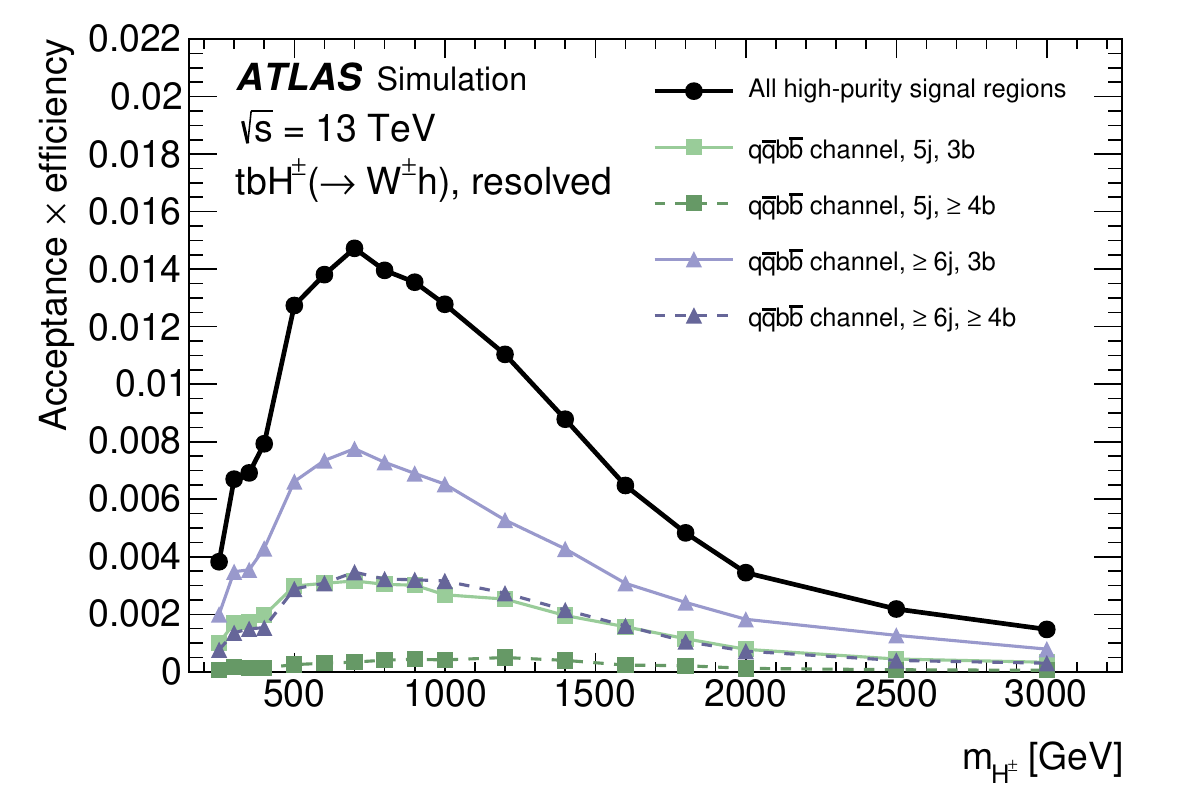}}\\

\subfloat[]{\includegraphics[width=0.455\textwidth]{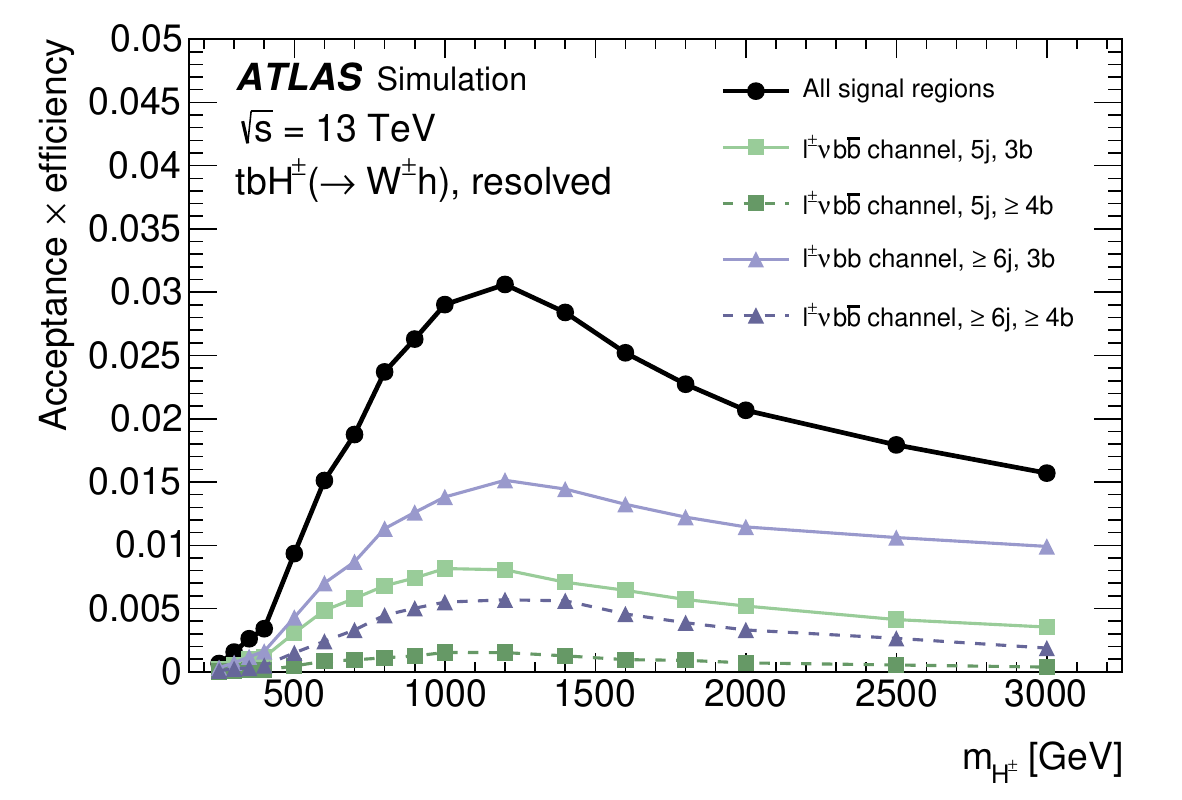}}
\caption{Product of acceptance and efficiency for $pp\rightarrow tb H^\pm(\rightarrow W^\pm h)$
as a function of the charged Higgs boson mass for (a) the resolved $q\bar{q} b\bar{b}$ low-purity signal regions,
(b) the resolved $q\bar{q} b\bar{b}$ high-purity signal regions, and (c) the resolved $\ell^\pm\nu b\bar{b}$ signal regions.}
\label{fig:SignalSelectionEfficienciesResolved}
\end{figure}

\begin{figure}[!h]
\centering
\subfloat[]{\includegraphics[width=0.455\textwidth]{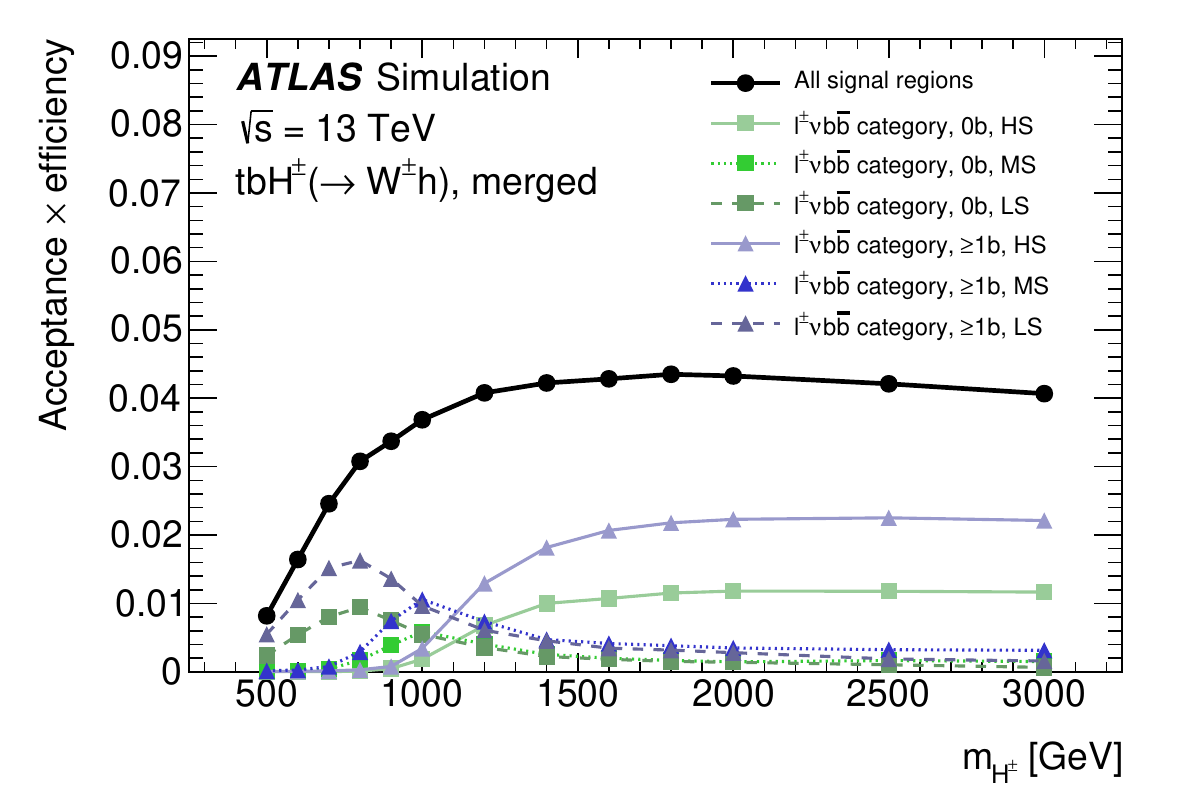}}
\subfloat[]{\includegraphics[width=0.455\textwidth]{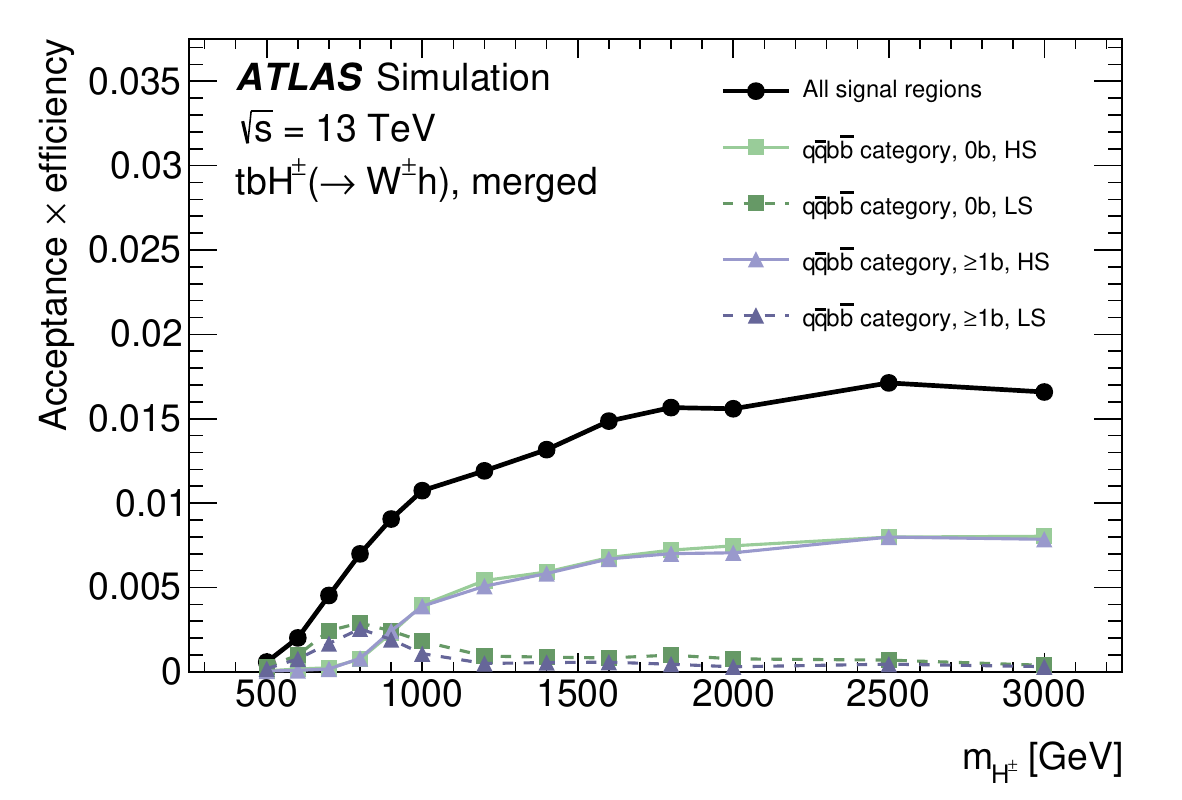}}
\caption{Product of acceptance and efficiency for $pp\rightarrow tb H^\pm(\rightarrow W^\pm h)$
as a function of the charged Higgs boson mass for (a) the merged $\ell^\pm\nu b\bar{b}$ high-NN score (HS), medium-NN score (MS)
and low-NN score (LS) signal regions, and (b) the merged $q\bar{q} b\bar{b}$ high-NN score and low-NN score signal regions.}
\label{fig:SignalSelectionEfficienciesMerged}
\end{figure}

The products of kinematic acceptance and reconstruction efficiency for $pp\rightarrow tb H^\pm(\rightarrow W^\pm h)$ is presented in
Figures~\ref{fig:SignalSelectionEfficienciesResolved} and ~\ref{fig:SignalSelectionEfficienciesMerged} separately for all signal regions
of the resolved and merged $\ell^\pm\nu b\bar{b}$ and $q\bar{q}b\bar{b}$ decay channels. In this context, the acceptance is defined as the
fraction of simulated signal events for which the expected final state particles satisfy all relevant object definition
requirements. The denominator of the acceptance is calculated considering simulated signal events with inclusive decays of
the $W$ boson, and $h\rightarrow b\bar{b}$ decays of the $125\,\GeV$ Higgs boson. The reconstruction efficiency is defined
as the ratio of simulated signal events that satisfy all selection criteria for a given signal region to the total number of
simulated signal events that satisfy the acceptance requirements.

{\def\arraystretch{1.425}
\begin{table}[!h]
\caption{Summary of signal and control regions considered in the statistical analysis for the resolved and merged channels.}
\label{tab:AnalysisRegionsFullTable}
\begin{center}
\begin{footnotesize}
\resizebox{1.0\textwidth}{!}{
\begin{tabular}{c | c | c l}
\hline\hline
Region & Requirement  & $\ell^\pm\nu b \bar{b}$  channel & $q\bar{q} b\bar{b}$ channel\\
\hline
\multicolumn{4}{c}{Resolved}   \\
\hline{\multirow{2}{*}{Signal regions}}
& Jet \& $b$-tag multiplicity  &  \multicolumn{2}{c}{$5j3b$, $5j\geq4b$, $\geq6j3b$, $\geq6j\geq4b$}         \\
& BDT score                 &  $w_\mathrm{BDT}^\mathrm{max} \geq 0.7$ & $w_\mathrm{BDT}^\mathrm{max} \geq 0.9$  \\
\hline{\multirow{3}{*}{Low-purity signal regions}}
& Jet \& $b$-tag multiplicity  &  \multicolumn{2}{c}{$5j3b$, $5j\geq4b$, $\geq6j3b$, $\geq6j\geq4b$}    \\
& {\multirow{2}{*}{BDT score}} &  {\multirow{2}{*}{--}}  & $0.0 \leq w_\mathrm{BDT}^\mathrm{max} < 0.9$ (for events with $5j3b$ or $\geq6j3b$)        \\
&                                 &                         & $0.6 \leq w_\mathrm{BDT}^\mathrm{max} < 0.9$ (for events with $5j\geq4b$ or $\geq6j\geq4b$) \\
\hline{\multirow{3}{*}{Control regions}}
& Jet \& $b$-tag multiplicity      &  \multicolumn{2}{c}{$5j3b$, $5j\geq4b$, $\geq6j3b$, $\geq6j\geq4b$}    \\
& {\multirow{2}{*}{BDT score}}  &  {\multirow{2}{*}{$ -0.5 \leq w_\mathrm{BDT}^\mathrm{max} < 0.5$}}  & $-0.5 \leq w_\mathrm{BDT}^\mathrm{max} < 0.0$ (for events with $5j3b$ or $\geq6j3b$) \\
&                                  &                                                                     & $-0.5 \leq w_\mathrm{BDT}^\mathrm{max} < 0.6$ (for events with $5j\geq4b$ or $\geq6j\geq4b$) \\
\hline
\multicolumn{4}{c}{Merged}   \\
\hline{\multirow{4}{*}{High-NN score signal region}}
& $b$-tag multiplicity   &  \multicolumn{2}{c}{$0b$, $\geq 1b$}     \\
& Mass window &  \multicolumn{2}{c}{$95\,\GeV \leq m_{J} < 140\,\GeV$}  \\
& {\multirow{2}{*}{NN score}} & {\multirow{2}{*}{$w_\mathrm{NN} \geq  0.83$}} &  $w_\mathrm{NN} \geq  0.2$ (for events with $0b$) \\
&                                &                                               &  $w_\mathrm{NN} \geq  0.1$ (for events with $\geq 1b$) \\
\hline{\multirow{3}{*}{Medium-NN score signal region}}
& $b$-tag multiplicity    &  \multicolumn{2}{c}{$0b$, $\geq 1b$}     \\
& Mass window &  \multicolumn{2}{c}{$95\,\GeV \leq m_{J} < 140\,\GeV$}  \\
& NN score &  $0.4 \leq w_\mathrm{NN} <  0.83$   & \hspace{2.05cm} --     \\
\hline{\multirow{4}{*}{Low-NN score signal region}}
& $b$-tag multiplicity    &  \multicolumn{2}{c}{$0b$, $\geq 1b$}     \\
& Mass window &  \multicolumn{2}{c}{$95\,\GeV \leq m_{J} < 140\,\GeV$}  \\
& {\multirow{2}{*}{NN score}} & {\multirow{2}{*}{$w_\mathrm{NN} <  0.4$}}  &  $w_\mathrm{NN} <  0.2$ (for events with $0b$) \\
&                                &                                            &  $w_\mathrm{NN} <  0.1$ (for events with $\geq 1b$) \\
\hline{\multirow{3}{*}{Low-mass control region}}
& $b$-tag multiplicity &  \multicolumn{2}{c}{$0b$, $\geq 1b$}     \\
& Mass window &  \multicolumn{2}{c}{$50\,\GeV \leq m_{J}< 95\,\GeV$}  \\
& NN score &  \multicolumn{2}{c}{\textit{--}}      \\
\hline{\multirow{3}{*}{High-mass control region}}
& $b$-tag multiplicity  &  \multicolumn{2}{c}{$0b$, $\geq 1b$}     \\
& Mass window &  \multicolumn{2}{c}{$140\,\GeV \leq m_{J}< 250\,\GeV$}  \\
& NN score &  \multicolumn{2}{c}{\textit{--}}      \\
\hline\hline
\end{tabular}}
\end{footnotesize}
\end{center}
\end{table}}


%
\section{Background modelling}\label{sec:BackgroundEst}


The background composition in the signal and control regions depends on the event categories
and $b$-jet multiplicities. In the resolved analysis categories, the dominant background source is
$t\bar{t}$ + jets production, contributing between $81\%$ and $95\%$ of the total background.
Other backgrounds such as $W/Z$ + jets or $Wt$ production are significantly smaller.
In the merged categories, the most significant background contributions stem from $t\bar{t}$ + jets and $W$ + jets production.
Contributions from diboson, SM $Vh$, $t\bar{t}h$, $t\bar{t}V$, and from the rare top-quark processes (i.e. $tZq$, $tWZ$,
$tHjb$, $tWh$, and $t\bar{t}t\bar{t}$) are small in all channels. Background contributions from events containing non-prompt
leptons are found to be negligible. These findings are consistent with previous studies in the same and similar final
states~\cite{HDBS-2018-51,HDBS-2020-19}.\par

Backgrounds are estimated from samples of simulated events, where the normalisation of the dominant
backgrounds is obtained from data. In addition, shape corrections are derived for the $t\bar{t}$ + jets production process.
These corrections are necessary because the additional jets in the $t\bar{t}$ + jets events are produced by the
parton shower, which leads to a mis-modelling of high jet multiplicities and the hardness of additional jet emissions.
Hence, in the resolved analysis channels, the simulated $t\bar{t}$ + jets events are reweighted as a
function of the $H^\mathrm{all}_\mathrm{T}$ observable.\footnote{The $H^\mathrm{all}_\mathrm{T}$
observable is defined as the scalar sum of the transverse momenta of all jets and the charged lepton in the event.}.
The reweighting function is obtained from fits to the data-to-simulation ratio in events with one charged lepton,
$\met > 30\,\GeV$, and at least five jets of which exactly two are $b$-tagged ($\geq5j2b$)
following the approach detailed in Ref.~\cite{HDBS-2018-51}. This reweighting procedure is performed separately for different jet
multiplicity intervals, distinguishing events with five, six, seven, or at least eight jets. The reweighting factors are expressed
as a function of the $H^\mathrm{all}_\mathrm{T}$ observable:

\begin{align}
r_i \left(H^\mathrm{all}_\mathrm{T}\right) = \frac{N^\mathrm{Data}_{i}\left(H^\mathrm{all}_\mathrm{T}\right) - N^{\mathrm{MC},
\mathrm{non}-t\bar{t}}_{i} \left(H^\mathrm{all}_\mathrm{T}\right)}{N^{\mathrm{MC}, t\bar{t}}_{i}\left(H^\mathrm{all}_\mathrm{T}\right)}, \nonumber
\end{align}

where $N_i$ is the number of events in a $H^\mathrm{all}_\mathrm{T}$ interval and the $i$-th
jet bin. The ensemble of $r_i \left(H^\mathrm{all}_\mathrm{T}\right)$ is fitted per jet bin with an exponential plus sigmoid
functional form,\footnote{In addition a hyperbola plus sigmoid functional form and a 2nd order polynomial plus first order
exponential functional form were tested. However, the exponential plus sigmoid functional form was found to fit the data
best.} which is then used as a correction function applied to the $5j3b$, $5j\geq4b$, $\geq6j3b$ and $\geq6j\geq4b$
regions of the $\ell^\pm\nu b \bar{b}$ and $q\bar{q} b\bar{b}$ analysis channels. In contrast, the merged analysis channels
are more inclusive in jet multiplicity, making them less dependent on the modelling of the additional jet activity.
While a similar correction procedure was tested for the merged analysis, it did not significantly improve the modelling of
relevant observables. Thus, no such correction procedure is applied in the merged analysis. \par

Events containing top-quark pairs are categorised according to the flavour of jets produced
in association with the $t\bar{t}$ system. This categorisation procedure is based on
generator-level information, where jets are reconstructed from stable particles (i.e.
particles with a mean lifetime $\tau > 3 \times 10^{-11}\,s$) using the anti-$k_{t}$
algorithm with the radius parameter set to $R = 0.4$. These jets are required to have
$p_\mathrm{T} > 15\,\GeV$ and $|\eta| < 2.5$, and their flavour is determined by
counting $b$- or $c$-hadrons with $p_\mathrm{T}$ larger than $5\,\GeV$
within a cone of $\Delta R = 0.4$ around the jet axis. Jets including $b$-hadrons
are labelled as $b$-jets, while jets that do not include any $b$-hadron, but do include one
or more $c$-hadrons are labelled as $c$-jets. Events that include at least one $b$- or $c$-jet,
not considering heavy-flavour jets from top-quark, $W$-boson, $Z$-boson or Higgs-boson decays,
are labelled as $t\bar{t} + \mathrm{HF}$ (with HF denoting `heavy flavour'). Events not containing
any heavy-flavour jets, aside from those from top-quark or $W$-boson decays, are labelled as
$t\bar{t} + \mathrm{LF}$ (with LF denoting `light flavour'). \par

The subdominant background processes are grouped into four components: single-top-quark production
(including $s$- and $t$-channels, $Wt$ production, $tZq$, and $tWZ$), $VV$ \& $V$ + jets (including $W/Z$ + jets, $WW$, $ZZ$, and $WZ$),
$t\bar{t} + X$ (including $t\bar{t}h$, $t\bar{t}W$, and $t\bar{t}Z$), and the remaining backgrounds referred to as `Others'
(including $tHjb$, $tWh$, $t\bar{t}t\bar{t}$, and SM $Vh$).


%
\section{Systematic uncertainties}\label{sec:Systematics}


The distributions of the $m_{W^\pm h}$ observable and the event yields in the signal and control regions are affected by both
experimental and modelling uncertainties, which enter the final fits as nuisance parameters (NPs)~\cite{Conway:2011in}. Uncertainties
in the modelling of physics objects are correlated between signal and background processes, channels, kinematic regions and distributions
of observables. The modelling uncertainties are evaluated separately for the signal and all relevant background processes.

\subsection{Experimental systematic uncertainties}

All experimental uncertainties, except for the luminosity uncertainty, impact both the normalisation and shape of the
simulated distributions. The dominant experimental uncertainties are uncertainties in the flavour-tagging performance, in the
jet energy scale and resolution calibration, and in the modelling of the pile-up activity.\par

Uncertainties in the trigger selection, the charged lepton reconstruction, identification and isolation criteria, as well
as the lepton momentum scale and resolution are measured in data applying tag-and-probe techniques to $Z\rightarrow \ell^+\ell^-$
events~\cite{EGAM-2018-01,MUON-2018-03}. These uncertainties have only a small impact on the charged Higgs boson search.\par

The uncertainties in the small-$R$ jet energy scale and resolution have contributions from \textit{in situ} calibration
measurements, from the dependency on the pile-up activity and from the flavour composition of the jets~\cite{JETM-2018-05}.
The uncertainty in the scale and resolution of the large-$R$ jet energy and mass is derived by comparing the ratio of
calorimeter-based to track-based measurements in dijet data and simulation, as described in Ref.~\cite{JETM-2018-02}. \par

Dedicated measurements are performed to calibrate the jet flavour-tagging efficiency for $b$-jets and the mis-tagging rates for
$c$- and light-flavour jets to the performance in data. The results of these measurements are expressed as $p_\mathrm{T}$
dependent correction factors that are derived for $b$-jets in $t\bar{t}$ events with dilepton final states~\cite{FTAG-2018-01},
for $c$-jets in $t\bar{t}$ events with single-lepton final states~\cite{FTAG-2020-08} and in $Z$+jets events for light-flavour
jets~\cite{FTAG-2019-02}. Uncertainties in these correction factors are decomposed into uncorrelated components and then
propagated to the charged Higgs boson search. Additional uncertainties are considered to extrapolate the measured
efficiencies to high jet $p_\mathrm{T}$~\cite{ATL-PHYS-PUB-2021-003}. These terms are calculated from simulated
events by considering variations of the quantities affecting the $b$-tagging performance, such as the impact parameter
resolution, percentage of tracks from random combinations of measurements in the ID, description of the detector material,
and track multiplicity per jet. \par

The $h\rightarrow b\bar{b}$ identification efficiencies and the corresponding mis-identification rates for boosted top-quark jets are
corrected to data using dedicated measurements. Correction factors and their corresponding uncertainties in the
$h\rightarrow b\bar{b}$ identification efficiencies are obtained from events containing $Z\rightarrow b\bar{b}$ decays~\cite{ATL-PHYS-PUB-2021-035}.
The correction factors are dependent on the large-$R$ jet $p_\mathrm{T}$ and vary from approximately 0.96 to 1.34
with uncertainties ranging from around $30\%$ to $35\%$. The mis-identification efficiency scale factors for boosted top quarks
are measured in top-quark pair events. They vary from $1.10 \pm 0.12$ at low $p_\mathrm{T}$ to $1.00 \pm 0.16$ at
high $p_\mathrm{T}$. Mis-identification efficiency uncertainties for light-quark and gluon jets were estimated in the sideband control regions of the
merged analysis by performing a three-component fit to data: one component describing events in which the
$h\rightarrow b\bar{b}$ tagged jet originates from the decay of a boosted top quark, one component
describing events in which the $h\rightarrow b\bar{b}$ tagged jet originates from a light-quark or gluon, and a third component
describing events in which the tagged jet stems from a boosted $Z$, $W$, or Higgs boson.
The large-$R$ jet mass distributions of these three components are fitted to data across different jet \pt intervals,
excluding the Higgs boson mass region. In these fits, the normalisation factors of the top-quark jet and light-quark or gluon jet components were
allowed to vary freely, while the third component was fixed to the SM prediction. The final uncertainties are determined by
summing the deviation of the central value from unity and the uncertainty of a given normalisation factor. These uncertainties range
from $7\%$ at low large-$R$ jet \pt to $41\%$ for jets with \pt exceeding $500\,$\GeV. \par

The uncertainties in the energy scale and resolution of the small-$R$ jets and leptons are propagated to the calculation of \met,
which also has additional uncertainties from the modelling of the underlying event and the momentum scale, momentum resolution
and reconstruction efficiency of the tracks used to compute the soft-term~\cite{JETM-2020-03}.\par

Finally, a global luminosity uncertainty of $0.83\%$ is applied to the normalisation of the simulated signal and
background samples~\cite{DAPR-2021-01,LUCID2}.\par

\subsection{Modelling systematic uncertainties}

Modelling uncertainties are taken into account for all simulated signal and relevant background processes and cover three areas:
shape uncertainties that account for uncertainties in the $m_{W^\pm h}$ template shapes used for the statistical analysis; absolute
normalisation uncertainties; and relative acceptance uncertainties (referred to as extrapolation uncertainties), that account for
differences in the acceptance between regions with a common floating normalisation factor, e.g. migration effects between the
signal and control regions, the $\ell^\pm\nu b\bar{b}$ and $q\bar{q} b\bar{b}$ channels, and the different jet multiplicity regions.
While normalisation and shape uncertainties may be described by a common nuisance parameter, extrapolation uncertainties
are always described by separate nuisance parameters.
The modelling uncertainties are assessed by comparing nominal and alternative event generators and the underlying event and parton
shower models. In general, the perturbative accuracy and the PDF sets used in these alternative
configurations match those of the nominal generators (unless explicitly stated). Uncertainties due to the modelling of PDF sets are
evaluated following the PDF4LHC recommendations~\cite{Butterworth:2015oua}, and uncertainties due to missing higher orders are evaluated
by varying the renormalisation and factorisation scales, $\mu_{R}$ and $\mu_{F}$, as described below.\par

For the signal processes, uncertainties in the acceptance are derived by comparing the predictions of the nominal PDF set with those
from the \CT[14]~\cite{CT14} and \MMHT~\cite{MMHT2014} PDF sets and then comparing the larger variation with the difference
from the root-mean-square spread of the nominal replica sets. The variation with the largest impact on the results is taken as the final
uncertainty. Further uncertainties are obtained by replacing \Pythia[8.244] by \Herwig[7.2.2] for the showering and hadronisation, and by
varying the renormalisation and factorisation scales. The total acceptance uncertainties show substantial variation across different
charged Higgs boson masses, decay channels, and event categories. \par

The modelling uncertainties of the top-quark pair production processes are derived as follows. To assess the uncertainties
in the matching of the matrix element to the parton shower, the nominal $t\bar{t} + \mathrm{HF}$ and
$t\bar{t} + \mathrm{LF}$ samples are compared with samples where \POWHEGBOX is replaced by \MGNLO. The parton shower
modelling uncertainties are determined by replacing \Pythia[8.230] by \Herwig[7.0.4] and switching to
the H7UE set of tuned parameters~\cite{HERWIG7}. The uncertainties in the modelling of the initial-state radiation (ISR)
and the final-state radiation (FSR) are addressed by varying the strong coupling constant $\alpha_{S}$
independently at the matrix element and the parton shower generation stage. In the matrix element,
the parameter $\alpha_{S}$ is increased (decreased) to 0.140 (0.115) instead of the nominal value 0.127, while
in the parton shower $\alpha_{S}$ is increased (decreased) to 0.142 (0.115) instead of the nominal value.
Uncertainties related to missing higher-order terms in the perturbative expansion are estimated by scaling
$\mu_{R}$ and $\mu_{F}$ independently up and down by a factor of two.
Additional uncertainties in the associated production of top-quark pairs and $b$-quarks are estimated by comparing the
predictions of the nominal \POWHEGBOX + \Pythia sample, in which the extra $b$-quarks arise from the PS, to the predictions of
an alternative \POWHEGBOX + \Pythia sample simulated using the four-flavour scheme. In the latter sample, an additional
$b$-quark pair is produced in association with the top-quark pair ($t\bar{t}b\bar{b}$) in the matrix element, which is
calculated at NLO accuracy in QCD. Furthermore, an uncertainty in the
relative composition of the additional $c$- and $b$-jet contributions in the $t\bar{t} + \mathrm{HF}$ sample is taken into
account. This uncertainty is evaluated by varying the relative normalisations of the $t\bar{t} + \geq 1b$ and
$t\bar{t} + \geq 1c$ contributions according to the predictions of the alternative MC generators relative to the predictions
of the nominal MC generator configuration. These variations range from $10\%$ to $30\%$ in the regions of the resolved analysis and
from $3\%$ to $30\%$ in the regions of the merged analysis. While varying the $t\bar{t} + \geq 1b$ and $t\bar{t} + \geq 1c$ contributions,
the overall normalisation of the $t\bar{t} + \mathrm{HF}$ background is kept constant. When calculating the modelling uncertainties for
the $t\bar{t} + \mathrm{HF}$ and $t\bar{t} + \mathrm{LF}$ processes, the same $H_\mathrm{T}$ reweighting is applied to the
nominal and alternative models.\par

A $5\%$ uncertainty is considered on the cross-section of the $Wt$ production mode and the $s$- and $t$-channel production of
single top quarks~\cite{SingleTopUncertainties}. Uncertainties associated with the matching of the matrix element and the parton shower
and the choice of the parton shower and hadronisation models, are estimated by comparing, for each of these three processes, the
nominal \POWHEGBOX + \Pythia[8.230] sample with alternative samples produced using \MGNLO[2.62] + \Pythia[8.140] and \POWHEGBOX + \Herwig[7.0.4], respectively.
Uncertainties associated with the interference between top-quark pair and $Wt$ production at NLO are evaluated by
comparing the predictions of the nominal $Wt$ sample with those from an alternative sample generated with \POWHEGBOX
and \Pythia[8.230] using the diagram subtraction (DS) scheme~\cite{DS_Wt_Paper} instead of the diagram removal
scheme. Finally, an $8\%$ uncertainty is considered on the cross-section of the $tZq$ process, and a $50\%$
uncertainty is considered on the $tWZ$ cross-section~\cite{HDBS-2018-51}. \par

An uncertainty of $^{+9\%}_{-12\%}$ in the $t\bar{t}h$ production cross-section is considered, including effects from varying the factorisation
and renormalisation scales, the PDF set, and $\alpha_{S}$~\cite{deFlorian:2016spz,Zhang:2014gcy}. The nominal \POWHEGBOX + \Pythia[8.230] sample is compared with a sample produced with
\MGNLO[2.60] + \Pythia[8.230] to assess the uncertainty associated with the matching of the matrix element and the parton shower, and to a
sample generated with \POWHEGBOX + \Herwig[7.0.4] for the uncertainty in the modelling of the parton shower.

The uncertainties in the normalisation of the $t\bar{t}Z$ and $t\bar{t}W^\pm$ backgrounds are $^{+10.4\%}_{-12.0\%}$ and $^{+13.4\%}_{-12.0\%}$,
respectively and include effects from varying the factorisation and renormalisation scales and the PDF set~\cite{Campbell:2012dh}.
In addition, uncertainties in the choice of the parton shower and hadronisation models and matching scheme for the $t\bar{t}Z$ and $t\bar{t}W^\pm$
processes are estimated by comparing the nominal \MGNLO[2.33] + \Pythia[8.210] samples to alternative samples simulated using \Sherpa[2.2.1] with the
\NNPDF[3.0nnlo] PDF set. For the alternative $t\bar{t}W^\pm$ event samples, diagrams with up to one additional parton emission are generated
at NLO accuracy in QCD, and diagrams with two, three or four additional parton emissions are generated at LO accuracy in QCD.
For the production of $t\bar{t}Z$ events, diagrams with up to four additional parton emission are generated at LO accuracy in QCD.
In both samples, the CKKW matching scale of the additional emissions was set to $30\,\GeV$. \par

An overall $50\%$ uncertainty is considered in the normalisation of the four-top-quarks background~\cite{MG5,Frederix:2017wme},
while uncertainties of $^{+7.5\%}_{-15.4\%}$ and $^{+9.1\%}_{-9.2\%}$ are assigned to the $thjb$ and $thW$
backgrounds~\cite{MG5}. These uncertainties cover effects from varying the renormalisation and factorisation scales, the PDF set and $\alpha_{S}$.\par

In the merged event categories, the modelling uncertainties in the $W$+jets background are determined as follows:
uncertainties related to missing higher order terms in the perturbative expansion are estimated by individually varying
$\mu_{R}$ and $\mu_{F}$ by a factor of $2$ or $0.5$. Six combinations are considered: $(\mu_{R}, \mu_{F})$ $=$ $(0.5, 0.5)$,
$(0.5, 1.0)$, $(1.0, 0.5)$, $(1.0, 2.0)$, $(2.0, 1.0)$, and $(2.0, 2.0)$ times their nominal value. The final uncertainty
is derived as the variation of $\mu_{R}$ and $\mu_{F}$ that has the largest impact on the results.
Additionally, variations of the \Sherpa merging scale are taken into account, where the nominal value, $20\,\GeV$, is changed to $30\,\GeV$
and $15\,\GeV$. Furthermore, uncertainties in the parton shower evolution's upper cutoff scale are assessed by varying the resummation scale
by factors of $1/\sqrt{2}$ and $\sqrt{2}$ from its nominal value. Additional modelling uncertainties are derived by comparing the nominal
$W$+jets sample to an alternative sample simulated using \MGNLO[2.6.5] with the \NNPDF[3.1nnlo] PDF set. The alternative sample
includes vector bosons with up to three additional partons at NLO accuracy in QCD. \Pythia[8.240] with the A14
set of tuned parameters is used for showering and hadronisation, and the different jet multiplicities were merged using the FxFx
procedure~\cite{Frederix:2012ps}. In the resolved event categories, where the $W$+jets background is nearly negligible, a flat $40\%$
uncertainty is considered on the normalisation of the $W$+jets background. \par

Both the resolved and merged analyses assign an uncertainty of $35\%$ to the $Z$+jets background normalisation, and a $50\%$ uncertainty in the
normalisation of the diboson backgrounds. These normalisation uncertainties account for effects from varying the renormalisation and factorisation
scales, the merging scale and the PDF set and $\alpha_{S}$.\par

Additional uncertainties in the $H_\mathrm{T}$ reweighting of the $t\bar{t}$ + jets background in the resolved analysis regions
(cf. Section~\ref{sec:BackgroundEst}) are evaluated by independently varying the fitted parameters of the exponential plus sigmoid
functional form within their uncertainties and reapplying the reweighting procedure. The resulting differences in the normalisation
and $m_{W^\pm h}$ shape of the $t\bar{t}$ + jets background compared with the nominal reweighting are assigned as a systematic
uncertainty. These uncertainties are evaluated separately for each jet multiplicity bin.
In addition to the nominal reweighting function, two other functional forms (cf. Section~\ref{sec:BackgroundEst}) were considered for the
reweighting procedure. The differences in results between the nominal and alternative functions are negligible compared to
the MC statistical uncertainties of the $t\bar{t}$ + jets sample. Therefore, no additional uncertainty due to the choice of
the reweighting function is assigned.\par

Residual mismodelling (i.e. non-closure in the comparison between data and simulation) of the transverse $W$-boson momentum,
$p_{\mathrm{T},W}$, distribution in the resolved analysis regions and the $p_{\mathrm{T},W}/m_{Wh}$, $\Delta\Phi\left(W,h\right)$
and $\met$ distributions in the merged analysis regions are observed in the respective control regions. The non-closure uncertainties are
derived by parameterising the data-to-simulation ratios of mis-modelled observables. These parametric functions are then used to reweight
the observables and assess the impact on the shape of the $m_{W^{\pm}h}$ distribution. The magnitude of the $m_{W^{\pm}h}$ shape differences
reach up to $30\%$ in the tails of the distributions. The final uncertainty is quantified as the variation between the reweighted and
nominal $m_{W^{\pm}h}$ distributions, and is incorporated as an additional nuisance parameter affecting all processes.\par

The relative modelling systematic uncertainties (at the pre-fit stage), impacting the normalisation,
cross-region extrapolation, and shape of the signal and background processes are summarised in Table~\ref{tab:syst1}.

\begin{table}[!h]
\caption{\label{tab:syst1}
Relative modelling systematic uncertainties (at the pre-fit stage) in the normalisation,
cross-region extrapolation, and shape of signal and the background processes included in the fits described
in the text. An `S' indicates that a shape variation is included for the listed sources. `$A \leftrightarrow B$'
indicates relative acceptance uncertainties that account for the relative normalisation differences between
two regions, $A$ and $B$, with a common floating normalisation factor (i.e. these are extrapolation uncertainties).
Furthermore, `norm.' is the product of cross-section and acceptance variations, and a value of `float' indicates that the
parameter is not constrained in the fit. A range of values means that the size of the uncertainties vary between the regions
included in the fit, where the minimum and maximum relative uncertainties in the predicted yields are stated.
No dedicated normalisation uncertainty is taken into account for the $t\bar{t}$ + LF background given that the normalisation of
this process is fixed by the $H_\mathrm{T}$ reweighting. Each of the listed uncertainties is treated as uncorrelated in the fits.
The uncertainties are listed separately for (a) the resolved analysis channels and (b) the merged analysis channels.
}
\begin{center}
\vspace{-0.5cm}
\subfloat[Relative acceptance and normalisation uncertainties for the resolved analysis channels]{\resizebox{1.0\textwidth}{!}{
\begin{tabular}{l c  c  c  c  c  c  c }
\hline
& $t\bar{t}$ + HF & $t\bar{t}$ + LF & $t\bar{t}$ + X & Single Top      & $VV$ \& $V$ + jets & Others & Signal          \\
\hline
Norm.                                                            &    float        &      fixed      &  0.8\%-- 12\%   & $1.1\%$ -- $4.7\%$ &   2\%--34\%     &  2.9\%--47\%    &     float       \\
PDF                                                              &      S          & 0.2\%--0.5\%, S &     --          &    --           &       --           &   --            &  0.5\%--6\%     \\
ISR                                                              &      S          & 0.7\%--5\%, S   &     --          &    --           &       --           &   --            &     --          \\
FSR                                                              &      S          & 1.3\%--19\%, S  &     --          &    --           &       --           &   --            &     --          \\
ME-PS matching                                                   &      S          & 2.0\%--40\%, S  &  0.6\%--7\%, S  & 0.6\%--53\%, S  &       --           &   --            &     --          \\
Parton shower                                                    &      S          & 3.3\%--22\%, S  &  1.7\%--13\%, S & 2.2\%--61\%, S  &       --           &   --            &  1.0\%--50\%, S \\
4FS vs. 5FS                                                      &      S          &       --        &     --          &    --           &       --           &   --            &     --          \\
Heavy-flavour composition                                        &      S          &       --        &     --          &    --           &       --           &   --            &     --          \\
DS vs. DR scheme ($Wt$)                                          &     --          &       --        &     --          & 3.3\%--68\%, S  &       --           &   --            &     --          \\
Renormalisation/factorisation scales                             &      S          & 0.5\%--3.0\%, S &     --          &    --           &       --           &   --            &  1.0\%--13\%, S \\
$H_\mathrm{T}$ reweighting                                       &      S          & 0.5\%--6\%, S   &     --          &    --           &       --           &   --            &     --          \\
$p_{\mathrm{T},W}$ non-closure                                   &      S          & 0.5\%--0.9\%, S & 0.7\%--1.6\%, S & 2.7\%--4\%, S   &  1.8\%--3.3\%, S   & 1.1\%--2.0\%, S &  3.0\%--30\%, S \\
$5j$ $\leftrightarrow$ $\geq 6j$                                 &  17\%--34\%     &       --        &     --          &    --           &       --           &   --            &     --          \\
SRs  $\leftrightarrow$ CRs                                       &  11\%--33\%     &       --        &     --          &    --           &       --           &   --            &     --          \\
SRs  $\leftrightarrow$ low-purity SRs                            &  6.8\%--17\%    &       --        &     --          &    --           &       --           &   --            &     --          \\
Low-purity SRs $\leftrightarrow$ CRs                             &  5.2\%--12\%    &       --        &     --          &    --           &       --           &   --            &     --          \\
$\ell^\pm\nu b\bar{b}\leftrightarrow q\bar{q} b\bar{b}$          &  5.1\%--17\%    &       --        &     --          &    --           &       --           &   --            &     --          \\
\hline
\end{tabular}}}

\subfloat[Relative acceptance and normalisation uncertainties for the merged analysis channels]{\resizebox{1.0\textwidth}{!}{
\begin{tabular}{l c  c  c  c  c  c  c }
\hline
& $t\bar{t}$ + HF & $t\bar{t}$ + LF & $t\bar{t}$ + X & Single Top       & $VV$ \& $V$ + jets & Others & Signal           \\ \hline
Norm.                                               & float           & float           &     0.5\%-11\%    & $5\%$ -- $50\%$     &      float            &  0.5\%- 50\%   & float         \\
PDF                                                 & S               & S               &         --        &        --           &      --               &    --          & $<$0.5\%      \\
ISR                                                 & S               & S               &         --        &        --           &      --               &    --          & --            \\
FSR                                                 & S               & S               &         --        &        --           &      --               &    --          & --            \\
ME-PS matching                                      & S               & S               &     0.5\%-28\%, S &  0.5\%-26\%, S      &  S                    &    --          & --            \\
Parton shower                                       & S               & S               &     0.5\%-18\%, S &  0.5\%-60\%, S      &  S                    &    --          & 0.8\%-60\%, S    \\
4FS vs. 5FS                                         & S               & --              & --                & --                  & --                    & --             & --               \\
Heavy-flavour composition                           & S               & --              & --                & --                  & --                    & --             & --               \\
DS vs. DR scheme ($Wt$)                             & --              & --              & --                & 13\%-86\%, S        & --                    & --             & --               \\
Renormalisation/factorisation scales                & S               & S               &         --        &        --           & S                     & --             & 0.5\%--11\%, S \\
$p_{\mathrm{T},W}/m_{Wh}$ non-closure               & S               & S               &    1.0\%-4\%, S   & 0.8\%-9\%, S        & S                     & 0.7\%-9\%, S   & 0.5\%-21\%, S  \\
$\Delta\Phi\left(W,h\right)$ non-closure            & S               & S               &    0.6\%-1.2\%, S & 0.5\%-1.2\%, S      & S                     & 0.5\%-0.6\%, S & 0.5\%-3.2\%, S \\
$\met$ non-closure                                  & S               & S               &    0.8\%-8\%, S   & 0.6\%-8\%, S        & S                     & 4\%-10\%, S    & 0.5\%-12\%, S \\
$0b$ $\leftrightarrow$ $\geq 1b$                    & 4\%--24\%       & 11\%--18\%      & --                & --                  & 8\%--70\%      & --     & --               \\
$\ell^\pm\nu b\bar{b}\leftrightarrow q\bar{q} b\bar{b}$ & 18\%--35\%  & 18\%--50\%      & --                & --                  & 16\%--60\%     & --     & --               \\
High-mass CRs $\leftrightarrow$ low-mass CRs        & 3.4\%--19\%     & 3.7\%--16\%     & --                & --                  & 10\%--32\%     & --     & --               \\
CRs $\leftrightarrow$ high-score SR                 & 13.1\%--22\%    & 6\%--30\%       & --                & --                  & 4\%--28\%      & --     & --               \\
CRs $\leftrightarrow$ medium-score SR               & 11\%--17\%      & 6\%--22\%       & --                & --                  & 4\%--12\%      & --     & --               \\
CRs $\leftrightarrow$ low-score SR                  & 6\%--13\%       & 2.7\%--23\%     & --                & --                  & 6\%--40\%      & --     & --               \\
Low-score SR $\leftrightarrow$ high-score SR        & 17\%--27\%      & 7\%--30\%       & --                & --                  & 11\%--35\%     & --     & --               \\
Low-score SR $\leftrightarrow$ medium-score SR      & 13\%--19\%      & 7\%--23\%       & --                & --                  & 11\%--14\%     & --     & --               \\
Medium-score SR $\leftrightarrow$ high-score SR     & 7\%--34\%       & 11\%--26\%      & --                & --                  & 14\%--20\%     & --     & --               \\
\hline
\end{tabular}}}
\end{center}
\end{table}



%
\section{Results}\label{sec:result}


To test for the presence of a massive charged Higgs boson in data, the $m_{W^\pm h}$ templates obtained from the
simulated signal and background event samples are fitted to data using a binned maximum-likelihood approach
based on the \textsc{RooFit/RooStats} framework~\cite{Moneta:1289965,RooFit,Baak2015}. Fits are
simultaneously performed on the signal and control regions defined in Section~\ref{sec:SRs_and_CRs} to constrain the
normalisation of the main backgrounds and the nuisance parameters describing the systematic uncertainties detailed in
Section~\ref{sec:Systematics}. Each systematic variation is incorporated in the fit as an individual nuisance parameter using a Gaussian
constraint, while nuisance parameters describing statistical uncertainties are incorporated using a Poisson constraint.
Systematic variations that are impacted by large statistical fluctuations are smoothed, while
systematic variations with negligible impact on the final results are pruned away.
Asimov data sets~\cite{Cowan:2010js} are used to evaluate the expected performance of each fit.\par

As no significant excess over the SM expectations is found, the results of this search are expressed
as upper limits at the $95\%$ confidence level on the production cross-section
times branching ratio of charged Higgs bosons for a wide range of resonance masses.
The largest deviation from the SM expectations is found for a charged Higgs boson mass of 900\,\GeV
and corresponds to a local significance of about 0.9 standard deviations (or equivalent a $p$-value, i.e. the probability
that the background can produce a fluctuation greater than the excess observed in data, of 0.184).
The limits are evaluated using the CL$_\text{s}$ method~\cite{Read:2002hq} and the
profile-likelihood-ratio test statistic in the asymptotic approximation~\cite{Cowan:2010js}. \par

Two separate likelihood fits are performed to search for charged Higgs bosons: one for the resolved analysis and the other for the
merged analysis. Each fit includes dedicated signal and control regions for $q\bar{q}b\bar{b}$ and $\ell^\pm\nu b\bar{b}$ decays (cf. Section~\ref{sec:SRs_and_CRs}) and
the fit models differ between the analyses. The resolved fit has the signal cross-section and the global normalisation factor of the
$t\bar{t} + \mathrm{HF}$ background as free parameters, while the merged fit has the signal cross-section and the global normalisation
factors of the $t\bar{t} + \mathrm{HF}$, $t\bar{t} + \mathrm{LF}$, and $VV$ \& $V$ + jets backgrounds as free parameters.
Common normalisation factors are applied across all regions used in a fit. However, dedicated nuisance parameters address extrapolation uncertainties
between the signal and control regions, decay channels, and the different jet multiplicity regions. The corresponding extrapolation uncertainties
are detailed in Table~\ref{tab:syst1}. Normalisation factors and uncertainties from background-only fits\footnote{When performing
signal-plus-background fits, the corresponding normalisation factors show minor deviations from those obtained in the background-only fits.
However, these differences are insignificant in comparison to the uncertainties in the normalisation factors.} in the resolved and merged analysis
categories are summarised in Table~\ref{tab:NormalisationFactors}. The normalisation factors for the $t\bar{t} + \mathrm{HF}$
background exceed unity in both resolved and merged analyses. The normalisation factors are consistent within
about one standard deviation across the two analyses. Furthermore, the normalisation factor for the $t\bar{t} + \mathrm{LF}$
background component in the merged analysis is below unity, consistent with the corrections obtained from the $H^\mathrm{all}_\mathrm{T}$
reweighting in the resolved analysis.
\clearpage

\begin{table}[t]
\caption{\label{tab:NormalisationFactors}
Post-fit normalisation factors and their uncertainties obtained from a combined background-only fit to the various
signal and control regions of the resolved and merged analyses. Numbers are presented for the background components that are
allowed to float in the likelihood fit. Uncertainties in the cross-section times branching ratio and the overall acceptance
are not considered in the denominators of these normalisation factors.}
\begin{center}
\begin{tabular}{c c c}
\hline
Background                   &  Resolved analysis  & Merged analysis \\
\hline
$t\bar{t} + \mathrm{HF}$     &  $1.39\pm0.18$ &  $1.20\pm 0.17$ \\
$t\bar{t} + \mathrm{LF}$     &    --          &  $0.75\pm 0.08$ \\
$VV$ \& $V$ + jets           &    --          &  $1.17\pm 0.13$ \\
\hline
\end{tabular}
\end{center}
\end{table}

In addition to constraining the normalisation of the dominant backgrounds, the fit also constrains several modelling uncertainties. Significant
constraints are observed in the resolved analysis for nuisance parameters associated with the flavour composition, parton shower modelling and
matrix element to parton shower matching of the $t\bar{t} + \mathrm{HF}$ background. Furthermore, nuisance parameters associated with
the parton shower modelling of the $t\bar{t} + \mathrm{LF}$ background and the comparison between the DS and DR schemes of the $Wt$ background
are constrained substantially. The nuisance parameter corresponding to the flavour composition of the $t\bar{t} + \mathrm{HF}$ background
is constrained to $20\%$ of its initial value, while the other nuisance parameters listed above are constrained to about $40\%$-$60\%$ of
their initial values. These constraints are evident in fits on both Asimov and real data. In addition to the constraints, a few
significantly pulled nuisance parameters are observed in both analyses. In the resolved analysis, the nuisance parameter associated with the
$p_{\mathrm{T},W}$ non-closure uncertainty is pulled by about one standard deviation. In the merged analyses, the nuisance parameters associated
with the $\met$ non-closure uncertainty, the uncertainties related to missing higher order terms in the perturbative expansion of the $W$ + jets
background, and the matrix element to parton shower matching of the $t\bar{t} + \mathrm{LF}$ background are each pulled by around 0.7 to 0.8
standard deviations.\par

\begin{figure}[t]
\centering
\subfloat[]{\includegraphics[width=1.05\textwidth]{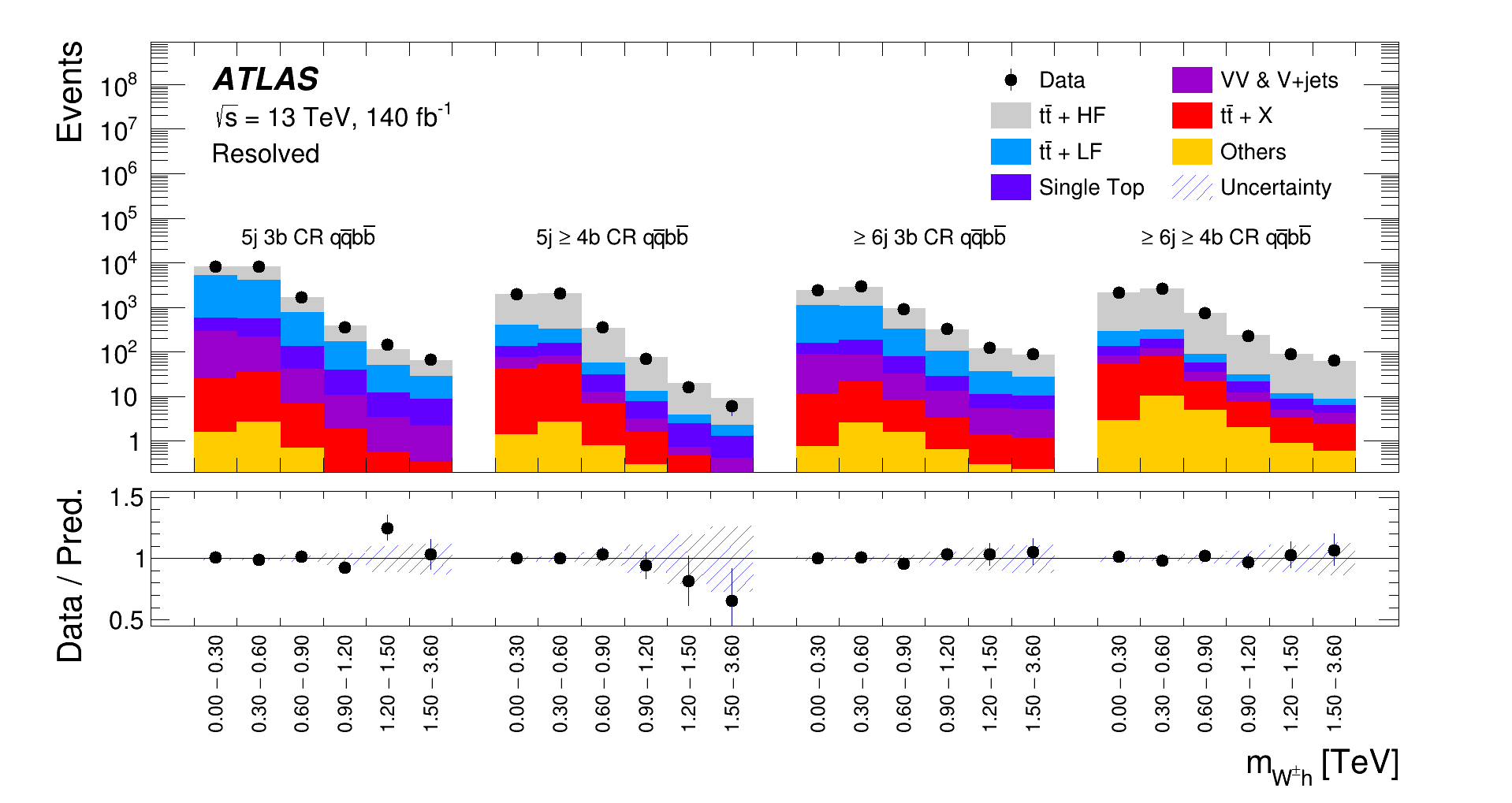}}\\
\subfloat[]{\includegraphics[width=1.05\textwidth]{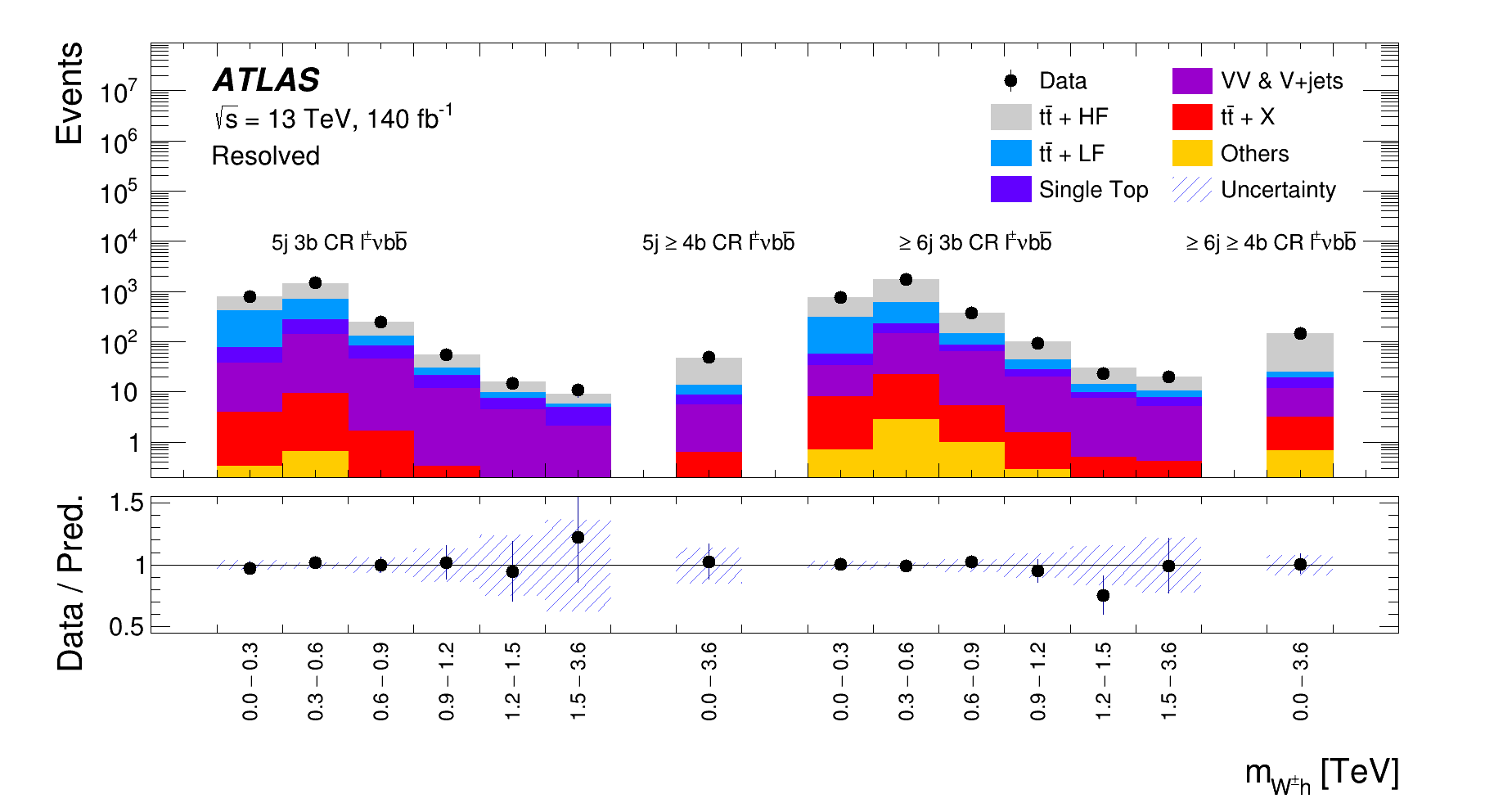}}
\caption{Distributions of the $m_{W^\pm h}$ observable in the control regions of the resolved (a) $q\bar{q} b\bar{b}$
and (b) $\ell^\pm\nu b\bar{b}$ event categories.
The term `Others' summarises events from $tHjb$, $tWh$, $t\bar{t}t\bar{t}$, and SM $Vh$ production.
The distributions are presented after a background-only maximum-likelihood fit to data. The lower panels show the
ratio of the observed to the estimated SM background. The shaded bands show the total post-fit uncertainty in
the background.}
\label{fig:CR_summary_plots_I}
\end{figure}

\begin{figure}[t]
\centering
\subfloat[]{\includegraphics[width=1.05\textwidth]{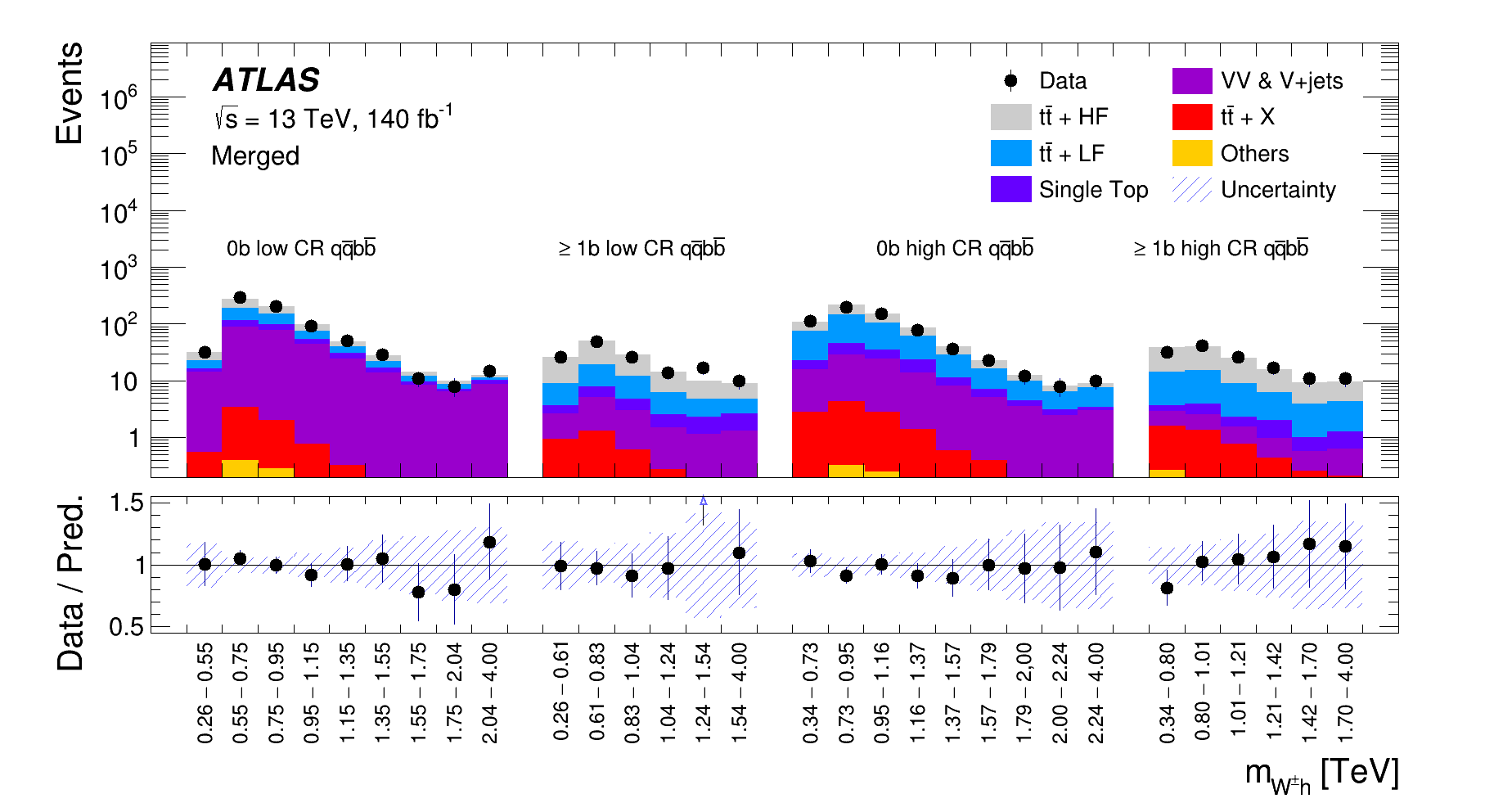}}\\
\subfloat[]{\includegraphics[width=1.05\textwidth]{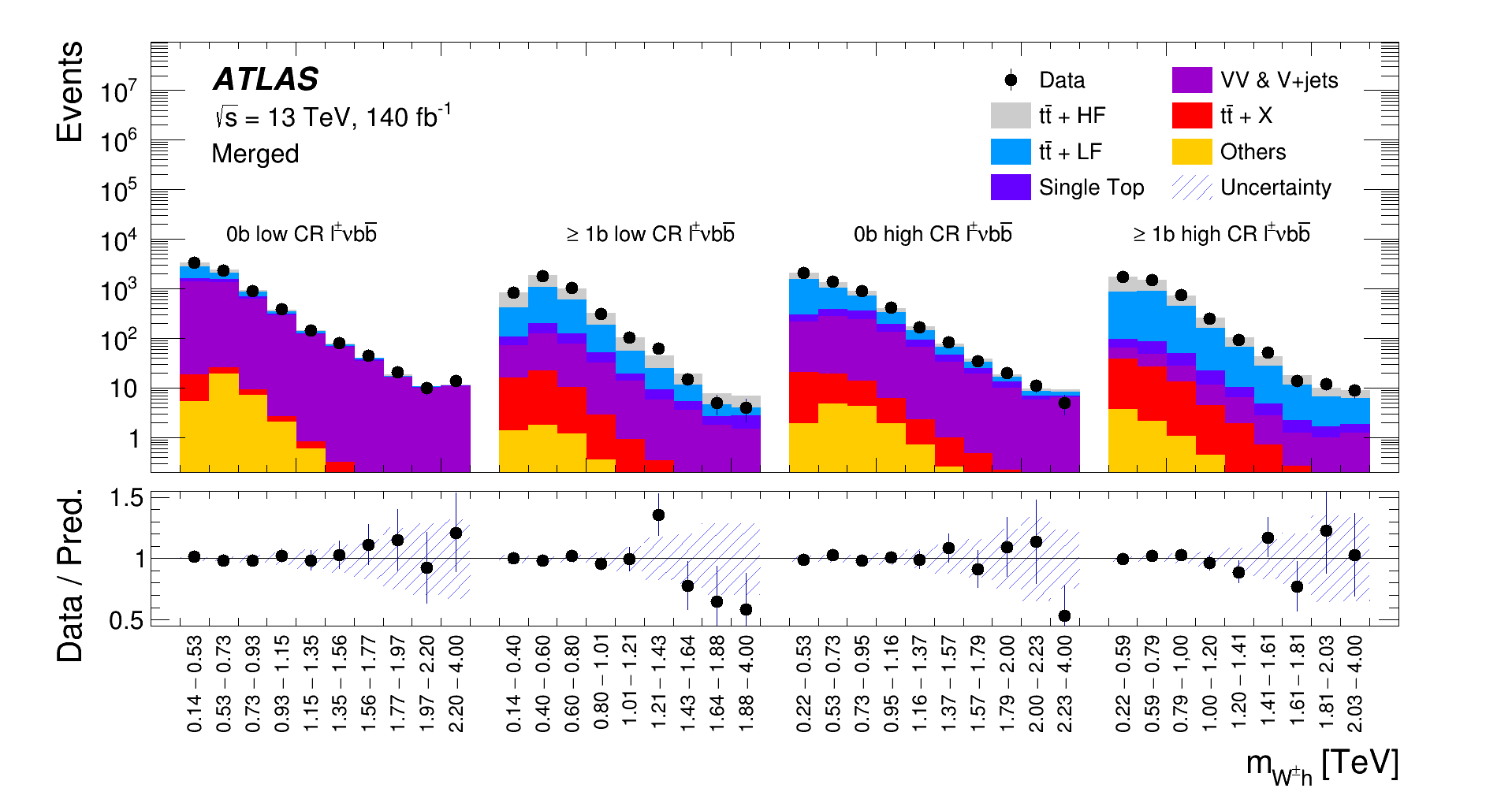}}
\caption{Distributions of the $m_{W^\pm h}$ observable in the control regions of the merged (a) $q\bar{q} b\bar{b}$ and
(b) $\ell^\pm\nu b\bar{b}$ event categories. The term `Others' summarises events from $tHjb$, $tWh$, $t\bar{t}t\bar{t}$,
and SM $Vh$ production. The distributions are presented after a background-only maximum-likelihood fit to data.
The lower panels show the ratio of the observed to the estimated SM background.
The shaded bands show the total post-fit uncertainty in the background.}
\label{fig:CR_summary_plots_II}
\end{figure}

\begin{figure}[h!]
\centering
\subfloat[]{\includegraphics[width=0.495\textwidth]{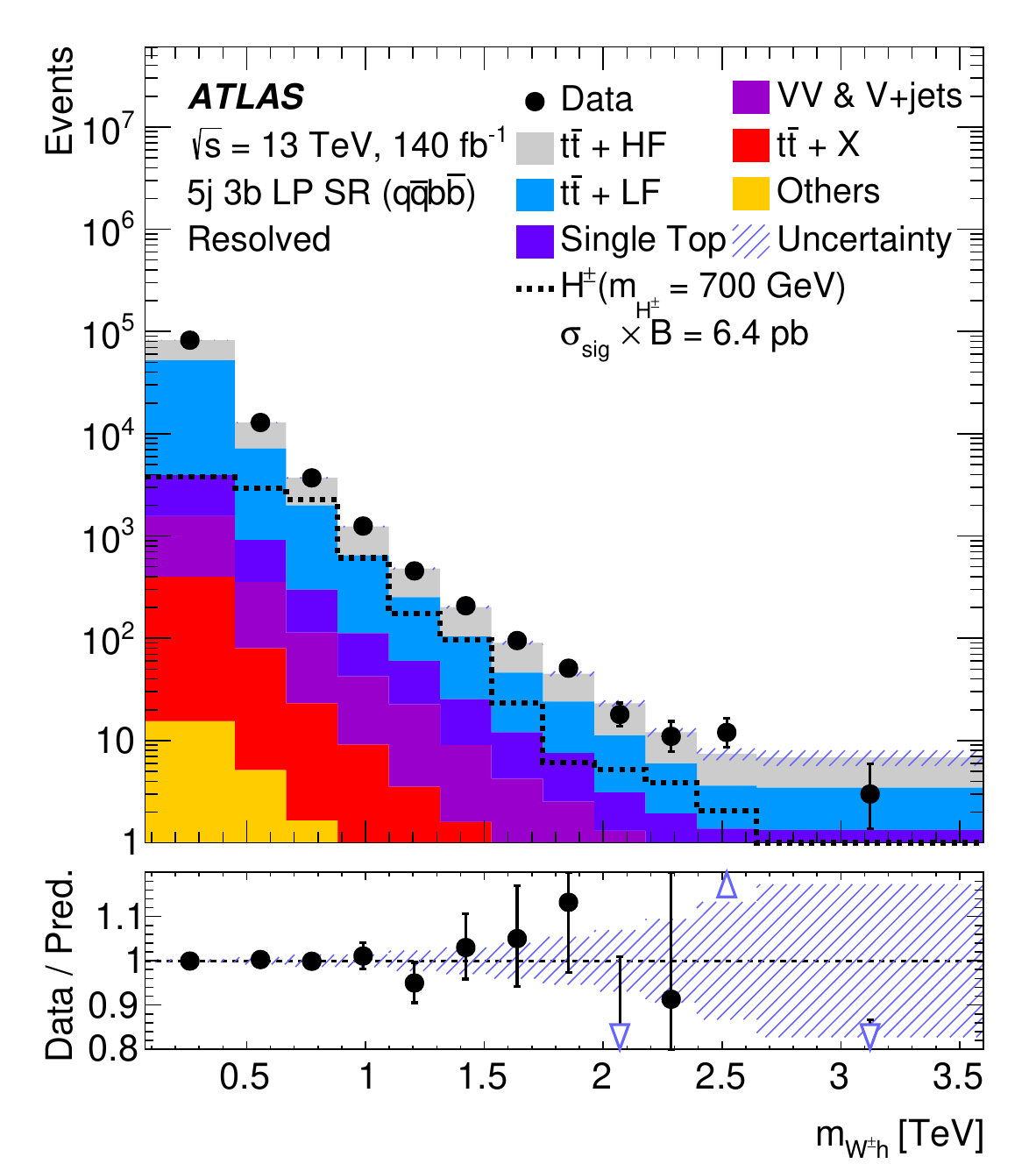}}
\subfloat[]{\includegraphics[width=0.495\textwidth]{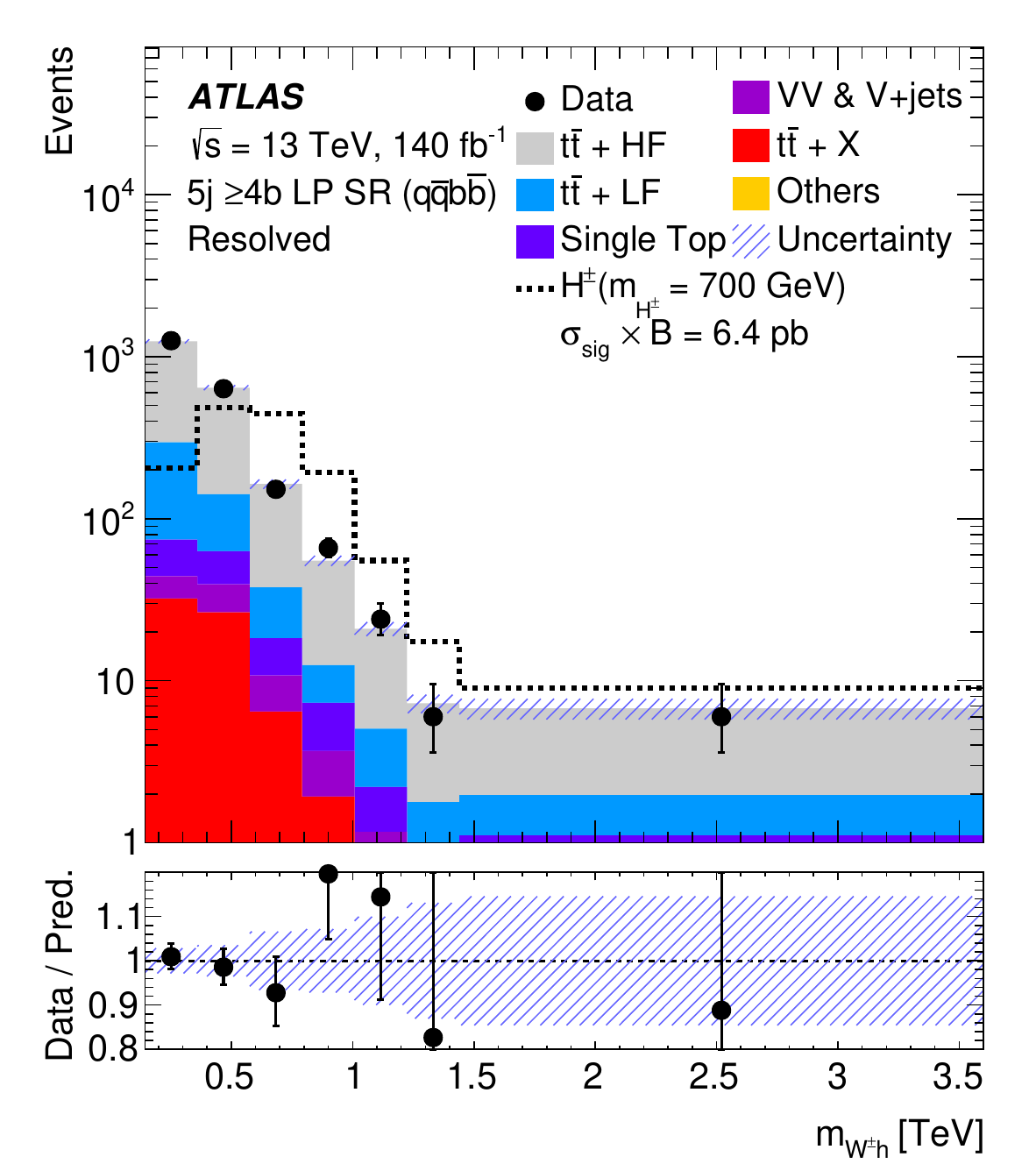}}\\
\subfloat[]{\includegraphics[width=0.495\textwidth]{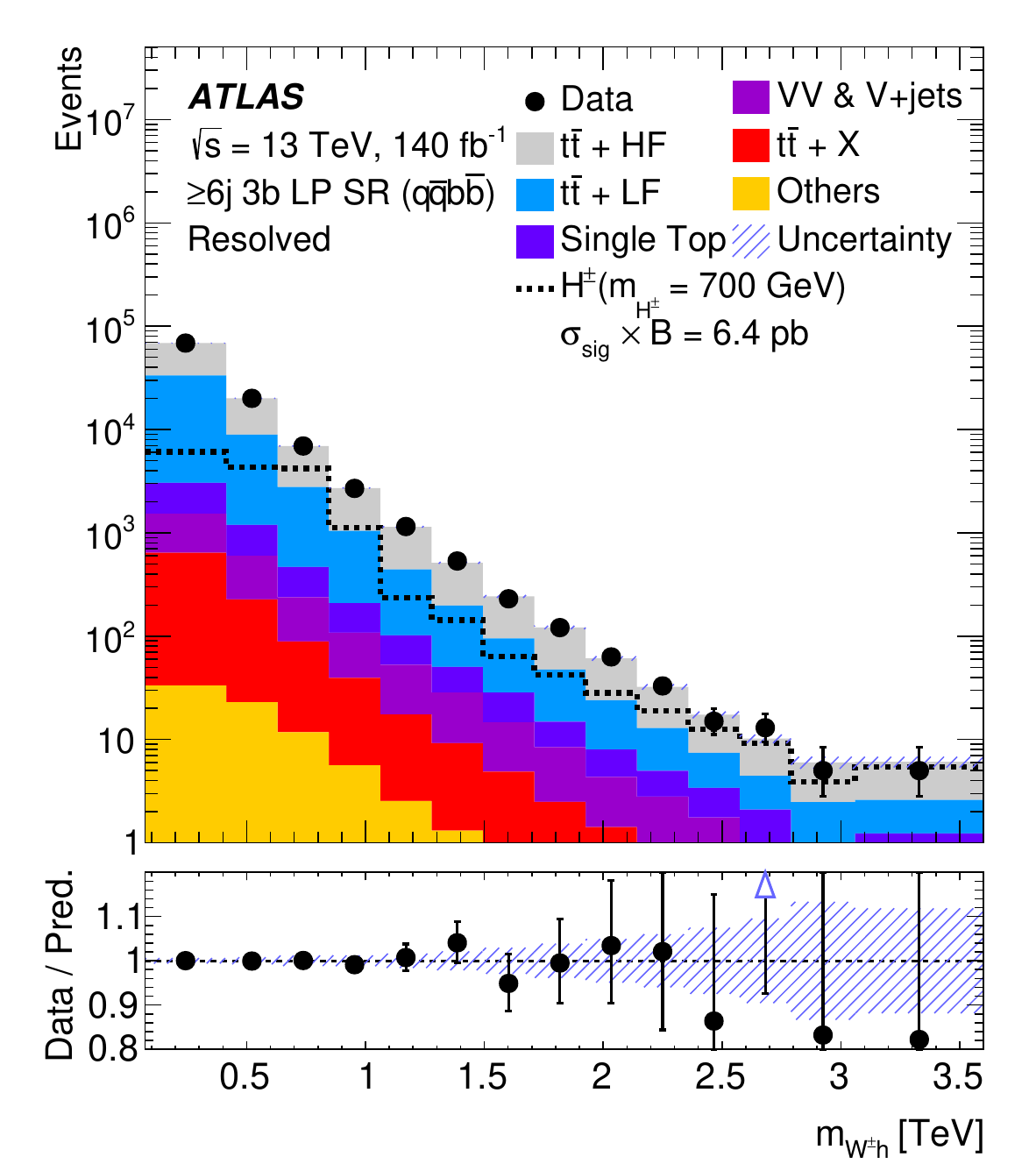}}
\subfloat[]{\includegraphics[width=0.495\textwidth]{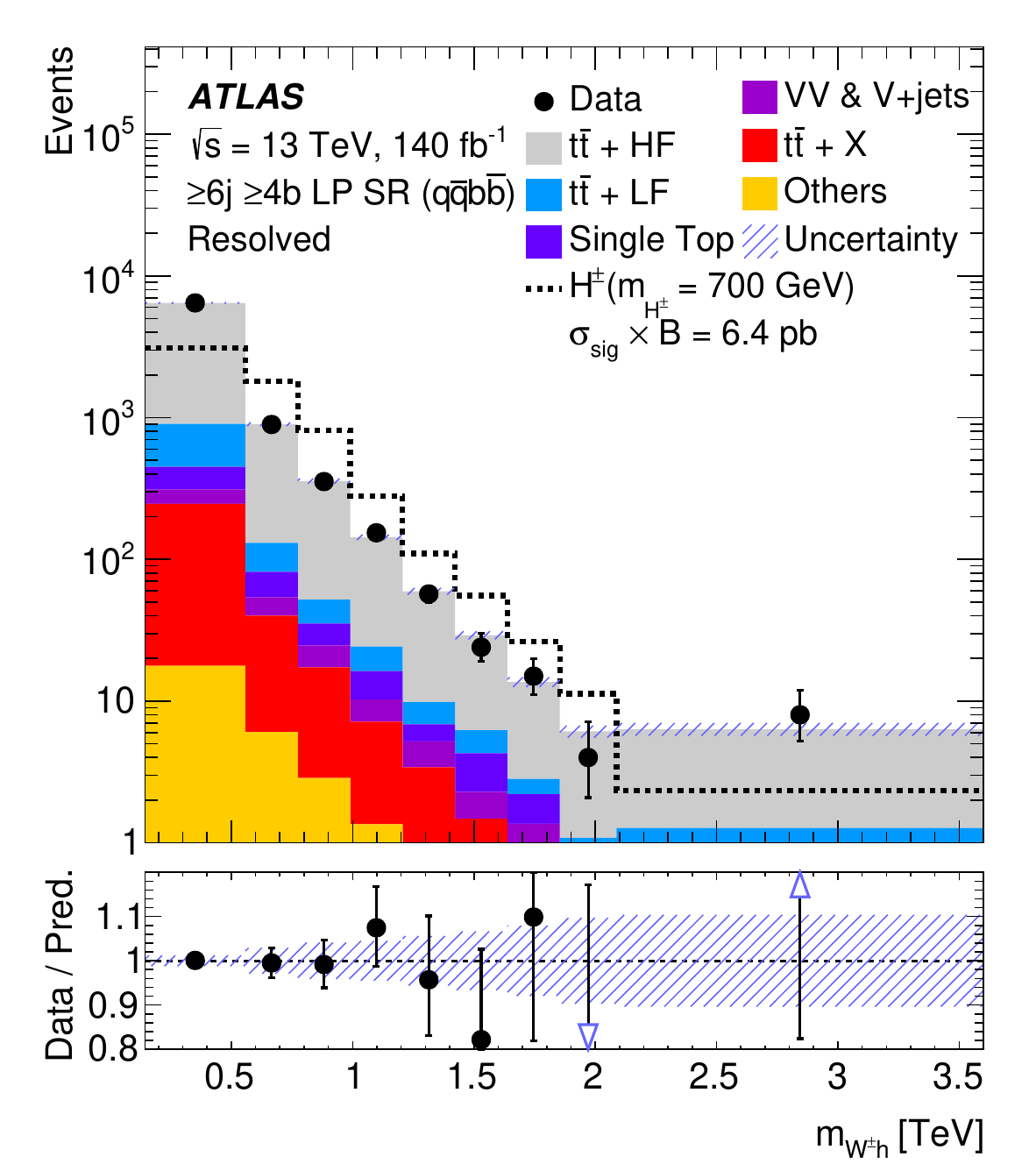}}
\caption{The $m_{W^\pm h}$ distributions in the low-purity signal regions (LP SR) of the resolved $q\bar{q}b\bar{b}$ event categories.
The term `Others' summarises events from $tHjb$, $tWh$, $t\bar{t}t\bar{t}$, and SM $Vh$ production.
The background prediction is shown after a background-only maximum-likelihood fit to data. The lower panels show the
ratio of the observed to the estimated SM background. The shaded bands show the total post-fit uncertainty in the background.
The expected signal contribution assuming $m_{H^\pm} = 700\,\GeV$,
normalised to a cross-section times branching ratio ($\sigma_\mathrm{sig} \times B$) of $6.4\,$pb, is shown as a dashed histogram.}
\label{fig:PostFit_mWH_resolved_qqbb_LPSR}
\end{figure}

\begin{figure}[h!]
\centering
\subfloat[]{\includegraphics[width=0.495\textwidth]{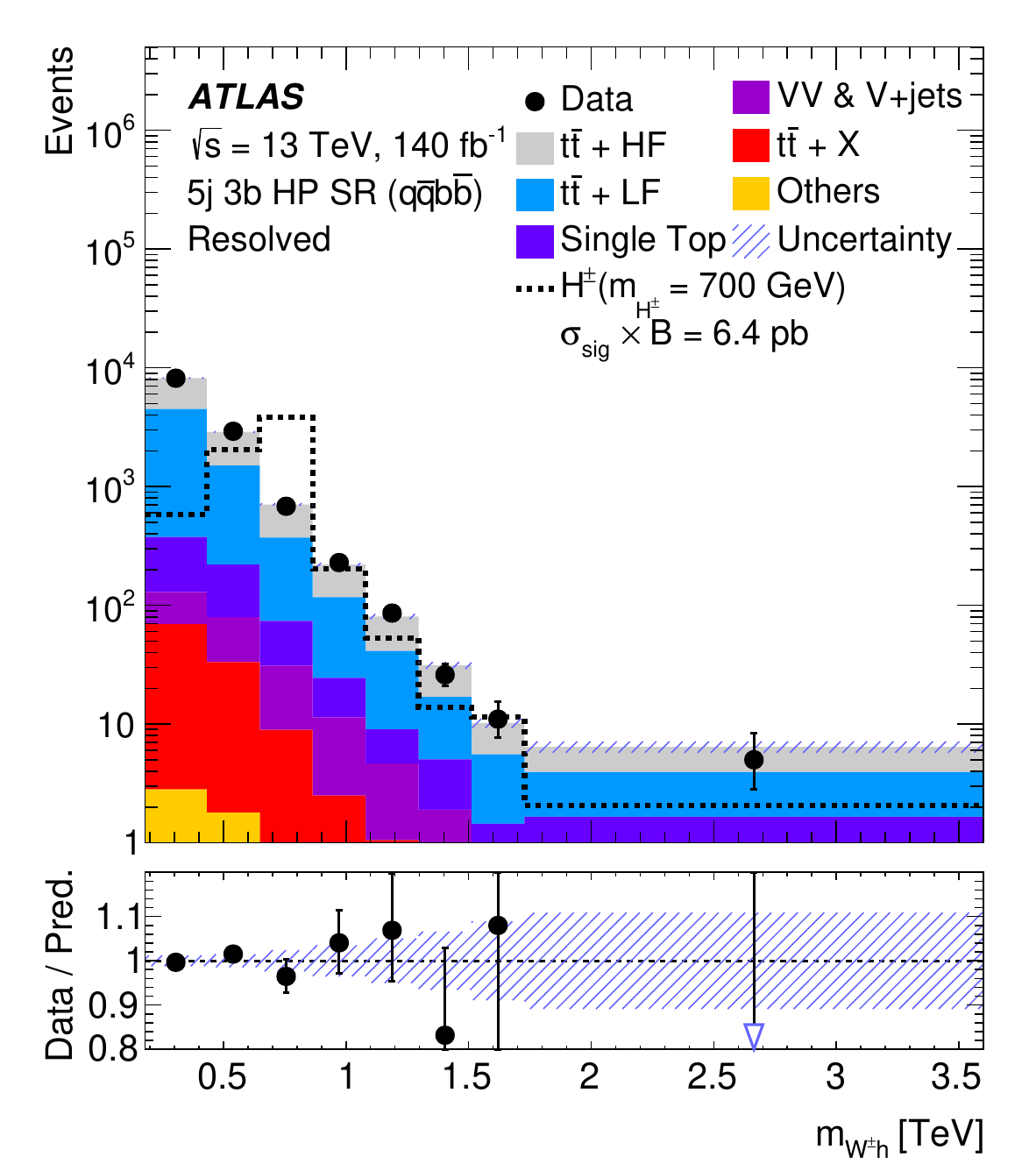}}
\subfloat[]{\includegraphics[width=0.495\textwidth]{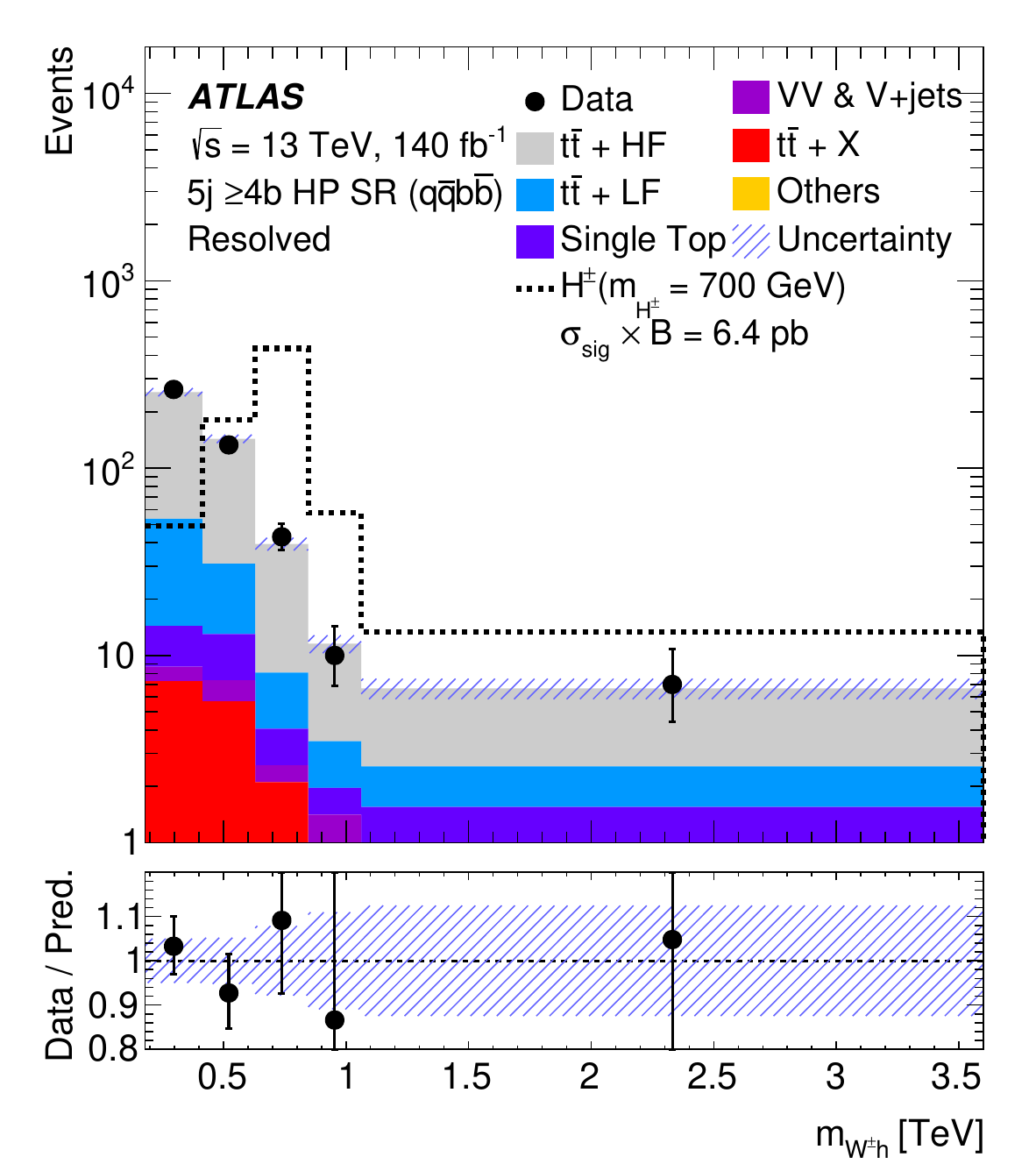}}\\
\subfloat[]{\includegraphics[width=0.495\textwidth]{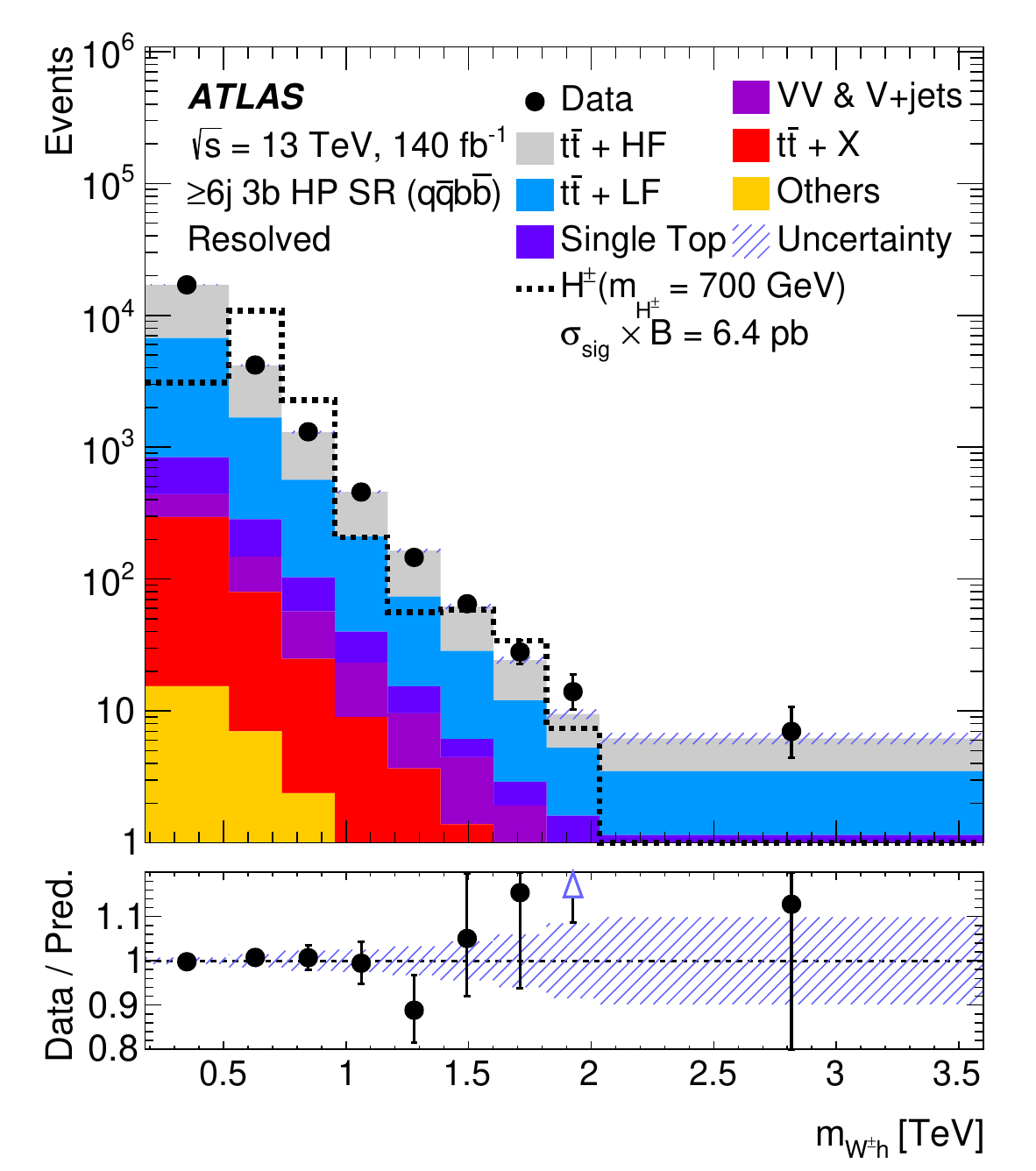}}
\subfloat[]{\includegraphics[width=0.495\textwidth]{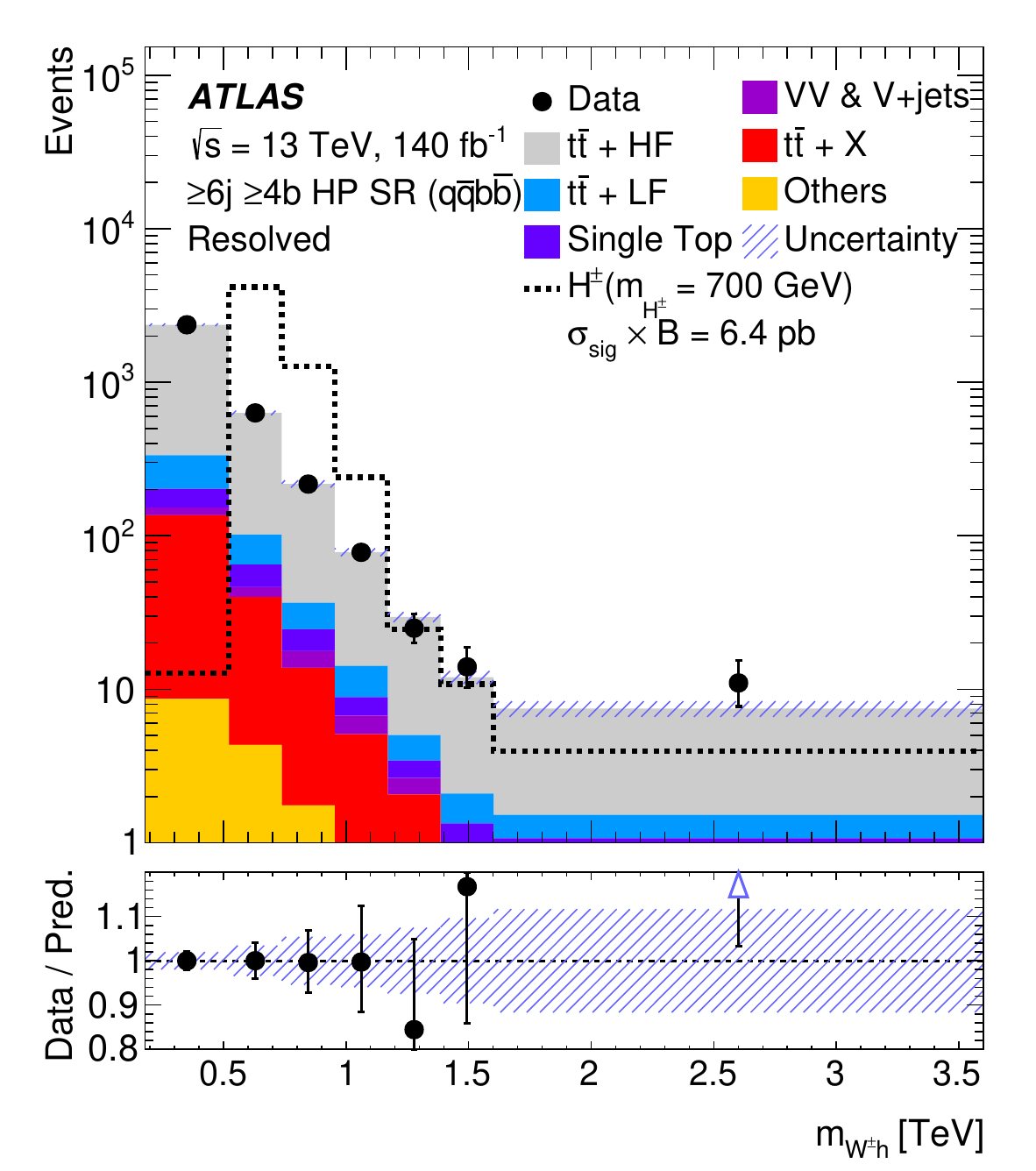}}
\caption{The $m_{W^\pm h}$ distributions in the high-purity signal regions (HP SR) of the resolved $q\bar{q}b\bar{b}$ event categories.
The term `Others' summarises events from $tHjb$, $tWh$, $t\bar{t}t\bar{t}$, and SM $Vh$ production.
The background prediction is shown after a background-only maximum-likelihood fit to data. The lower panels show the
ratio of the observed to the estimated SM background. The shaded bands show the total post-fit uncertainty in the
background. The expected signal contribution assuming $m_{H^\pm} = 700\,\GeV$, normalised to a cross-section
times branching ratio ($\sigma_\mathrm{sig} \times B$) of $6.4\,$pb, is shown as a dashed histogram.}
\label{fig:PostFit_mWH_resolved_qqbb_HPSR}
\end{figure}

\begin{figure}[h!]
\centering
\subfloat[]{\includegraphics[width=0.495\textwidth]{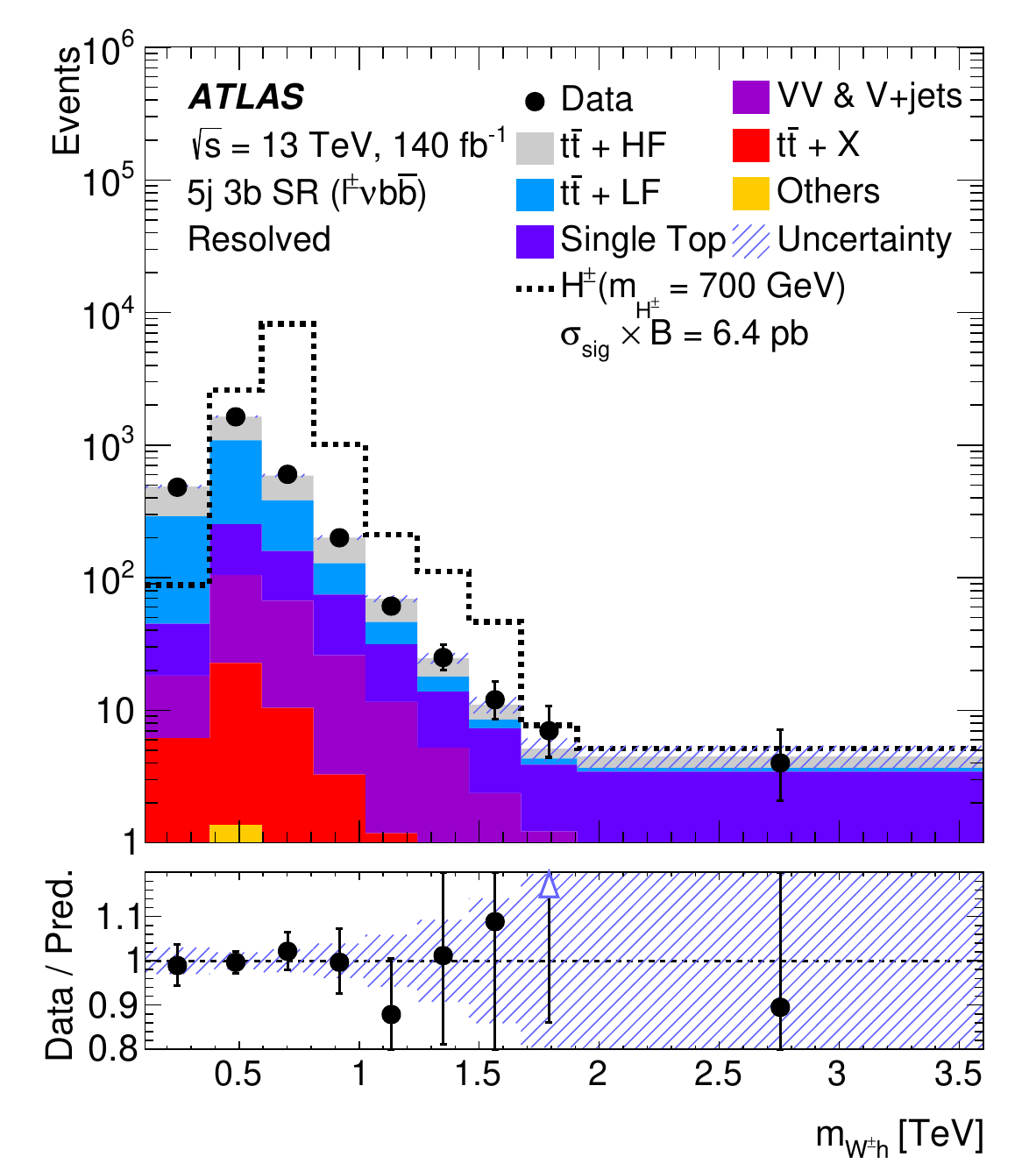}}
\subfloat[]{\includegraphics[width=0.495\textwidth]{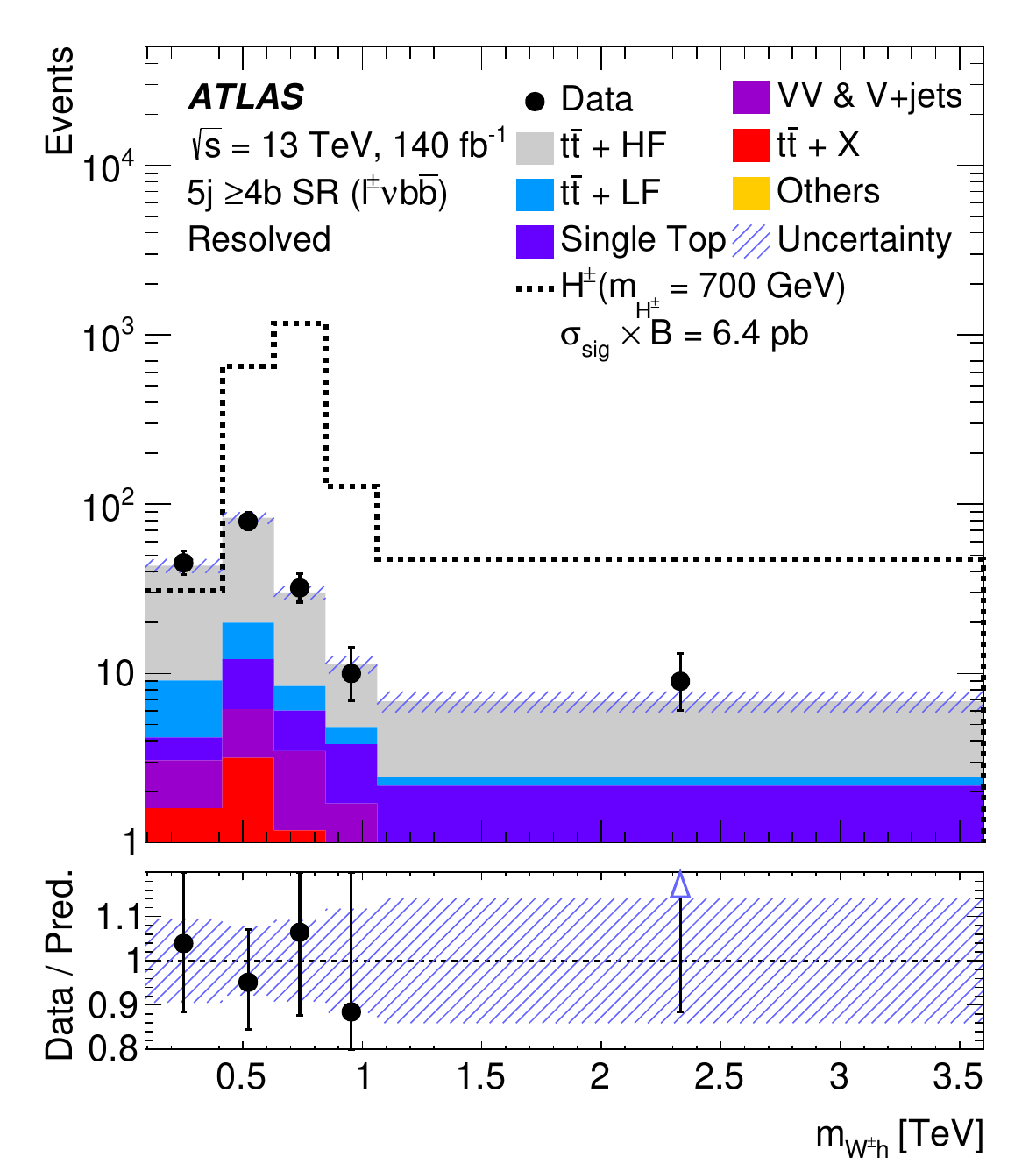}}\\
\subfloat[]{\includegraphics[width=0.495\textwidth]{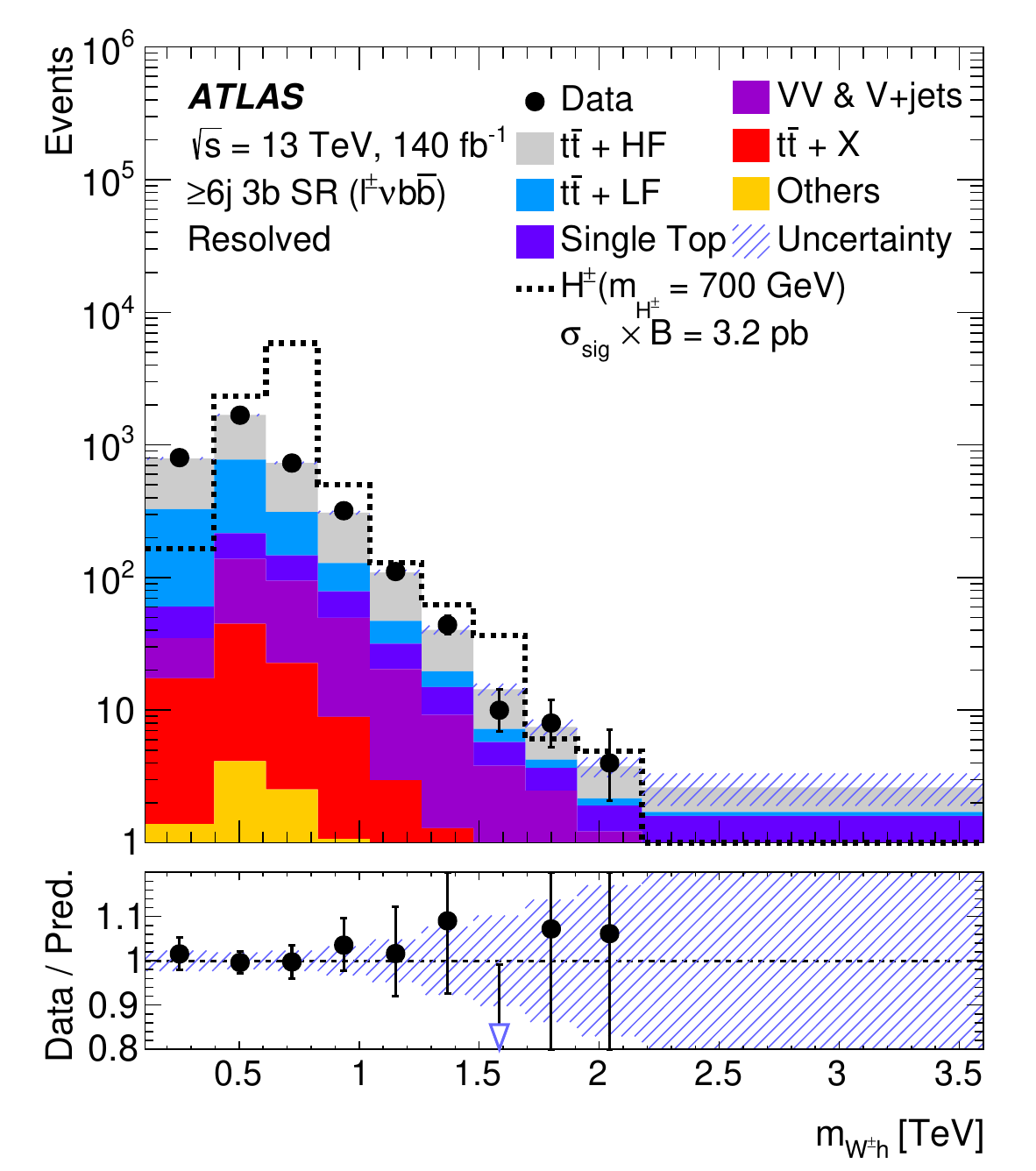}}
\subfloat[]{\includegraphics[width=0.495\textwidth]{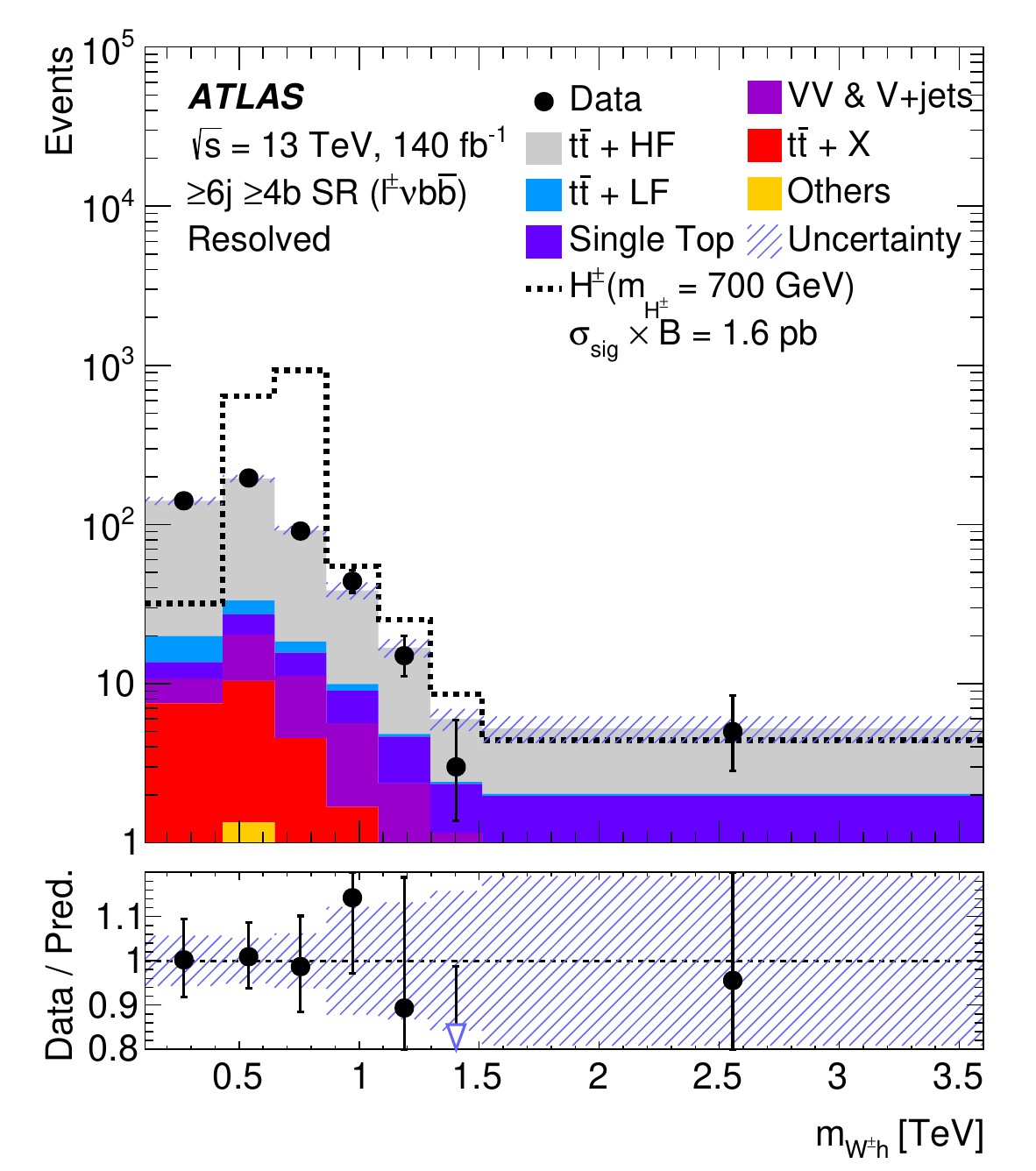}}
\caption{The $m_{W^\pm h}$ distributions in the signal regions (SRs) of the resolved $\ell^\pm\nu b\bar{b}$ event categories.
The term `Others' summarises events from $tHjb$, $tWh$, $t\bar{t}t\bar{t}$, and SM $Vh$ production.
The background prediction is shown after a background-only maximum-likelihood fit to data. The lower panels show the
ratio of the observed to the estimated SM background. The shaded bands show the total post-fit uncertainty in the
background. The expected signal contribution assuming $m_{H^\pm} = 700\,\GeV$, normalised to different values of the
cross-section times branching ratio ($\sigma_\mathrm{sig} \times B$), is shown as a dashed histogram.}
\label{fig:PostFit_mWH_resolved_lvbb_SR}
\end{figure}

\begin{figure}[h!]
\centering
\subfloat[]{\includegraphics[width=0.495\textwidth]{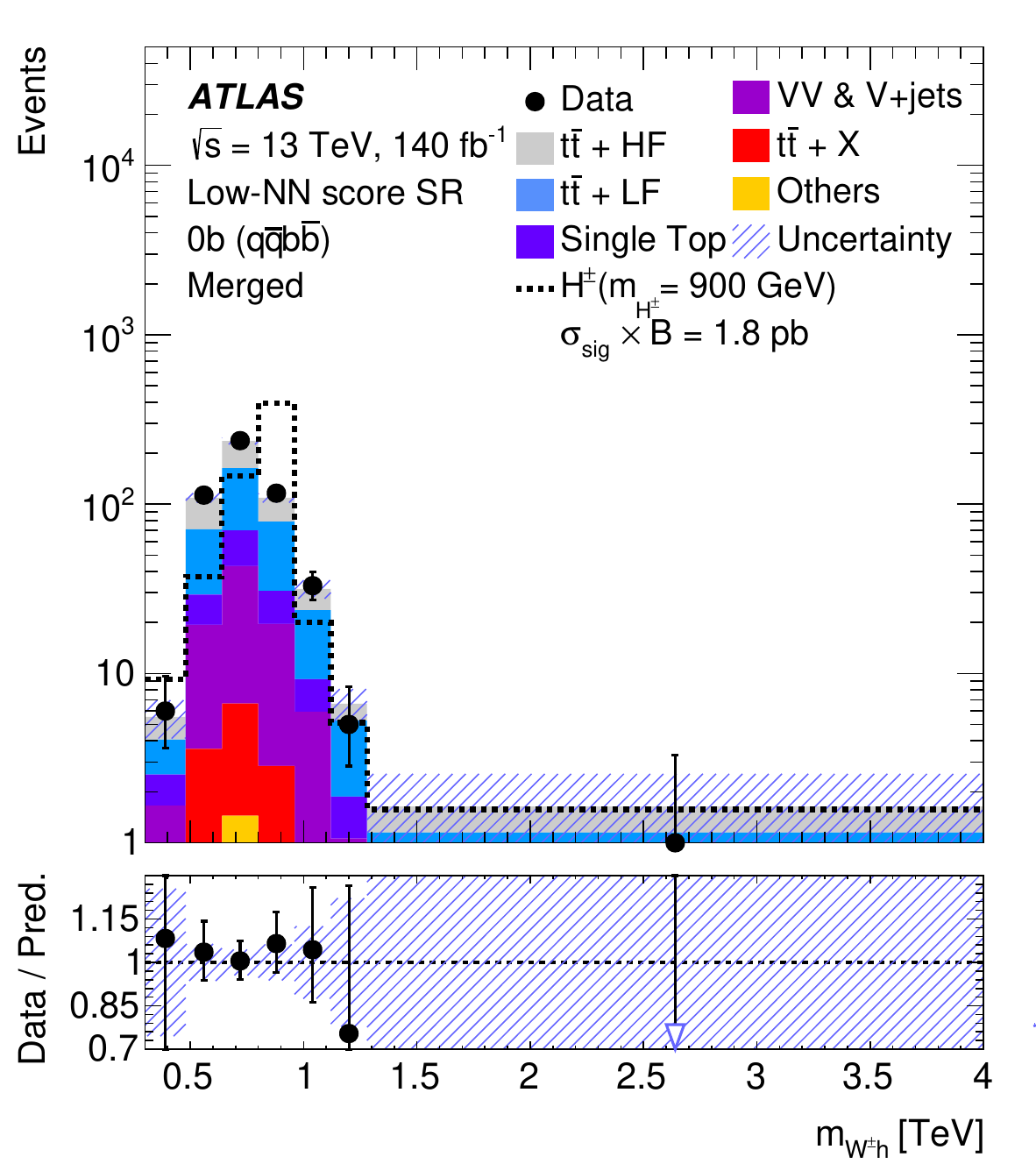}}
\subfloat[]{\includegraphics[width=0.495\textwidth]{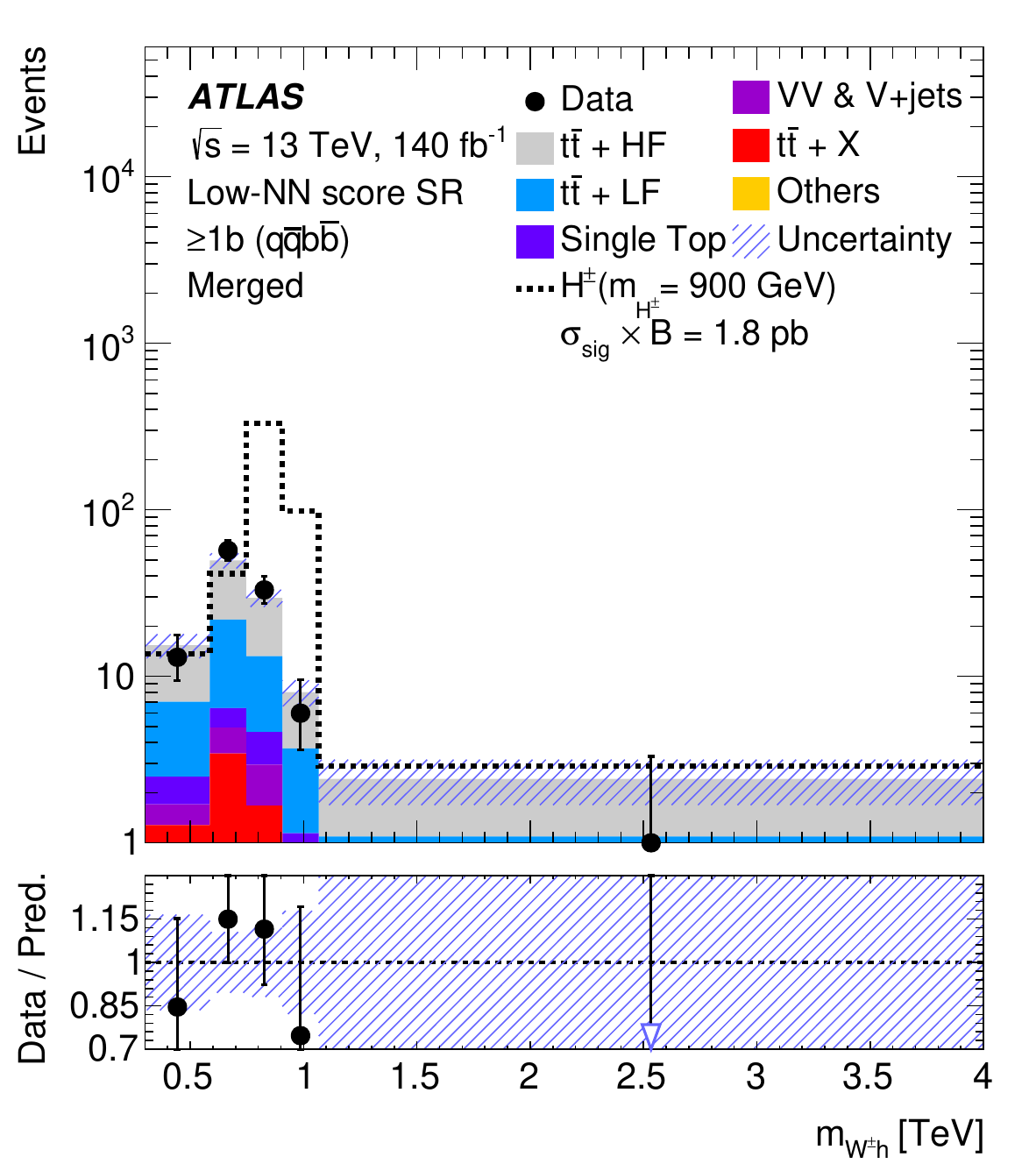}}\\
\subfloat[]{\includegraphics[width=0.495\textwidth]{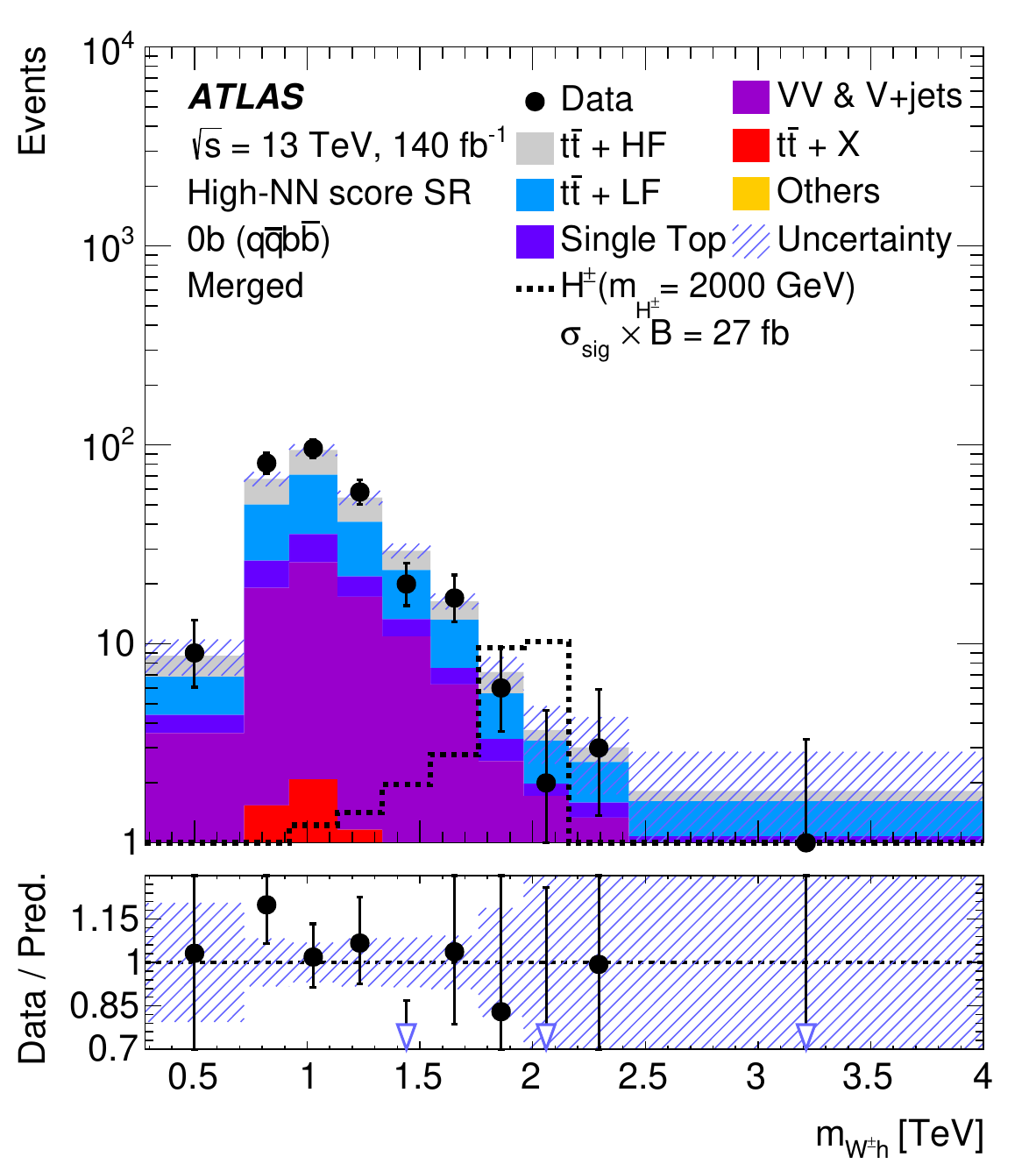}}
\subfloat[]{\includegraphics[width=0.495\textwidth]{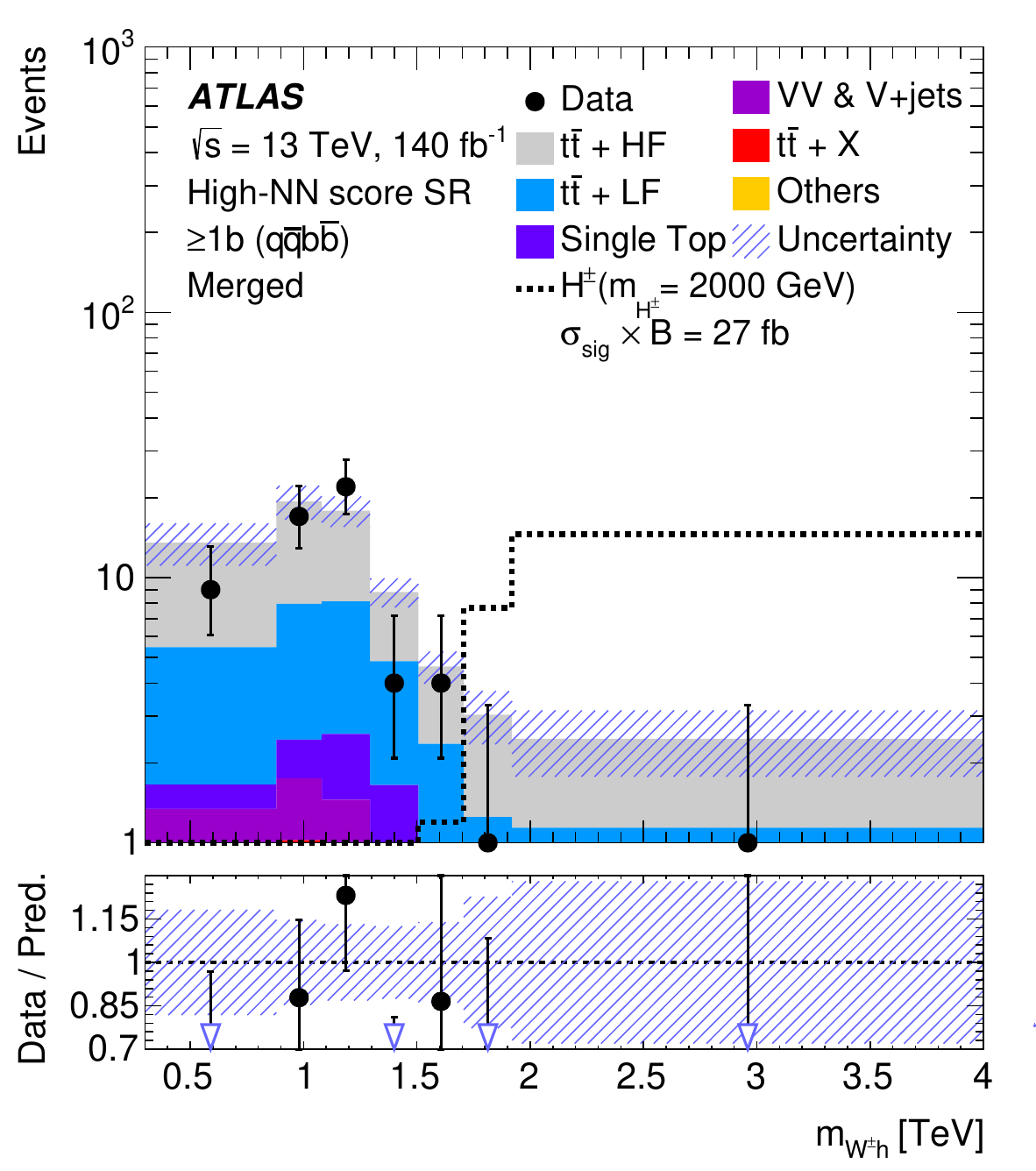}}
\caption{The $m_{W^\pm h}$ distributions in the low-NN-score signal regions (low-NN-score SRs) and high-NN-score signal regions
(high-NN-score SRs) of the merged $q\bar{q}b\bar{b}$ event categories.
The term `Others' summarises events from $tHjb$, $tWh$, $t\bar{t}t\bar{t}$, and SM $Vh$ production.
The background prediction is shown after a background-only maximum-likelihood fit to data. The lower panels show the
ratio of the observed to the estimated SM background. The shaded bands show the total post-fit uncertainty in
the background. The expected signal contributions assuming $m_{H^\pm} = 900\,\GeV$ and $m_{H^\pm} = 2000\,\GeV$,
normalised to cross-section times branching ratio ($\sigma_\mathrm{sig} \times B$) values of $1.8\,$pb and $27\,$fb
respectively, are shown as dashed histograms.}
\label{fig:PostFit_mWH_merged_qqbb_LS_HS_SRs}
\end{figure}

\begin{figure}[h!]
\centering
\subfloat[]{\includegraphics[width=0.495\textwidth]{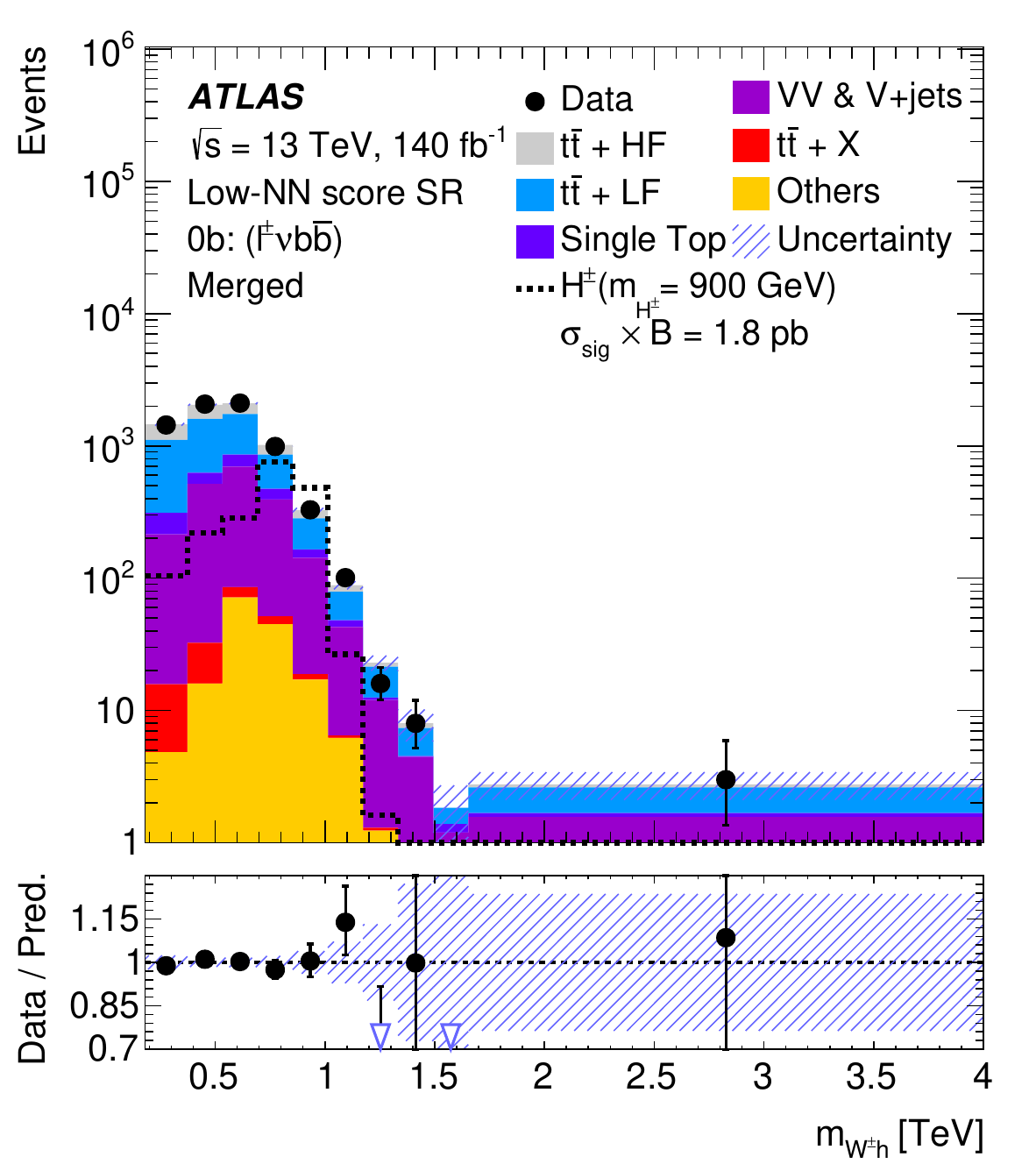}}
\subfloat[]{\includegraphics[width=0.495\textwidth]{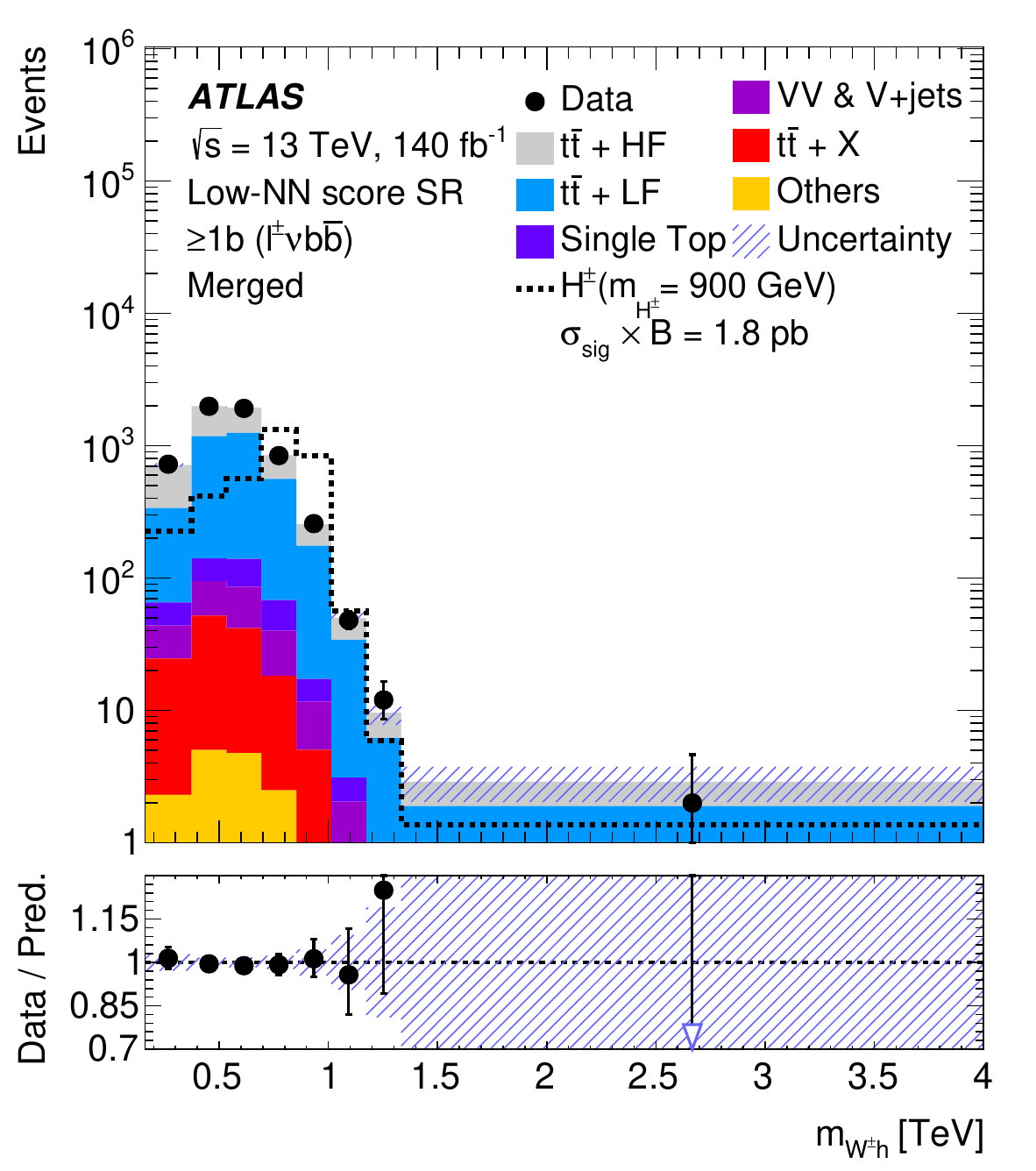}}\\
\subfloat[]{\includegraphics[width=0.495\textwidth]{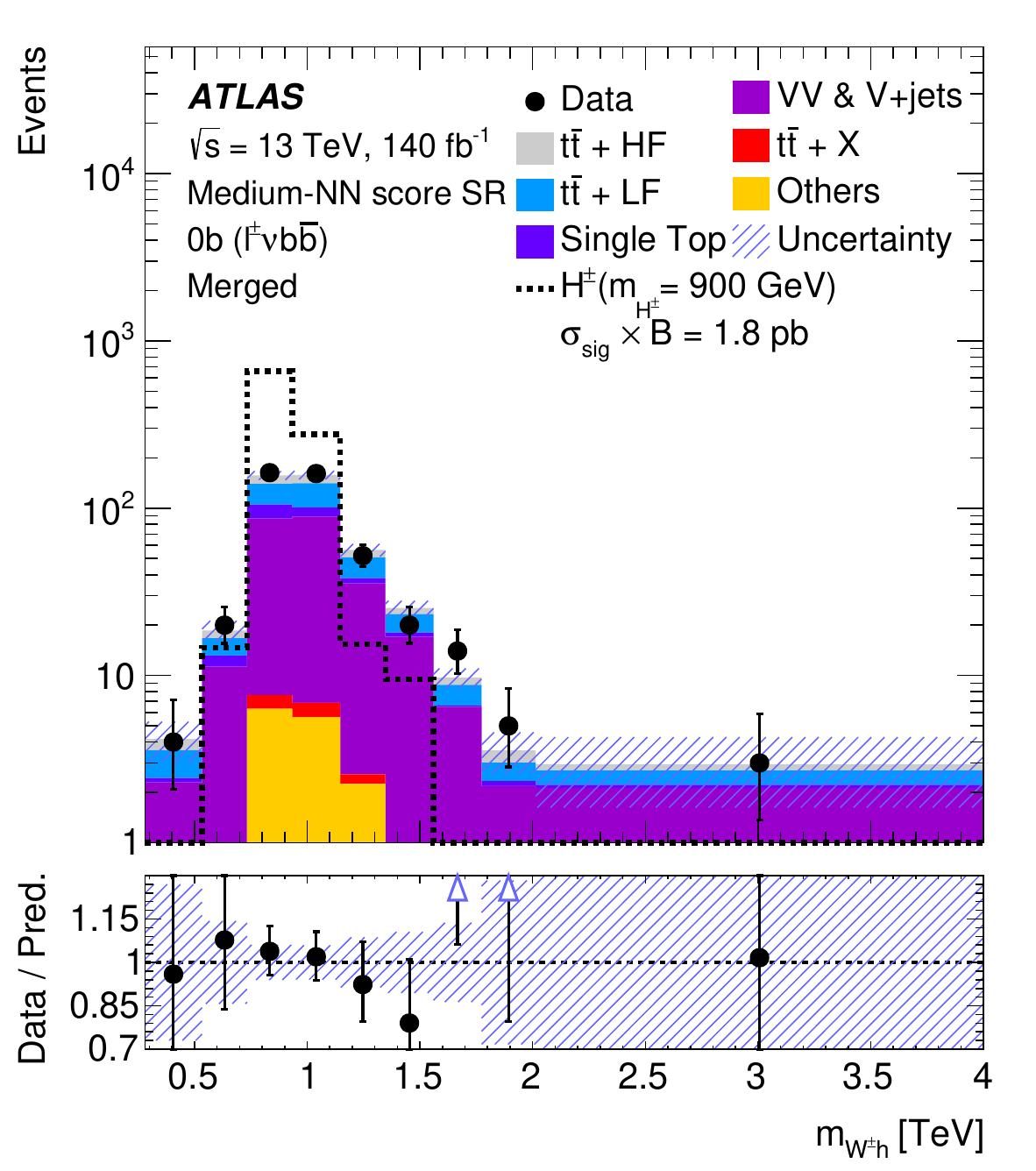}}
\subfloat[]{\includegraphics[width=0.495\textwidth]{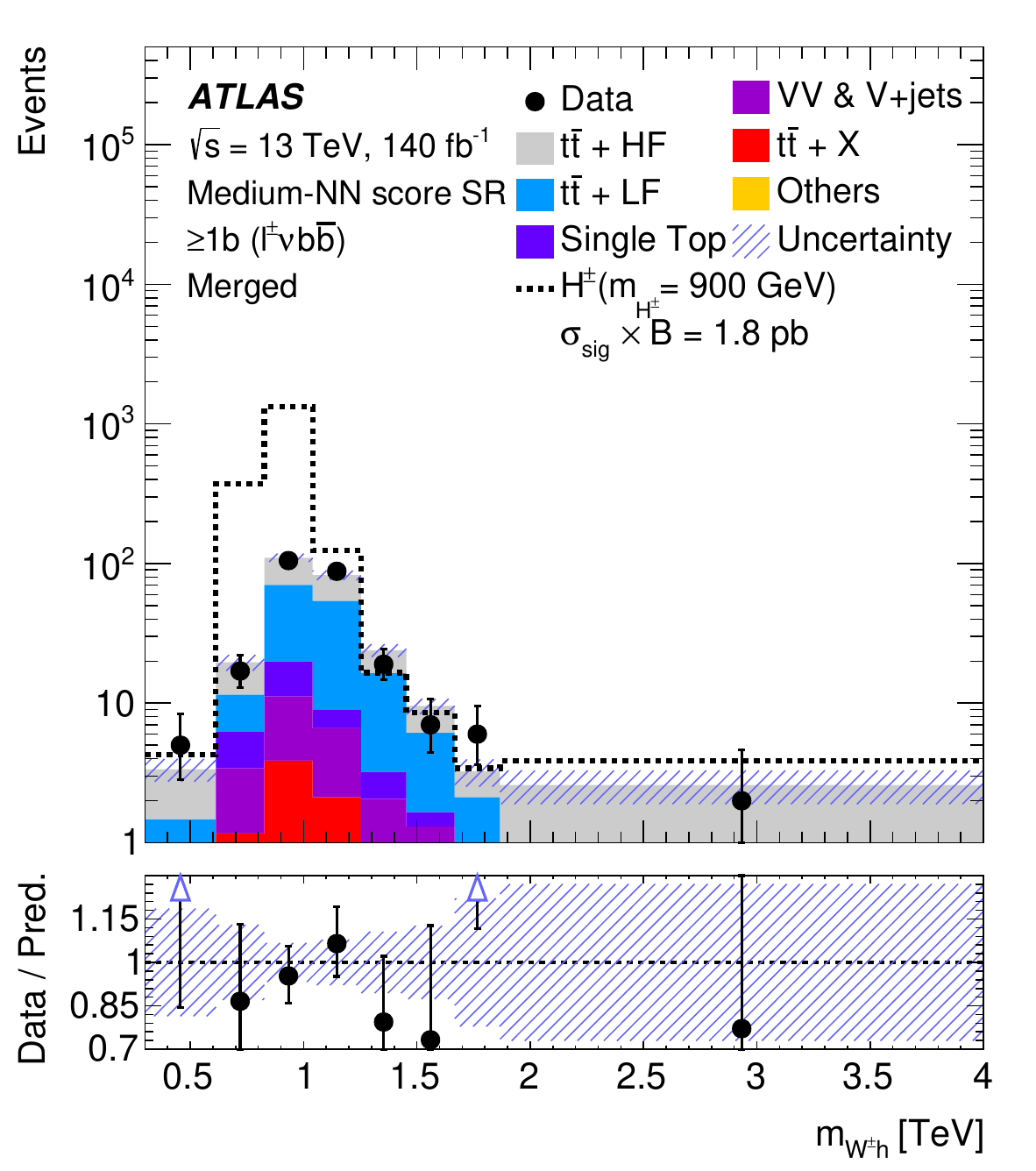}}
\caption{The $m_{W^\pm h}$ distributions in the low-NN-score signal regions (low-NN-score SRs) and medium-NN-score signal regions
(medium-NN-score SRs) of the merged $\ell^\pm\nu b\bar{b}$ event
categories. The term `Others' summarises events from $tHjb$, $tWh$, $t\bar{t}t\bar{t}$, and SM $Vh$ production.
The background prediction is shown after a background-only maximum-likelihood fit to data. The lower panels show the
ratio of the observed to the estimated SM background. The shaded bands show the total post-fit uncertainty
in the background. The expected signal contribution assuming $m_{H^\pm} = 900\,\GeV$, normalised to a cross-section times
branching ratio ($\sigma_\mathrm{sig} \times B$) values of $1.8\,$pb, is shown as a dashed histogram.}
\label{fig:PostFit_mWH_merged_lvbb_LS_MS_SRs}
\end{figure}

\begin{figure}[h!]
\centering
\subfloat[]{\includegraphics[width=0.495\textwidth]{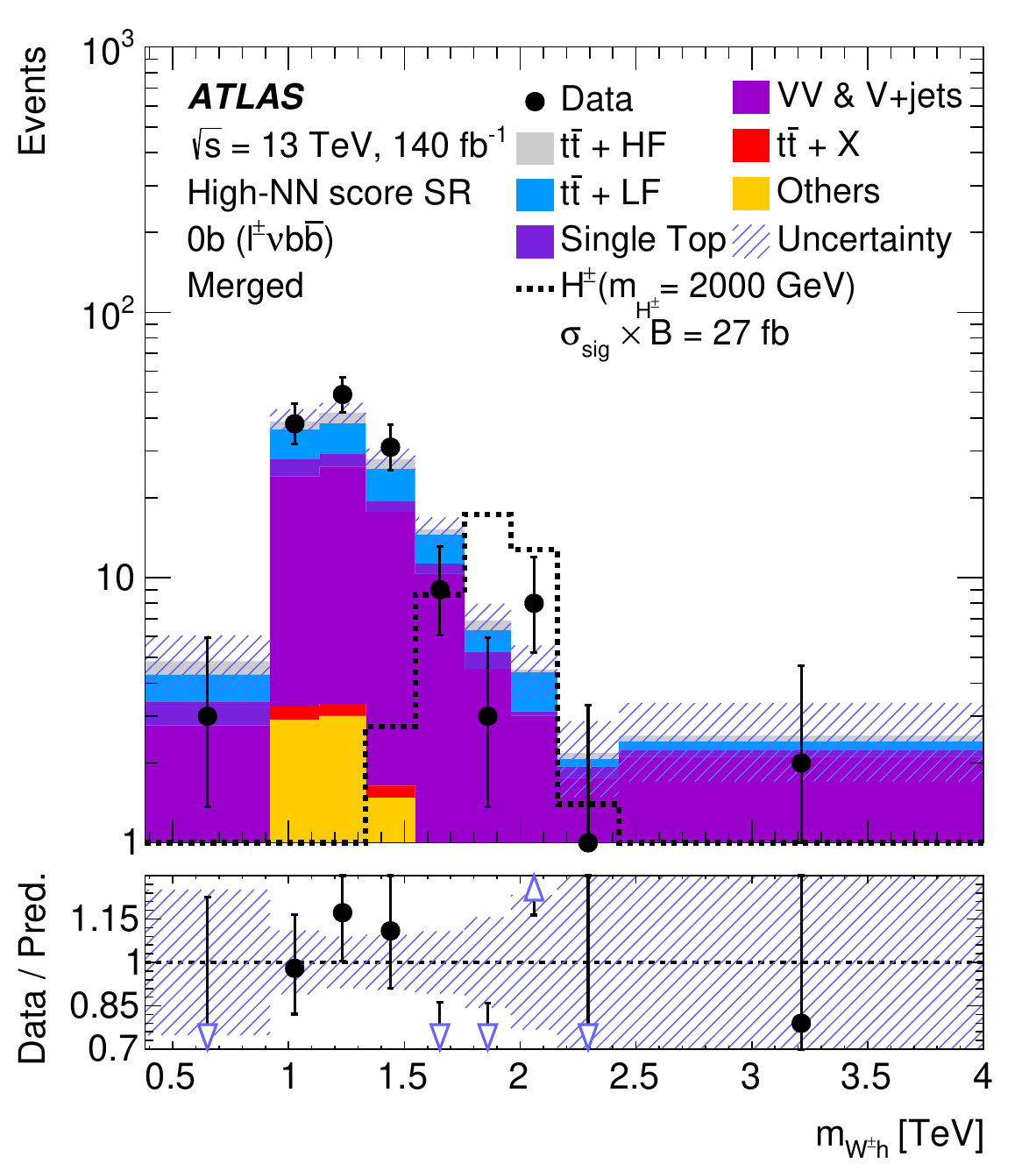}}
\subfloat[]{\includegraphics[width=0.495\textwidth]{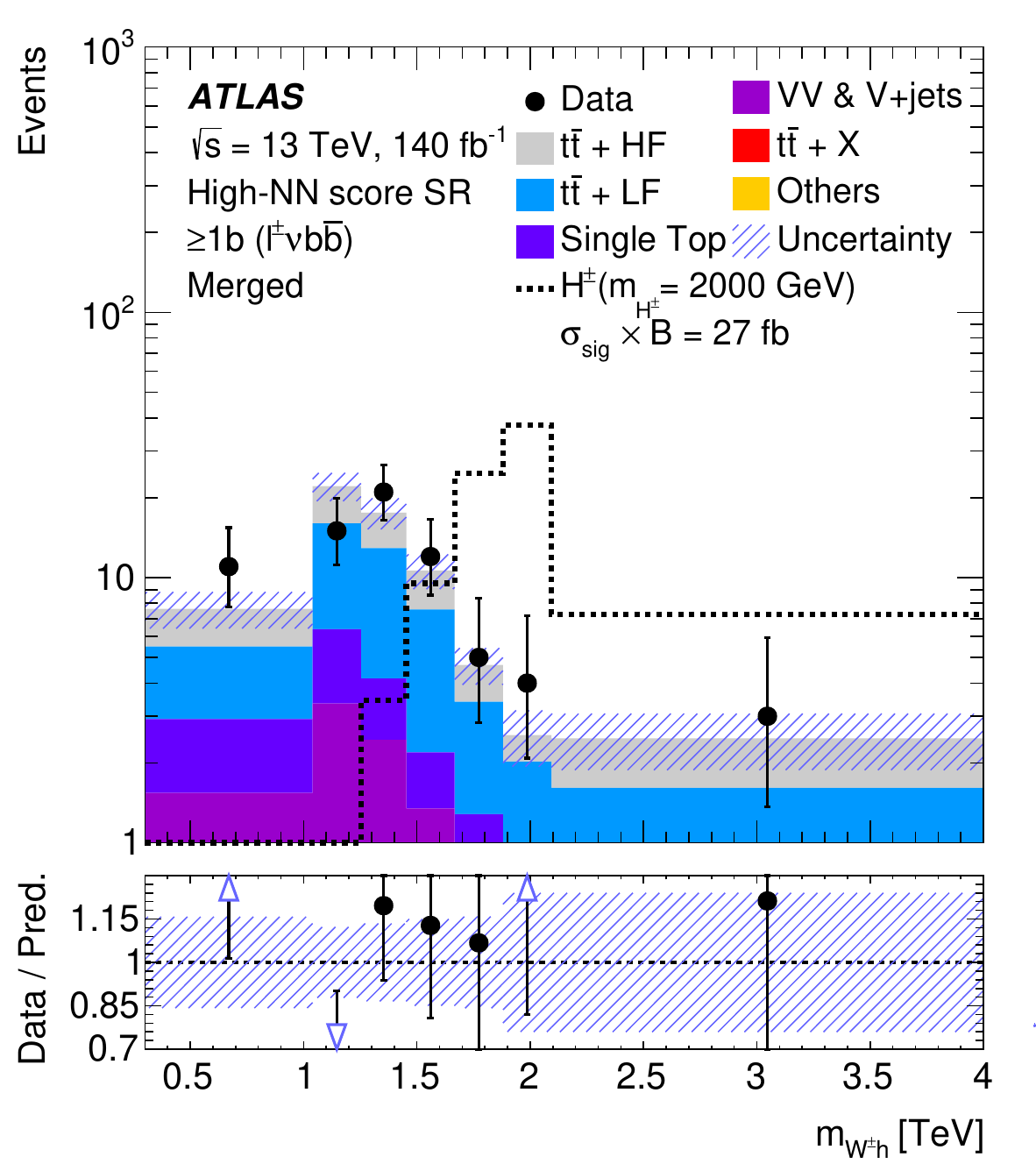}}
\caption{The $m_{W^\pm h}$ distributions in the high-NN-score signal regions (high-NN-score SRs) of the merged $\ell^\pm\nu b\bar{b}$ event
categories. The term `Others' summarises events from $tHjb$, $tWh$, $t\bar{t}t\bar{t}$, and SM $Vh$ production.
The background prediction is shown after a background-only maximum-likelihood fit to data. The lower panels show the
ratio of the observed to the estimated SM background. The shaded bands show the total post-fit uncertainty
in the background. The expected signal contribution assuming $m_{H^\pm} = 2000\,\GeV$, normalised to a cross-section times
branching ratio ($\sigma_\mathrm{sig} \times B$) of $27\,$fb, is shown as a dashed histogram.}
\label{fig:PostFit_mWH_merged_lvbb_HS_SRs}
\end{figure}

The $m_{W^\pm h}$ distributions after a background-only fit to data are shown in Figures~\ref{fig:CR_summary_plots_I} to~\ref{fig:CR_summary_plots_II}
for the control regions of the resolved and merged analyses. The corresponding signal region distributions are shown in
Figures~\ref{fig:PostFit_mWH_resolved_qqbb_LPSR} to~\ref{fig:PostFit_mWH_resolved_lvbb_SR}
for the signal regions of the resolved analysis and in Figures~\ref{fig:PostFit_mWH_merged_qqbb_LS_HS_SRs} to~\ref{fig:PostFit_mWH_merged_lvbb_HS_SRs}
for the signal regions of the merged analysis. Additionally, the expected and observed event yields
after fits of the backgrounds to data performed under the background-only hypothesis are shown in Table~\ref{Tab:EventYieldsPostFit}.

\clearpage

\clearpage

\clearpage

\clearpage

\clearpage

\begin{table}[!h]
\begin{center}
\caption{
Event yields in the various signal and control regions of the (a) resolved and (b) merged analyses after a background-only
fit to data. The quoted uncertainties are the total post-fit uncertainties. The uncertainties
in the individual background predictions are larger than the total background uncertainty due to correlations resulting from the fit to data.}
\subfloat[Post-fit event yields of the resolved event categories]{\resizebox{1.0\textwidth}{!}{
\begin{tabular}{ |c |c c c c c c |c |c |}
\hline
& $t\bar{t}$ + LF & $t\bar{t}$ + HF & Single top & $VV$ \& $V$ + jets & $t\bar{t}$ + X & Others & Total bkg. & Data \\
\hline
$5j3b$ CR ($q\bar{q}b\bar{b}$)                     &  $9000 \pm 500$  &  $8400 \pm 600$  &  $770 \pm 190$  &  $500 \pm 150$  &  $64 \pm 11$  &  $5.0 \pm 0.6$  &  $18720 \pm 150$  &  $18737$\\
$5j\geq4b$ CR ($q\bar{q}b\bar{b}$)                 &  $470 \pm 80$  &  $3640 \pm 130$  &  $160 \pm 50$  &  $69 \pm 20$  &  $100 \pm 15$  &  $5.3 \pm 0.6$  &  $4450 \pm 70$  &  $4449$\\
$\geq6j3b$ CR ($q\bar{q}b\bar{b}$)                 &  $2200 \pm 400$  &  $4100 \pm 400$  &  $250 \pm 80$  &  $180 \pm 60$  &  $41 \pm 8$  &  $6.0 \pm 2.4$  &  $6790 \pm 90$  &  $6788$\\
$\geq6j\geq4b$ CR ($q\bar{q}b\bar{b}$)             &  $330 \pm 110$  &  $5140 \pm 150$  &  $160 \pm 60$  &  $87 \pm 25$  &  $147 \pm 18$  &  $22 \pm 9$  &  $5890 \pm 80$  &  $5889$\\
$5j3b$ low-purity SR ($q\bar{q}b\bar{b}$)          &  $57000 \pm 4000$  &  $38000 \pm 4000$  &  $3300 \pm 900$  &  $1600 \pm 500$  &  $500 \pm 80$  &  $23.4 \pm 1.4$  &  $100970 \pm 330$  &  $100957$\\
$5j\geq4b$ low-purity SR ($q\bar{q}b\bar{b}$)      &  $330 \pm 60$  &  $1650 \pm 90$  &  $67 \pm 24$  &  $32 \pm 9$  &  $66 \pm 10$  &  $2.44 \pm 0.22$  &  $2140 \pm 50$  &  $2145$\\
$\geq6j3b$ low-purity SR ($q\bar{q}b\bar{b}$)      &  $42000 \pm 6000$  &  $53000 \pm 6000$  &  $2500 \pm 700$  &  $1600 \pm 500$  &  $960 \pm 160$  &  $79 \pm 31$  &  $100500 \pm 500$  &  $100485$\\
$\geq6j\geq4b$ low-purity SR ($q\bar{q}b\bar{b}$)  &  $530 \pm 140$  &  $6830 \pm 200$  &  $190 \pm 80$  &  $94 \pm 27$  &  $290 \pm 40$  &  $29 \pm 13$  &  $7960 \pm 90$  &  $7963$\\
$5j3b$ high-purity SR ($q\bar{q}b\bar{b}$)         &  $5800 \pm 700$  &  $5600 \pm 700$  &  $450 \pm 160$  &  $140 \pm 40$  &  $110 \pm 19$  &  $5.53 \pm 0.29$  &  $12130 \pm 140$  &  $12130$\\
$5j\geq4b$ high-purity SR ($q\bar{q}b\bar{b}$)     &  $64 \pm 16$  &  $357 \pm 28$  &  $14 \pm 11$  &  $4.5 \pm 1.7$  &  $16.0 \pm 2.9$  &  $0.58 \pm 0.16$  &  $456 \pm 21$  &  $456$\\
$\geq6j3b$ high-purity SR ($q\bar{q}b\bar{b}$)     &  $8000 \pm 1400$  &  $14000 \pm 1400$  &  $610 \pm 220$  &  $270 \pm 80$  &  $390 \pm 60$  &  $26 \pm 10$  &  $23260 \pm 160$  &  $23258$\\
$\geq6j\geq4b$ high-purity SR ($q\bar{q}b\bar{b}$) &  $190 \pm 60$  &  $2850 \pm 100$  &  $80 \pm 40$  &  $30 \pm 10$  &  $182 \pm 27$  &  $16 \pm 7$  &  $3340 \pm 60$  &  $3343$ \\
\hline
$5j3b$ CR ($\ell^\pm\nu b\bar{b}$)                     &  $820 \pm 80$  &  $1310 \pm 130$  &  $230 \pm 90$  &  $220 \pm 70$  &  $14 \pm 4$  &  $1.22 \pm 0.16$  &  $2600 \pm 50$  &  $2599$\\
$5j\geq4b$ CR ($\ell^\pm\nu b\bar{b}$)                 &  $5.2 \pm 2.2$  &  $34 \pm 6$  &  $3.0 \pm 1.9$  &  $5.0 \pm 1.7$  &  $0.60 \pm 0.15$  &  $0.05 \pm 0.05$  &  $48 \pm 6$  &  $49$\\
$\geq6j3b$ CR ($\ell^\pm\nu b\bar{b}$)                 &  $710 \pm 120$  &  $1870 \pm 180$  &  $140 \pm 70$  &  $240 \pm 70$  &  $33 \pm 7$  &  $5.0 \pm 2.2$  &  $3000 \pm 50$  &  $2991$\\
$\geq6j\geq4b$ CR ($\ell^\pm\nu b\bar{b}$)             &  $6.2 \pm 2.7$  &  $121 \pm 14$  &  $7 \pm 5$  &  $8.9 \pm 2.7$  &  $2.5 \pm 0.4$  &  $0.68 \pm 0.32$  &  $146 \pm 12$  &  $147$\\
$5j3b$ SR ($\ell^\pm\nu b\bar{b}$)                     &  $1380 \pm 140$  &  $1060 \pm 180$  &  $350 \pm 80$  &  $190 \pm 60$  &  $42 \pm 8$  &  $2.72 \pm 0.24$  &  $3030 \pm 50$  &  $3026$\\
$5j\geq4b$ SR ($\ell^\pm\nu b\bar{b}$)                 &  $16 \pm 4$  &  $130 \pm 15$  &  $13 \pm 6$  &  $8.8 \pm 2.7$  &  $6.2 \pm 0.9$  &  $0.43 \pm 0.08$  &  $175 \pm 13$  &  $175$\\
$\geq6j3b$ SR ($\ell^\pm\nu b\bar{b}$)                 &  $1060 \pm 170$  &  $2060 \pm 220$  &  $210 \pm 100$  &  $260 \pm 80$  &  $89 \pm 16$  &  $10 \pm 4$  &  $3690 \pm 60$  &  $3703$\\
$\geq6j\geq4b$ SR ($\ell^\pm\nu b\bar{b}$)             &  $16 \pm 5$  &  $403 \pm 32$  &  $23 \pm 18$  &  $27 \pm 8$  &  $21.7 \pm 2.9$  &  $3.5 \pm 1.7$  &  $494 \pm 22$  &  $495$\\
\hline
\end{tabular}}}

\subfloat[Post-fit event yields of the merged event categories]{\resizebox{1.0\textwidth}{!}{
\begin{tabular}{ |c |c c c c c c |c |c |}
\hline
& $t\bar{t}$ + LF & $t\bar{t}$ + HF & Single top & $VV$ \& $V$ + jets & $t\bar{t}$ + X & Others & Total bkg. & Data \\
\hline
High-$m_{h}$ $\geq1b$ CR ($q\bar{q}b\bar{b}$)    &$   39 \pm   9  $&$    86 \pm  14  $&$   5.0 \pm 3.5  $&$   4.6 \pm  0.8  $&$  4.0 \pm 1.5  $&$   0.6   \pm 0.3    $&$ 139  \pm 10  $&$ 138$ \\
Low-$m_{h}$  $\geq1b$ CR ($q\bar{q}b\bar{b}$)    &$   33 \pm   7  $&$    82 \pm  11  $&$   9   \pm 5    $&$  11.4 \pm  1.9  $&$  3.2 \pm 0.9  $&$    0.24 \pm 0.11   $&$ 139  \pm  9  $&$ 142 $\\
High-$m_{h}$ $0b$ CR ($q\bar{q}b\bar{b}$)        &$  300 \pm  50  $&$   202 \pm  35  $&$  50   \pm 30   $&$  92   \pm   16  $&$ 12.0 \pm 2.5  $&$   0.9   \pm 0.3    $&$ 659  \pm 22  $&$ 632$\\
Low-$m_{h}$ $0b$ CR ($q\bar{q}b\bar{b}$)         &$  180 \pm  30  $&$   187 \pm  28  $&$  70   \pm 40   $&$ 281   \pm  42   $&$  6.4 \pm 1.2  $&$   1.0   \pm 0.4    $&$ 724  \pm 24  $&$ 728 $\\
Low-NN-score $0b$ SR ($q\bar{q}b\bar{b}$)        &$  203 \pm  29  $&$   152 \pm 24   $&$  53   \pm 27   $&$ 77    \pm  16   $&$ 11.1 \pm 2.2  $&$   3.0   \pm 1.2    $&$ 499 \pm 17   $&$ 511$\\
Low-NN-score $\geq1b$ SR ($q\bar{q}b\bar{b}$)    &$   32 \pm  10  $&$    58 \pm 11   $&$   4.6 \pm 1.3  $&$ 3.6   \pm   1.8 $&$ 6.2  \pm 1.3  $&$   0.64  \pm 0.20   $&$ 105 \pm  8   $&$ 110 $\\
High-NN-score $0b$ SR ($q\bar{q}b\bar{b}$)       &$  102 \pm  16  $&$    68 \pm 17   $&$  28   \pm 17   $&$ 83    \pm  22   $&$ 4.4  \pm 0.9  $&$   1.8   \pm 0.8    $&$ 286 \pm  14  $&$ 293$ \\
High-NN-score $\geq1b$ SR ($q\bar{q}b\bar{b}$)   &$   21 \pm   4  $&$    39 \pm 7    $&$   3.5 \pm 2.7  $&$ 3.0   \pm   0.8 $&$ 2.9  \pm 0.8  $&$   0.27  \pm 0.08   $&$  70 \pm  6   $&$ 58$\\
\hline
High-$m_{h}$ $\geq1b$ CR ($\ell^\pm\nu b\bar{b}$)    &$ 2230 \pm 240  $&$  1870 \pm 250  $&$ 110   \pm 40   $&$  78   \pm  13   $&$ 79   \pm 16   $&$   7.6   \pm 2.8    $&$ 4380 \pm  60 $&$ 4414$\\
Low-$m_{h}$ $\geq1b$ CR ($\ell^\pm\nu b\bar{b}$)     &$ 1840 \pm 200  $&$  1860 \pm 220  $&$ 190   \pm 60   $&$ 278   \pm   53  $&$ 48   \pm 9    $&$   5.0   \pm 1.2    $&$ 4220 \pm  60 $&$ 4202$ \\
High-$m_{h}$ $0b$ CR ($\ell^\pm\nu b\bar{b}$)        &$ 2480 \pm 240  $&$  1160 \pm 170  $&$ 420   \pm 160  $&$ 940   \pm  240  $&$ 50   \pm 9    $&$  14     \pm 7      $&$ 5060 \pm  70 $&$ 5063$\\
Low-$m_{h}$ $0b$ CR ($\ell^\pm\nu b\bar{b}$)         &$ 1950 \pm 250  $&$   930 \pm 130  $&$ 560   \pm 160  $&$ 3840  \pm 280   $&$ 23   \pm 5    $&$  34     \pm 17     $&$ 7340 \pm 80  $&$ 7357$\\
Low-NN-score $0b$ SR ($\ell^\pm\nu b\bar{b}$)        &$ 3210 \pm 310  $&$  1360 \pm 200  $&$ 480   \pm 130  $&$ 1820  \pm 250   $&$ 50   \pm 9    $&$ 160     \pm 80     $&$ 7070 \pm 80  $&$ 7066$\\
Low-NN-score $\geq1b$ SR ($\ell^\pm\nu b\bar{b}$)    &$ 3120 \pm 310  $&$  2260 \pm 310  $&$ 160   \pm 50   $&$ 137   \pm  19   $&$ 127  \pm 25   $&$  15.3   \pm 3.0    $&$ 5810 \pm 70  $&$ 5785$\\
Medium-NN-score $0b$ SR ($\ell^\pm\nu b\bar{b}$)     &$  101 \pm  28  $&$    46 \pm 12   $&$  36   \pm 22   $&$ 230   \pm  30   $&$ 3.3  \pm 0.7  $&$  16     \pm 8      $&$ 435 \pm 20   $&$ 442 $\\
Medium-NN-score $\geq1b$ SR ($\ell^\pm\nu b\bar{b}$) &$  121 \pm  16  $&$    92 \pm 18   $&$  16   \pm 8    $&$ 18    \pm   6   $&$ 6.9  \pm 1.4  $&$   1.2   \pm 0.4    $&$ 255 \pm 14   $&$ 249$\\
High-NN-score $0b$ SR ($\ell^\pm\nu b\bar{b}$)       &$   30 \pm  12  $&$    11 \pm 5    $&$  11   \pm 11   $&$ 82    \pm  18   $&$ 1.1  \pm 0.2  $&$   9     \pm 5      $&$ 145 \pm 11   $&$ 144$\\
High-NN-score $\geq1b$ SR ($\ell^\pm\nu b\bar{b}$)   &$   30 \pm  13  $&$    19 \pm 6    $&$   8   \pm 7    $&$ 8     \pm   6   $&$ 2.1  \pm 0.3  $&$   0.62  \pm 0.25   $&$  68 \pm  8   $&$ 71$\\
\hline
\end{tabular}}}
\label{Tab:EventYieldsPostFit}
\end{center}
\end{table}

%


This search probes for charged Higgs bosons in the mass range from $250\,\GeV$ to $3\,\TeV$. In total, 17 signal
mass hypotheses are tested. The resolved and merged analyses show complementary
sensitivities to different charged Higgs boson mass regions. The resolved analysis is more sensitive for masses up to (and including)
$900\,\GeV$, while the merged analysis dominates at higher masses. Instead of performing a statistical combination of the two analyses,
the analysis with the more stringent expected upper limit on the production cross-section times branching ratio is used at
each mass point.\par

\begin{figure}[!h]
\centering
\includegraphics[width=0.725\textwidth]{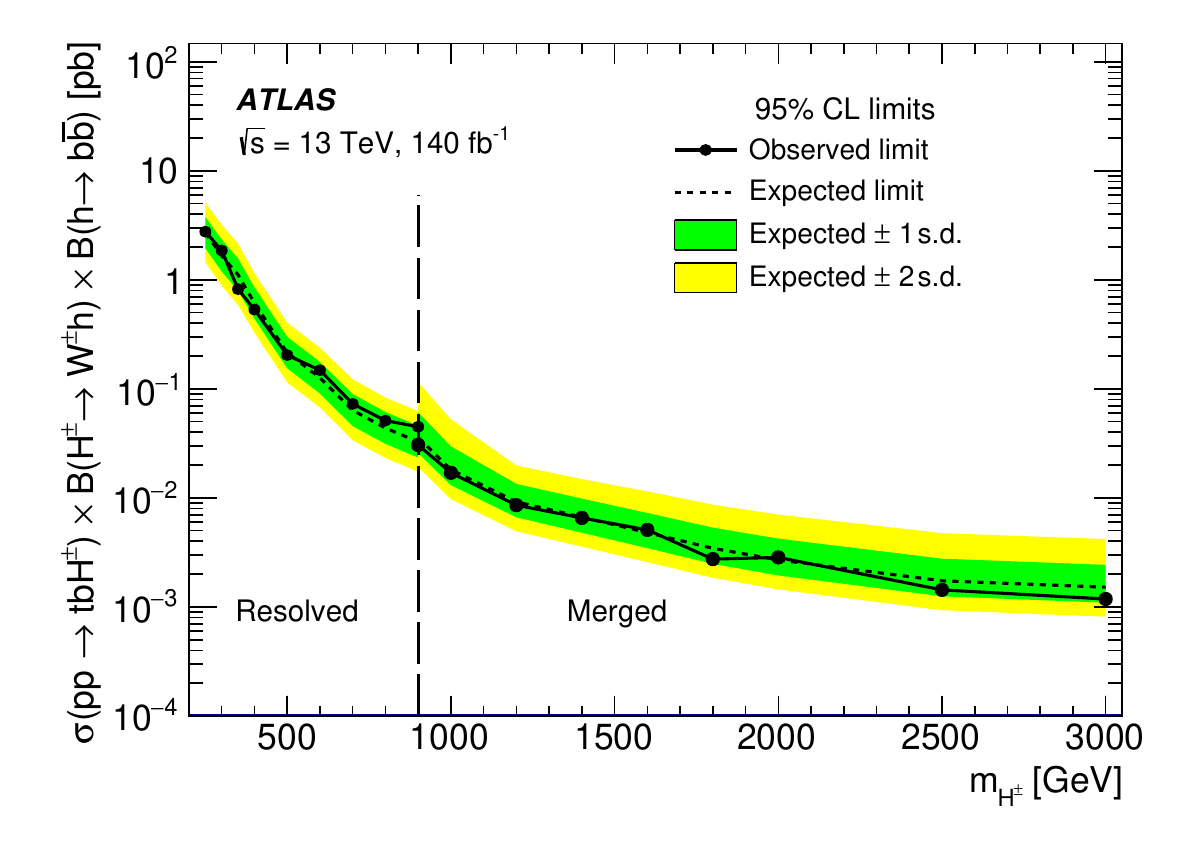}
\caption{Upper limits at $95\%$ CL on the product of the cross-section for $pp\rightarrow tb H^\pm$ and
the branching ratio $\mathcal{B}\left(W^\pm h\right) \times \mathcal{B}\left(h\rightarrow b\bar{b}\right)$ from the combined fit to all signal and
control regions of the resolved and merged analyses. The bands surrounding the expected limit correspond to the $\pm1$ and $\pm2$
standard deviation (s.d.) intervals around the expected limit. The results of the resolved analysis are used up to a mass of 900\,\GeV and those of the
merged analysis are used at higher masses.}
\label{fig:limits_Hplus}
\end{figure}

The $95\%$ CL upper limits on the production cross-section times branching ratio for a
charged Higgs boson decaying via $H^\pm\rightarrow W^\pm h$ are presented in Figure \ref{fig:limits_Hplus} as a function
of the charged Higgs boson mass. The observed limits range from $2.8\,$pb for a mass value of $250\,\GeV$ to
$1.2\,$fb for a mass of 3\,\TeV. The limits on the signal hypotheses with $m_{H^\pm} \geq 0.7\,\TeV$ were also
calculated using pseudo-experiments to validate the asymptotic approximation approach in a phase-space region
strongly limited by a low number of data events. It was found that the asymptotic approximation is valid
within $10\%$.\par

The expected exclusion limits are dominated by the $\ell^\pm\nu b\bar{b}$ channel except for the $250\,\GeV$ mass hypothesis, where the
$q\bar{q}b\bar{b}$ channel is stronger. In the resolved analysis, the $\ell^\pm\nu b\bar{b}$ channel provides a cross-section times
branching ratio limit that is 1.1 times lower at $350\,\GeV$ and 2.5 times lower at $800\,\GeV$. In the merged analysis, the
$\ell^\pm\nu b\bar{b}$ channel provides a limit that is 1.4 times lower at $900\,\GeV$ and 2.5 times lower at $3000\,\GeV$.
\clearpage

The dominant uncertainties for low charged Higgs boson masses (e.g. $m_{H^\pm}=0.4\,\TeV$) are related to the
modelling of the $t\bar{t} + \mathrm{HF}$ and $t\bar{t} + \mathrm{LF}$ backgrounds, while at high charged Higgs boson masses
(e.g. $m_{H^\pm}=2.0\,\TeV$) the dominant uncertainties are related to the data statistical uncertainty.
The relative impact of the dominant uncertainties on the best-fit signal-strength parameter $\mu$,
i.e. the ratio of the extracted to injected signal cross-section times branching ratio, are detailed in Table~\ref{tab:BreakdownHplus}.

{\def\arraystretch{1.275}
\begin{table}[!h]
\caption{
Breakdown of the relative contributions to the uncertainty in the best-fit signal-strength parameter $\mu$ of the
hypothesised production of a charged Higgs boson for two signal mass hypotheses: $m_{H^{\pm}} = 0.4\,\TeV$ and $m_{H^{\pm}} = 2.0\,\TeV$.
The contributions are obtained by fixing the relevant nuisance parameters to their post-fit values in the likelihood fit to data. The relative impact is determined
as the square-root of the difference of the squares of the nominal uncertainty and the varied uncertainty, divided by the nominal uncertainty.
The sum in quadrature of the individual components differs from the total uncertainty due to correlations between
uncertainties in the different groups. The uncertainty from data statistical uncertainties is determined from fits
with all nuisance parameters fixed to their post-fit values. The breakdown of uncertainties for the $m_{H^{\pm}} = 0.4\;\TeV$ mass point
corresponds to the resolved analysis, while for the $m_{H^{\pm}} = 2.0\;\TeV$ mass point, the breakdown of uncertainties
corresponds to the merged analysis. The signal cross-section times branching ratio is assumed to be $0.6\,$pb at $400\,\GeV$ and $2.7\,$fb at
$2\,\TeV$, corresponding to the expected upper limits for these two mass hypotheses.
\label{tab:BreakdownHplus}}
\begin{center}
\resizebox{1.0\textwidth}{!}{
\begin{tabular}{c  c | c  c }
\hline
\hline
\multicolumn{2}{c}{$m_{H^\pm} = 0.4\;\TeV$} & \multicolumn{2}{c}{$m_{H^\pm} = 2\;\TeV$}\\
\hline
Category        & Relative contribution   & Category  &  Relative contribution  \\
\hline
\multicolumn{4}{c}{Modelling uncertainties}                                      \\
\hline
$t\bar{t} + \mathrm{HF}$ modelling      & 74\%     & Extrapolation/migration                & 25\%\\
$t\bar{t} + \mathrm{LF}$ modelling      & 34\%     & $t\bar{t} + \mathrm{LF}$ modelling     & 17\%\\
Extrapolation/migration                 & 20\%     & Non-$t\bar{t}$ modelling               & 16\%\\
Non-$t\bar{t}$ modelling                &  9\%     & MC statistical uncertainty             & 12\%\\
Signal modelling                        &  5\%     & $t\bar{t} + \mathrm{HF}$ normalisation & 10\%\\
$t\bar{t} + \mathrm{HF}$ normalisation  &  5\%     & Non-closure                            &  7\%\\
Non-closure                             &  5\%     & $t\bar{t} + \mathrm{LF}$ normalisation &  6\%\\
MC statistical uncertainty              &  4\%     & $t\bar{t} + \mathrm{HF}$ modelling     &  6\%\\
&          & $VV$ modelling                         &  4\%\\
&          & Signal modelling                       &  3\%\\
&          & $VV$ normalisation                     &  1\%\\
\hline
\multicolumn{4}{c}{Experimental uncertainties}                                \\
\hline
Small-$R$ jets                &  15\%   &  $h \rightarrow b\bar{b}$ tagging & 12\%\\
Flavour tagging               &  14\%   &  Larger-$R$-jets                 &  4\%\\
Pile-up                       &   7\%   &  Small-$R$ jets                  &  4\%\\
Electrons                     &   4\%   &  Flavour tagging                 &  3\%\\
Muons                         &   1\%   &  Pile-up                         &  1\%\\
Luminosity                    &  0.3\%  &  Electrons                       &  1\%\\
Missing transverse momentum   & <0.1\%  &  Missing transverse momentum     &  0.1\%\\
&         &  Muons                           &  <0.1\%\\
&         &  Luminosity                      &  <0.1\%\\
\hline
Total systematic uncertainty  & 93\%    &   Total systematic uncertainty   & 49\% \\
Data statistical uncertainty  & 37\%    &   Data statistical uncertainty   & 87\%  \\
\hline
\hline
\end{tabular}}
\end{center}
\end{table}}
\clearpage



%

\section{Conclusion}\label{sec:conclusion}


A search for a heavy charged Higgs boson produced in association with a top quark and a bottom quark and decaying into a
$W$ boson and a 125\,\GeV Higgs boson is performed in the mass range from $250~\GeV$ to $3000~\GeV$.
This search uses $pp$ collision data at $\sqrt{s}=13\,\TeV$ collected with the ATLAS detector at the LHC from 2015 to 2018,
corresponding to an integrated luminosity of 140~fb$^{-1}$. \par

Two different analysis strategies are applied to ensure high sensitivity to both low- and high-mass charged Higgs bosons.
The mass range up to $900\,\GeV$ is probed in final states with exactly one charged electron or muon, missing transverse momentum and at
least five small-$R$ jets. Events are classified based on kinematic requirements and the multiplicity of $b$-tagged jets per event.
Finally, two sets of boosted decision trees are used to reconstruct the four-momentum of the charged Higgs boson candidate.
The mass range above $900\,\GeV$ is probed in final states with exactly one charged electron or muon,
missing transverse momentum and at least one large-$R$ jet. A recently developed boosted $h\rightarrow b\bar{b}$ tagging technique
is used to identify the decay of boosted Higgs bosons and sets of neural networks are employed to further separate between the signal
and the background processes. \par

Both analyses search for a localised excess in the distribution of the reconstructed invariant mass of the charged Higgs boson.
Neither analyses observe a significant excess of events above the SM prediction and upper limits at 95\% CL are set
on the production cross-section times branching ratio. The upper limits range from $2.8\,$pb for $m_{H^\pm} = 250~\GeV$ to $1.2~\mathrm{fb}$
for $m_{H^\pm} = 3000~\GeV$. \par

This search is performed for the first time at the LHC, complementing previous searches for $H^{\pm}\rightarrow tb$ and
$H^{\pm}\rightarrow\tau^\pm\nu$ decays by the ATLAS and CMS collaborations. While the $H^{\pm}\rightarrow W^\pm h$ decay
mode is predicted to be subdominant in common two-Higgs-doublet models, its branching ratio is predicted to be significant in other
extended scalar sector models such as the Georgi-Machacek model, or the three-Higgs-doublet model.\par


%

\section*{Acknowledgements}

%
%

%
%

We thank CERN for the very successful operation of the LHC and its injectors, as well as the support staff at
CERN and at our institutions worldwide without whom ATLAS could not be operated efficiently.

The crucial computing support from all WLCG partners is acknowledged gratefully, in particular from CERN, the ATLAS Tier-1 facilities at TRIUMF/SFU (Canada), NDGF (Denmark, Norway, Sweden), CC-IN2P3 (France), KIT/GridKA (Germany), INFN-CNAF (Italy), NL-T1 (Netherlands), PIC (Spain), RAL (UK) and BNL (USA), the Tier-2 facilities worldwide and large non-WLCG resource providers. Major contributors of computing resources are listed in Ref.~\cite{ATL-SOFT-PUB-2023-001}.

We gratefully acknowledge the support of ANPCyT, Argentina; YerPhI, Armenia; ARC, Australia; BMWFW and FWF, Austria; ANAS, Azerbaijan; CNPq and FAPESP, Brazil; NSERC, NRC and CFI, Canada; CERN; ANID, Chile; CAS, MOST and NSFC, China; Minciencias, Colombia; MEYS CR, Czech Republic; DNRF and DNSRC, Denmark; IN2P3-CNRS and CEA-DRF/IRFU, France; SRNSFG, Georgia; BMBF, HGF and MPG, Germany; GSRI, Greece; RGC and Hong Kong SAR, China; ICHEP and Academy of Sciences and Humanities, Israel; INFN, Italy; MEXT and JSPS, Japan; CNRST, Morocco; NWO, Netherlands; RCN, Norway; MNiSW, Poland; FCT, Portugal; MNE/IFA, Romania; MSTDI, Serbia; MSSR, Slovakia; ARIS and MVZI, Slovenia; DSI/NRF, South Africa; MICIU/AEI, Spain; SRC and Wallenberg Foundation, Sweden; SERI, SNSF and Cantons of Bern and Geneva, Switzerland; NSTC, Taipei; TENMAK, T\"urkiye; STFC/UKRI, United Kingdom; DOE and NSF, United States of America.

Individual groups and members have received support from BCKDF, CANARIE, CRC and DRAC, Canada; CERN-CZ, FORTE and PRIMUS, Czech Republic; COST, ERC, ERDF, Horizon 2020, ICSC-NextGenerationEU and Marie Sk{\l}odowska-Curie Actions, European Union; Investissements d'Avenir Labex, Investissements d'Avenir Idex and ANR, France; DFG and AvH Foundation, Germany; Herakleitos, Thales and Aristeia programmes co-financed by EU-ESF and the Greek NSRF, Greece; BSF-NSF and MINERVA, Israel; NCN and NAWA, Poland; La Caixa Banking Foundation, CERCA Programme Generalitat de Catalunya and PROMETEO and GenT Programmes Generalitat Valenciana, Spain; G\"{o}ran Gustafssons Stiftelse, Sweden; The Royal Society and Leverhulme Trust, United Kingdom.

In addition, individual members wish to acknowledge support from Armenia: Yerevan Physics Institute (FAPERJ); CERN: European Organization for Nuclear Research (CERN DOCT); Chile: Agencia Nacional de Investigaci\'on y Desarrollo (FONDECYT 1230812, FONDECYT 1230987, FONDECYT 1240864); China: Chinese Ministry of Science and Technology (MOST-2023YFA1605700, MOST-2023YFA1609300), National Natural Science Foundation of China (NSFC - 12175119, NSFC 12275265, NSFC-12075060); Czech Republic: Czech Science Foundation (GACR - 24-11373S), Ministry of Education Youth and Sports (FORTE CZ.02.01.01/00/22\_008/0004632), PRIMUS Research Programme (PRIMUS/21/SCI/017); EU: H2020 European Research Council (ERC - 101002463); European Union: European Research Council (ERC - 948254, ERC 101089007, ERC, BARD, 101116429), European Union, Future Artificial Intelligence Research (FAIR-NextGenerationEU PE00000013), Italian Center for High Performance Computing, Big Data and Quantum Computing (ICSC, NextGenerationEU); France: Agence Nationale de la Recherche (ANR-20-CE31-0013, ANR-21-CE31-0013, ANR-21-CE31-0022, ANR-22-EDIR-0002); Germany: Baden-Württemberg Stiftung (BW Stiftung-Postdoc Eliteprogramme), Deutsche Forschungsgemeinschaft (DFG - 469666862, DFG - CR 312/5-2); Italy: Istituto Nazionale di Fisica Nucleare (ICSC, NextGenerationEU), Ministero dell'Università e della Ricerca (PRIN - 20223N7F8K - PNRR M4.C2.1.1); Japan: Japan Society for the Promotion of Science (JSPS KAKENHI JP22H01227, JSPS KAKENHI JP22H04944, JSPS KAKENHI JP22KK0227, JSPS KAKENHI JP23KK0245); Norway: Research Council of Norway (RCN-314472); Poland: Ministry of Science and Higher Education (IDUB AGH, POB8, D4 no 9722), Polish National Agency for Academic Exchange (PPN/PPO/2020/1/00002/U/00001), Polish National Science Centre (NCN 2021/42/E/ST2/00350, NCN OPUS 2023/51/B/ST2/02507, NCN OPUS nr 2022/47/B/ST2/03059, NCN UMO-2019/34/E/ST2/00393, NCN \& H2020 MSCA 945339, UMO-2020/37/B/ST2/01043, UMO-2021/40/C/ST2/00187, UMO-2022/47/O/ST2/00148, UMO-2023/49/B/ST2/04085, UMO-2023/51/B/ST2/00920); Spain: Generalitat Valenciana (Artemisa, FEDER, IDIFEDER/2018/048), Ministry of Science and Innovation (MCIN \& NextGenEU PCI2022-135018-2, MICIN \& FEDER PID2021-125273NB, RYC2019-028510-I, RYC2020-030254-I, RYC2021-031273-I, RYC2022-038164-I); Sweden: Carl Trygger Foundation (Carl Trygger Foundation CTS 22:2312), Swedish Research Council (Swedish Research Council 2023-04654, VR 2018-00482, VR 2022-03845, VR 2022-04683, VR 2023-03403, VR grant 2021-03651), Knut and Alice Wallenberg Foundation (KAW 2018.0458, KAW 2019.0447, KAW 2022.0358); Switzerland: Swiss National Science Foundation (SNSF - PCEFP2\_194658); United Kingdom: Leverhulme Trust (Leverhulme Trust RPG-2020-004), Royal Society (NIF-R1-231091); United States of America: U.S. Department of Energy (ECA DE-AC02-76SF00515), Neubauer Family Foundation.

%
%


%
\clearpage
\printbibliography
\clearpage
\input{atlas_authlist}
%

%
%

%

%
%
%
%

\end{document}

%% file: atlas_authlist.tex
 
\begin{flushleft}
\hypersetup{urlcolor=black}
{\Large The ATLAS Collaboration}

\bigskip

\AtlasOrcid[0000-0002-6665-4934]{G.~Aad}$^\textrm{\scriptsize 104}$,
\AtlasOrcid[0000-0001-7616-1554]{E.~Aakvaag}$^\textrm{\scriptsize 17}$,
\AtlasOrcid[0000-0002-5888-2734]{B.~Abbott}$^\textrm{\scriptsize 123}$,
\AtlasOrcid[0000-0002-0287-5869]{S.~Abdelhameed}$^\textrm{\scriptsize 119a}$,
\AtlasOrcid[0000-0002-1002-1652]{K.~Abeling}$^\textrm{\scriptsize 56}$,
\AtlasOrcid[0000-0001-5763-2760]{N.J.~Abicht}$^\textrm{\scriptsize 50}$,
\AtlasOrcid[0000-0002-8496-9294]{S.H.~Abidi}$^\textrm{\scriptsize 30}$,
\AtlasOrcid[0009-0003-6578-220X]{M.~Aboelela}$^\textrm{\scriptsize 46}$,
\AtlasOrcid[0000-0002-9987-2292]{A.~Aboulhorma}$^\textrm{\scriptsize 36e}$,
\AtlasOrcid[0000-0001-5329-6640]{H.~Abramowicz}$^\textrm{\scriptsize 155}$
% ...
\bigskip
\\

% ...
$^{17}$Department for Physics and Technology, University of Bergen, Bergen; Norway.\\
% ...
$^{30}$Physics Department, Brookhaven National Laboratory, Upton NY; United States of America.\\
% ...
$^{36}$$^{(a)}$Facult\'e des Sciences Ain Chock, Universit\'e Hassan II de Casablanca;$^{(b)}$Facult\'{e} des Sciences, Universit\'{e} Ibn-Tofail, K\'{e}nitra;$^{(c)}$Facult\'e des Sciences Semlalia, Universit\'e Cadi Ayyad, LPHEA-Marrakech;$^{(d)}$LPMR, Facult\'e des Sciences, Universit\'e Mohamed Premier, Oujda;$^{(e)}$Facult\'e des sciences, Universit\'e Mohammed V, Rabat;$^{(f)}$Institute of Applied Physics, Mohammed VI Polytechnic University, Ben Guerir; Morocco.\\
% ...
$^{46}$Physics Department, Southern Methodist University, Dallas TX; United States of America.\\
% ...
$^{50}$Fakult\"{a}t Physik , Technische Universit{\"a}t Dortmund, Dortmund; Germany.\\
% ...
$^{56}$II. Physikalisches Institut, Georg-August-Universit\"{a}t G\"ottingen, G\"ottingen; Germany.\\
% ...
$^{104}$CPPM, Aix-Marseille Universit\'e, CNRS/IN2P3, Marseille; France.\\
% ...
$^{119}$$^{(a)}$New York University Abu Dhabi, Abu Dhabi;$^{(b)}$United Arab Emirates University, Al Ain; United Arab Emirates.\\
% ...
$^{123}$Homer L. Dodge Department of Physics and Astronomy, University of Oklahoma, Norman OK; United States of America.\\
% ...
$^{155}$Raymond and Beverly Sackler School of Physics and Astronomy, Tel Aviv University, Tel Aviv; Israel.\\
% ...
$^{a}$ Also Affiliated with an institute covered by a cooperation agreement with CERN.\\
$^{b}$ Also at An-Najah National University, Nablus; Palestine.\\
$^{c}$ Also at Borough of Manhattan Community College, City University of New York, New York NY; United States of America.\\
$^{d}$ Also at Center for High Energy Physics, Peking University; China.\\
$^{e}$ Also at Center for Interdisciplinary Research and Innovation (CIRI-AUTH), Thessaloniki; Greece.\\
$^{f}$ Also at CERN, Geneva; Switzerland.\\
% ...

\end{flushleft}
